\documentclass[prd,amsmath,amssymb,superscriptaddress,showpacs,preprint]{revtex4}
\usepackage{graphicx}% Include figure files
\usepackage{dcolumn}% Align table columns on decimal point
\usepackage{bm}% bold math
\usepackage{lineno}
\usepackage{multirow}
\newcommand{\inmath}[1] {\ifmmode#1\else$#1$\fi}
\newcommand{\definmath}[2] {\def#1{\ifmmode#2\else$#2$\fi}}

\newcommand{\ppbar}{p\bar{p}}

\newcommand{\ttbar}{t\bar{t}}
\newcommand{\bbbar}{b\bar{b}}
\newcommand{\ccbar}{c\bar{c}}
\newcommand{\jpsi}{J/\psi}
\newcommand{\dstar}{\mbox{$D^{*}$}}

\newcommand{\dzero}{\mbox{$D^{0}$}}
\newcommand{\lambdaz}{\mbox{$\Lambda^{0}$}}
\newcommand{\tptm}{\mbox{$\tau^+\tau^-$}}

\newcommand{\GeVc}{\mbox{$\mathrm{GeV}\!/\!c$}}
\newcommand{\GeVcc}{\mbox{$\mathrm{GeV}\!/\!c^2$}}
\newcommand{\MeVcc}{\mbox{$\mathrm{MeV}\!/\!c^2$}}
\newcommand{\GeV}{\mbox{$\mathrm{GeV}$}}
\newcommand{\TeV}{\mbox{$\mathrm{TeV}$}}
\definmath{\Pt}      {P_{\mathrm T}}
\definmath{\Et}      {E_{\mathrm T}}
\definmath{\Ht}      {H_{\mathrm T}}
\newcommand{\met}{E\!\!\!/_T }
\newcommand{\sltmu}{\mbox{$\mathrm{SLT}\mu$}}
\newcommand{\Pythia}{\mbox{{\sc pythia}}}
\newcommand{\Alpgen}{\mbox{{\sc alpgen}}}
\newcommand{\Evtgen}{\mbox{{\sc evtgen}}}
\newcommand{\Madevent}{\mbox{{\sc madevent}}}
\newcommand{\Herwig}{\mbox{{\sc herwig}}}
\newfont{\sans}{cmr8 at 10pt}

\begin{document}
%\pagewiselinenumbers

\title{Measurement of the $\ttbar$ Production Cross Section in 2~fb$^{-1}$ of $\ppbar$ Collisions at $\sqrt{s}=1.96$ TeV
Using Lepton Plus Jets Events with Soft Muon b-Tagging}
\vspace*{2.0cm}

% \author{The CDF Collaboration}
\affiliation{Institute of Physics, Academia Sinica, Taipei, Taiwan
11529, Republic of China} \affiliation{Argonne National Laboratory,
Argonne, Illinois 60439} \affiliation{University of Athens, 157 71
Athens, Greece} \affiliation{Institut de Fisica d'Altes Energies,
Universitat Autonoma de Barcelona, E-08193, Bellaterra (Barcelona),
Spain} \affiliation{Baylor University, Waco, Texas  76798}
\affiliation{Istituto Nazionale di Fisica Nucleare Bologna,
$^x$University of Bologna, I-40127 Bologna, Italy}
\affiliation{Brandeis University, Waltham, Massachusetts 02254}
\affiliation{University of California, Davis, Davis, California
95616} \affiliation{University of California, Los Angeles, Los
Angeles, California  90024} \affiliation{University of California,
San Diego, La Jolla, California  92093} \affiliation{University of
California, Santa Barbara, Santa Barbara, California 93106}
\affiliation{Instituto de Fisica de Cantabria, CSIC-University of
Cantabria, 39005 Santander, Spain} \affiliation{Carnegie Mellon
University, Pittsburgh, PA  15213} \affiliation{Enrico Fermi
Institute, University of Chicago, Chicago, Illinois 60637}
\affiliation{Comenius University, 842 48 Bratislava, Slovakia;
Institute of Experimental Physics, 040 01 Kosice, Slovakia}
\affiliation{Joint Institute for Nuclear Research, RU-141980 Dubna,
Russia} \affiliation{Duke University, Durham, North Carolina  27708}
\affiliation{Fermi National Accelerator Laboratory, Batavia,
Illinois 60510} \affiliation{University of Florida, Gainesville,
Florida  32611} \affiliation{Laboratori Nazionali di Frascati,
Istituto Nazionale di Fisica Nucleare, I-00044 Frascati, Italy}
\affiliation{University of Geneva, CH-1211 Geneva 4, Switzerland}
\affiliation{Glasgow University, Glasgow G12 8QQ, United Kingdom}
\affiliation{Harvard University, Cambridge, Massachusetts 02138}
\affiliation{Division of High Energy Physics, Department of Physics,
University of Helsinki and Helsinki Institute of Physics, FIN-00014,
Helsinki, Finland} \affiliation{University of Illinois, Urbana,
Illinois 61801} \affiliation{The Johns Hopkins University,
Baltimore, Maryland 21218} \affiliation{Institut f\"{u}r
Experimentelle Kernphysik, Universit\"{a}t Karlsruhe, 76128
Karlsruhe, Germany} \affiliation{Center for High Energy Physics:
Kyungpook National University, Daegu 702-701, Korea; Seoul National
University, Seoul 151-742, Korea; Sungkyunkwan University, Suwon
440-746, Korea; Korea Institute of Science and Technology
Information, Daejeon, 305-806, Korea; Chonnam National University,
Gwangju, 500-757, Korea} \affiliation{Ernest Orlando Lawrence
Berkeley National Laboratory, Berkeley, California 94720}
\affiliation{University of Liverpool, Liverpool L69 7ZE, United
Kingdom} \affiliation{University College London, London WC1E 6BT,
United Kingdom} \affiliation{Centro de Investigaciones Energeticas
Medioambientales y Tecnologicas, E-28040 Madrid, Spain}
\affiliation{Massachusetts Institute of Technology, Cambridge,
Massachusetts  02139} \affiliation{Institute of Particle Physics:
McGill University, Montr\'{e}al, Qu\'{e}bec, Canada H3A~2T8; Simon
Fraser University, Burnaby, British Columbia, Canada V5A~1S6;
University of Toronto, Toronto, Ontario, Canada M5S~1A7; and TRIUMF,
Vancouver, British Columbia, Canada V6T~2A3} \affiliation{University
of Michigan, Ann Arbor, Michigan 48109} \affiliation{Michigan State
University, East Lansing, Michigan  48824} \affiliation{Institution
for Theoretical and Experimental Physics, ITEP, Moscow 117259,
Russia} \affiliation{University of New Mexico, Albuquerque, New
Mexico 87131} \affiliation{Northwestern University, Evanston,
Illinois  60208} \affiliation{The Ohio State University, Columbus,
Ohio  43210} \affiliation{Okayama University, Okayama 700-8530,
Japan} \affiliation{Osaka City University, Osaka 588, Japan}
\affiliation{University of Oxford, Oxford OX1 3RH, United Kingdom}
\affiliation{Istituto Nazionale di Fisica Nucleare, Sezione di
Padova-Trento, $^y$University of Padova, I-35131 Padova, Italy}
\affiliation{LPNHE, Universite Pierre et Marie Curie/IN2P3-CNRS,
UMR7585, Paris, F-75252 France} \affiliation{University of
Pennsylvania, Philadelphia, Pennsylvania 19104}
\affiliation{Istituto Nazionale di Fisica Nucleare Pisa,
$^z$University of Pisa, $^{aa}$University of Siena and $^{bb}$Scuola
Normale Superiore, I-56127 Pisa, Italy} \affiliation{University of
Pittsburgh, Pittsburgh, Pennsylvania 15260} \affiliation{Purdue
University, West Lafayette, Indiana 47907} \affiliation{University
of Rochester, Rochester, New York 14627} \affiliation{The
Rockefeller University, New York, New York 10021}
\affiliation{Istituto Nazionale di Fisica Nucleare, Sezione di Roma
1, $^{cc}$Sapienza Universit\`{a} di Roma, I-00185 Roma, Italy}

\affiliation{Rutgers University, Piscataway, New Jersey 08855}
\affiliation{Texas A\&M University, College Station, Texas 77843}
\affiliation{Istituto Nazionale di Fisica Nucleare Trieste/Udine,
I-34100 Trieste, $^{dd}$University of Trieste/Udine, I-33100 Udine,
Italy} \affiliation{University of Tsukuba, Tsukuba, Ibaraki 305,
Japan} \affiliation{Tufts University, Medford, Massachusetts 02155}
\affiliation{Waseda University, Tokyo 169, Japan} \affiliation{Wayne
State University, Detroit, Michigan  48201} \affiliation{University
of Wisconsin, Madison, Wisconsin 53706} \affiliation{Yale
University, New Haven, Connecticut 06520}
\author{T.~Aaltonen}
\affiliation{Division of High Energy Physics, Department of Physics,
University of Helsinki and Helsinki Institute of Physics, FIN-00014,
Helsinki, Finland}
\author{J.~Adelman}
\affiliation{Enrico Fermi Institute, University of Chicago, Chicago,
Illinois 60637}
\author{T.~Akimoto}
\affiliation{University of Tsukuba, Tsukuba, Ibaraki 305, Japan}
\author{B.~\'{A}lvarez~Gonz\'{a}lez$^s$}
\affiliation{Instituto de Fisica de Cantabria, CSIC-University of
Cantabria, 39005 Santander, Spain}
\author{S.~Amerio$^y$}
\affiliation{Istituto Nazionale di Fisica Nucleare, Sezione di
Padova-Trento, $^y$University of Padova, I-35131 Padova, Italy}

\author{D.~Amidei}
\affiliation{University of Michigan, Ann Arbor, Michigan 48109}
\author{A.~Anastassov}
\affiliation{Northwestern University, Evanston, Illinois  60208}
\author{A.~Annovi}
\affiliation{Laboratori Nazionali di Frascati, Istituto Nazionale di
Fisica Nucleare, I-00044 Frascati, Italy}
\author{J.~Antos}
\affiliation{Comenius University, 842 48 Bratislava, Slovakia;
Institute of Experimental Physics, 040 01 Kosice, Slovakia}
\author{G.~Apollinari}
\affiliation{Fermi National Accelerator Laboratory, Batavia,
Illinois 60510}
% ...

\collaboration{CDF Collaboration\footnote{With visitors from
$^a$University of Massachusetts Amherst, Amherst, Massachusetts
01003, $^b$Universiteit Antwerpen, B-2610 Antwerp, Belgium,
$^c$University of Bristol, Bristol BS8 1TL, United Kingdom,
$^d$Chinese Academy of Sciences, Beijing 100864, China, $^e$Istituto
Nazionale di Fisica Nucleare, Sezione di Cagliari, 09042 Monserrato
(Cagliari), Italy, $^f$University of California Irvine, Irvine, CA
92697, $^g$University of California Santa Cruz, Santa Cruz, CA
95064, $^h$Cornell University, Ithaca, NY  14853, $^i$University of
Cyprus, Nicosia CY-1678, Cyprus, $^j$University College Dublin,
Dublin 4, Ireland, $^k$University of Edinburgh, Edinburgh EH9 3JZ,
United Kingdom, $^l$University of Fukui, Fukui City, Fukui
Prefecture, Japan 910-0017 $^m$Kinki University, Higashi-Osaka City,
Japan 577-8502 $^n$Universidad Iberoamericana, Mexico D.F., Mexico,
$^o$Queen Mary, University of London, London, E1 4NS, England,
$^p$University of Manchester, Manchester M13 9PL, England,
$^q$Nagasaki Institute of Applied Science, Nagasaki, Japan,
$^r$University of Notre Dame, Notre Dame, IN 46556, $^s$University
de Oviedo, E-33007 Oviedo, Spain, $^t$Texas Tech University,
Lubbock, TX  79609, $^u$IFIC(CSIC-Universitat de Valencia), 46071
Valencia, Spain, $^v$University of Virginia, Charlottesville, VA
22904, $^w$Bergische Universit\"at Wuppertal, 42097 Wuppertal,
Germany, $^{ee}$On leave from J.~Stefan Institute, Ljubljana,
Slovenia, }} \noaffiliation

%\affiliation{URL
%http://www-cdf.fnal.gov/physics/new/top/public/ljets/slt/public\_200.html}

\date{\today}

\begin{abstract}
% remove the space for publication
\vspace*{3.0cm}

We present a measurement of the $\ttbar$ production cross section in
$\ppbar$ collisions at $\sqrt{s}=1.96$ TeV using events containing a
high transverse momentum electron or muon, three or more jets, and
missing transverse energy. Events consistent with $\ttbar$ decay are
found by identifying jets containing candidate heavy-flavor
semileptonic decays to muons.
%Backgrounds are computed from a combination of Run
%II data and simulation. Signal acceptance is determined from Run II
%$\Pythia$ Monte Carlo.
The measurement uses a CDF Run II data sample corresponding to
$2\,\mathrm{fb^{-1}}$ of integrated luminosity.  Based on 248
candidate events with three or more jets and an expected background
of $79.5\pm5.3$ events, we measure a production cross section of
$9.1\pm 1.6 \,\mathrm{pb}$.
\end{abstract}

% activate the following line for publication

\maketitle

\newpage
\section{\label{sec:Intro}Introduction}

Top quark pair production in hadronic collisions in the standard
model proceeds via either quark-antiquark annihilation or through
gluon-gluon fusion. At the Fermilab Tevatron collider, with a
center-of-mass energy of 1.96 TeV, the production is expected to be
dominated by quark-antiquark annihilation.  For a top mass of 175
$\GeVcc$ the theoretical cross section is calculated to be $6.6\pm
0.6$~pb~\cite{theory} and decreases by approximately 0.2~pb for each
1~$\GeVcc$ increase in the top mass over the range $170~\GeVcc
<$M$_{top}<190~\GeVcc$.

Measurements of the cross section for top quark pair production
provide a test of the expected QCD production mechanism as well as
of the standard model decay into a $W$-boson and a bottom quark,
$t\rightarrow Wb$.  Non-standard model production mechanisms could
enhance the measured cross section, and non-standard model decays
could suppress the measured value, which assumes a branching
fraction of $t\rightarrow Wb$ of 100\%.

In this paper we describe a measurement of the $\ttbar$ production
cross section in $\ppbar$ collisions at $\sqrt{s}=1.96$ TeV with the
CDF II detector at the Fermilab Tevatron.  The measurement assumes
the standard model decay $t\rightarrow Wb$ of the top quark,
providing a final state from $\ttbar$ production that includes two
$W$ bosons and two bottom quarks.  We select events where one of the
$W$ bosons decays to an electron or muon which has large momentum
transverse to the beam direction ($\Pt$) plus a neutrino. The
neutrino is undetected and results in an imbalance in transverse
momentum.  This imbalance is labeled ``missing $\Et$" ($\met$)
because it is reconstructed based on the flow of energy in the
calorimeter~\cite{MET}.  The other $W$ boson in the event decays
hadronically to a pair of quarks.  The two quarks from the $W$
boson, and the two $b$ quarks from the top decays, hadronize and are
observed as jets of charged and neutral particles.  We take
advantage of the semileptonic decay of $b$-hadrons to muons to
identify final-state jets that result from hadronization of the
bottom quarks expected in the top decay. This ``soft-lepton tagging"
with muons, or $\sltmu$, is effective in reducing the background to
the $\ttbar$ signal from $W$ plus multijet production.  This
technique is complementary to measurements that take advantage of
the long lifetime of $b$-hadrons to identify jets from bottom quark
hadronization through the presence of a decay vertex displaced from
the primary interaction~\cite{SecVtxPRD}.

This measurement is an update of the measurement described
in~\cite{SLTPRD}, which was made with approximately one tenth of the
integrated luminosity used here. Full details of this analysis are
presented in~\cite{UlyThesis}. In addition to the larger dataset, we
report here on a new method for evaluating the background from
``mistags", i.e. those $\sltmu$s that do not arise from
semi-leptonic decays of heavy-flavor (HF) quarks.  This is described
in Section~\ref{sec:FakeMatrix}.

\section{The CDF Detector}
\begin{figure*}[htbp]
\begin{center}
\includegraphics[width=4in]{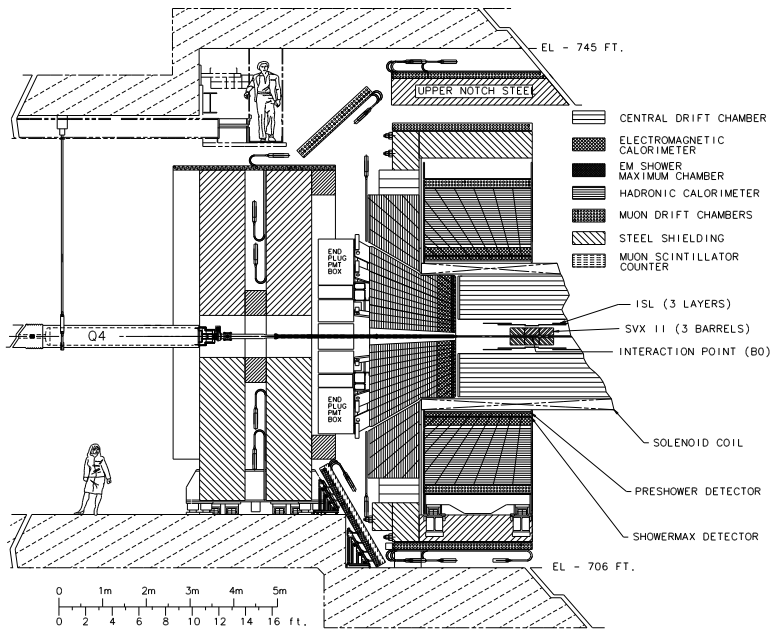}
\caption{Elevation view of the CDF II detector.} \label{fig:cdfel}
\end{center}
\end{figure*}

The CDF II detector is described in detail in~\cite{CDF}. We
describe briefly here those elements of the detector that are
central to this analysis. CDF II is a nearly azimuthally and
forward-backward symmetric detector designed to study $\ppbar$
interactions at the Fermilab Tevatron.  It consists of a magnetic
spectrometer surrounded by calorimeters and muon chambers. An
elevation view of the CDF~II detector is shown in
Fig.~\ref{fig:cdfel}.

%%%%%%%%%%%%%%%%%
%%%%%%%%%%%%%%%%%%

Charged particles are tracked inside a 1.4~T solenoidal magnetic
field by an 8-layer silicon strip detector, covering radii from
1.5~cm to 28~cm, followed by the central outer tracker (COT), an
open-cell drift chamber that provides up to 96 measurements of
charged particle position over the radial region from 40~cm to
137~cm. The 96 COT measurements are arranged in 8 ``superlayers" of
12 sense wires each, that alternate between axial and 2$^\circ$
stereo orientations.  The silicon detector tracks charged particles
with high efficiency for $\mid\!\eta\!\mid<$2.0, and the COT for
$\mid\!\eta\!\mid<$1.0~\cite{MET}.

Surrounding the tracking system, and outside the magnet coil, are
the electromagnetic and hadronic calorimeters, used to measure
charged and neutral particle energies. The electromagnetic
calorimeter is a lead-scintillator sandwich and the hadronic
calorimeter is an iron-scintillator sandwich.  Both calorimeters are
segmented in azimuth and polar angle to provide directional
information for the energy deposition. The segmentation varies with
position on the detector and is 15$^\circ$ in azimuth by 0.1 units
of $\eta$ in the central region ($\mid\!\eta\!\mid<1.1$).
Segmentation in the plug region ($1.1<\mid\!\eta\!\mid<3.6$) is
7.5$^\circ$ ($\mid\!\eta\!\mid<2.1$) or 15$^\circ$
($\mid\!\eta\!\mid >2.1$)~in azimuth and ranges from 0.1 to 0.64
units of $\eta$ (corresponding to a nearly constant 2.7$^\circ$
change in polar angle).  The electromagnetic calorimeters are
instrumented with proportional and scintillating strip detectors
that measure the transverse profile of electromagnetic showers at a
depth corresponding to the shower maximum.

Behind the central calorimeter are four layers of central muon drift
chambers covering $\mid\!\eta\!\mid<0.6$ (CMU).  The calorimeter
provides approximately one meter of steel shielding. Behind an
additional 60~cm of steel in the central region sit an additional
four layers of muon drift chambers (CMP) arranged in a box-shaped
layout around the central detector. Central muon extension (CMX)
chambers, which are arrayed in a conical geometry, provide muon
detection for the region $0.6<\mid\!\eta\!\mid<1.0$ with between
four and six layers of drift chamber, depending on zenith angle. The
CMX chambers covering from 225$^\circ$ to 315$^\circ$ in azimuth are
known as the `miniskirt' while those covering from 75$^\circ$ to
105$^\circ$ in azimuth are known as the `keystone'. The remainder of
the CMX chambers are referred to as the `arches'. The muon chambers
measure the coordinate of hits in the drift direction, $x$, via a
drift time measurement and a calibrated drift velocity, and for CMU
and CMX, the longitudinal coordinate, $z$. The longitudinal
coordinate is measured in CMU by comparing the pulse heights,
encoded in time-over-threshold, of pulses at opposite ends of the
sense wire. In CMX, the conical geometry provides a small stereo
angle from which the $z$ coordinate of track segments can be
measured. Reconstructed track segments in CMU and CMP have a maximum
of 4 hits, and in CMX a maximum of 6 hits.

\section{\label{sec:Data}Data Sample and Event Selection}

%\subsection{\label{sec:CollBeam}${\boldmath \ppbar}$ Collision Data}
This analysis is based on an integrated luminosity of
$2034\pm120$~pb$^{-1}$~\cite{klimenko} (1993~pb$^{-1}$ with the CMX
detector operational) collected with the CDF II detector between
March 2002 and May 2007.

\subsection{\label{sec:EvSel}Kinematic Selection} The triggering
and offline event selection used in this analysis are nearly
identical to that used in the previous analysis described
in~\cite{SLTPRD}.  For completeness we reproduce the basic trigger
and selection criteria here and highlight the few differences.

CDF II employs a three level trigger system, the first two
consisting of special purpose hardware and the third consisting of a
farm of commodity computers.  Triggers for this analysis are based
on selecting high transverse momentum electrons and muons.  The
electron sample is triggered as follows:  At the first trigger
level, events are selected by requiring a track with $\Pt>8~\GeVc$
matched to an electromagnetic calorimeter tower with $\Et>8$~GeV and
little energy in the hadronic calorimeter behind it.  At the second
trigger level, calorimeter energy clusters are assembled, and the
track found at the first level must be matched to an electromagnetic
cluster with $\Et>16$~GeV.  At the third level, offline
reconstruction is performed and an electron candidate with
$\Et>18$~GeV is required. The muon sample trigger begins at the
first trigger level with a track with $\Pt>4~\GeVc$ matched to hits
in the CMU and CMP chambers or a track with $\Pt>8~\GeVc$ matched to
hits in the CMX chambers. At the second level a track with
$\Pt>8~\GeVc$ is required in the event for all but the first few
percent of the integrated luminosity, for which triggers at the
first level were fed directly to the third level trigger.  At the
third trigger level a reconstructed track with $\Pt>18~\GeVc$ is
required to be matched to the muon chamber hits.

From the inclusive lepton dataset produced by the electron and muon
triggers described above, we select offline an inclusive $W$ plus
jets candidate sample by requiring a reconstructed isolated electron
with $\Et>20$~GeV or muon with $\Pt>20~\GeVc$, $\met
>30$~GeV and at least 1 jet with $\Et>20$~GeV and
$\mid\!\eta\!\mid<2.0$. We define an isolation parameter, $I$, as
the calorimeter energy in a cone of $\Delta
R\equiv\sqrt{\Delta\eta^2+\Delta\phi^2}<0.4$ around the lepton (not
including the lepton energy itself) divided by the $\Et$ ($\Pt$) of
the lepton. We select isolated electrons (muons) by requiring
$I<0.1$.   Electrons and muons satisfying these criteria are called
the ``primary lepton". Jets are identified using a fixed-cone
algorithm with a cone size of $\Delta R=0.4$ and are constrained to
originate at the $\ppbar$ collision vertex. Their energies are
corrected to account for detector response variations in $\eta$,
drifts in calorimeter gain, nonlinearity of calorimeter energy
response, multiple interactions in an event and for energy loss in
un-instrumented regions of the detector.  These corrections bring
the jet energies, on average, back to the sum $\Pt$ of the particles
in the jet cone, but not all the way back to the parton energy. This
is slightly different from the previous analysis~\cite{SLTPRD} where
the correction was done only for response variations in $\eta$, gain
drifts and multiple interactions. The jet counting threshold in that
analysis was $\Et
>15$~GeV, which corresponds roughly to the $\Et>20$~GeV used
here with the additional corrections.  The missing transverse energy
is corrected to account for the shifts in jet energies due to the
jet corrections above, and the $\met$ threshold has been raised from
20 GeV in the previous analysis to 30 GeV here, consistent with the
change in jet corrections. $Z$ boson candidate events are rejected
by removing events in which a second, same flavor, opposite sign
isolated lepton, together with the primary lepton, makes an
invariant mass between 76 $\GeVcc$ and 106~$\GeVcc$. The acceptance
of these selection criteria for $\ttbar$ events is discussed in
Section~\ref{sec:AccEff} below.

The $\ttbar$ signal region consists of $W$ candidate events with 3
or more jets, while the $W$+1 and $W$+2 jet events provide a control
sample with little signal contamination.

\begin{figure}[htbp]
\begin{center}
\includegraphics[width=3.375in]{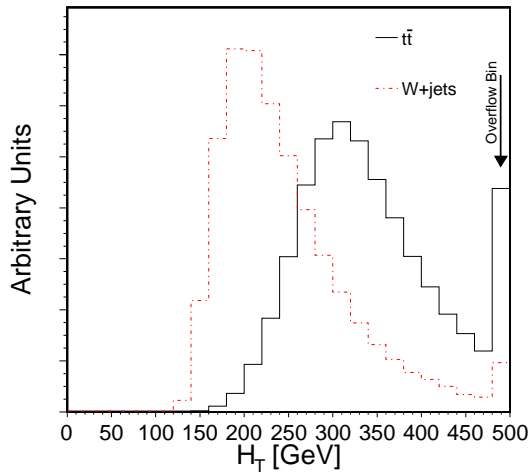}
\caption{The distribution of $\Ht$ for simulated $\ttbar$ and $W+$
jets events with at least three jets.} \label{fig:ht}
\end{center}
\end{figure}

The dataset selected above is dominated by QCD production of $W$
bosons with multiple jets.  As a first stage of background
reduction, we define a total event transverse energy, $\Ht$, as the
scalar sum of the electron $\Et$ or muon $\Pt$,  $\met$ and jet
$\Et$ for jets with $\Et>8$~GeV and $\mid\!\eta\!\mid<2.4$.
Figure~\ref{fig:ht} shows a comparison between the $\Ht$
distributions for simulated $\ttbar$ and $W+$ jets events with at
least 3 jets.  For 3 or more jet events, we require $\Ht>200$~GeV.
This requirement rejects approximately 30\% of the background while
retaining approximately 99\% of the $\ttbar$ signal.  No $\Ht$
requirement is made for the control region of 1- and 2-jet events.
%%%%%%%%%%%%%%%%%
%%%%%%%%%%%%%%%%%%

\subsection{\label{sec:SLT}Muon Tagging} Even after the $\Ht$
requirement is imposed, the expected signal to background ratio in
$W+\ge3$ jet events is about 1:7. To further improve the signal to
background ratio, events with one or more $b$-jets are identified by
searching inside jets for semileptonic decays of $b$-hadrons into
muons. The algorithm for identifying such candidate $b$-jets is
known as ``soft lepton tagging" or ``$\sltmu$" and a jet with a
candidate semileptonic $b$ decay to a muon is a ``tagged" jet.  The
$\sltmu$ algorithm is described in detail in
Reference~\cite{SLTPRD}.  We review here only its basic features.

%In general, muon identification at CDF relies on the presence of a
%track segment~(``stub") in the muon chambers, matched to a track in
%the central tracking systems and energy deposition in the
%calorimeters consistent with minimum ionization.  For the $\sltmu$,
%we are interested in identifying muons in jets and therefore
%selected candidate muons based solely on track-based quantities and
%without any requirement based on calorimeter energy.  Muons from
%semileptonic $b$ decay in $\ttbar$ range from a few GeV/c to more
%than 50 GeV/c in $\Pt$.  The $\sltmu$ algorithm must therefore be
%efficient in selecting muons over a broad $\Pt$ range, and at the
%same time have good rejection of `fake' tags from hadronic
%punch-through and meson decays-in-flight.

Muon identification at CDF relies on the presence of a track segment
(``stub") in the muon chambers, matched to a track in the central
tracking system. The soft muon tagger is based on a
$\chi^2$~function that uses all available information about the
match between the extrapolated COT track and the muon stub to
require that the deviations be consistent with the multiple Coulomb
scattering expected for a muon traversing the CDF calorimeter.

The algorithm begins by selecting ``taggable" tracks.  A track is
declared taggable if it contains at least 3 axial and 2 stereo COT
superlayers that have at least 5 hits each.  To obtain some
rejection for decays-in-flight (DIF), the impact parameter, $d_0$,
of the track with respect to the beamline, is required to be less
than 2~mm. The track is further required to originate within 60~cm
of the center of the detector along the beam direction.  Finally,
the track must have a $\Pt$ above an approximate range-out threshold
of 3.0 $\GeVc$ and extrapolate to within a fiducial volume at the
muon chambers that extends 3$\sigma_{MS}$ outside of the physical
edges of the chambers, where $\sigma_{MS}$ is the deviation expected
from multiple Coulomb scattering at the track $\Pt$.

Matching between the extrapolated COT track and the muon stub is
done using the following observables (``matching variables"): The
extrapolated position along the muon chamber drift direction~($x$),
the longitudinal coordinate along the chamber wires~($z$) when such
information is available, and the extrapolated slope~($\phi_L$).
Tracks in the COT are paired with stubs based on the best match
in~$x$, which must be less than 50~cm for a track-stub pair to
become a muon candidate.  We refer to the difference between the
extrapolated and measured positions in~$x$ and~$z$ as~$\Delta x$
and~$\Delta z$, respectively, and between the extrapolated and
measured slope as~$\Delta\phi_L$. The distributions of these
variables over an ensemble of events are referred to as the
``matching distributions''.

Candidate muons are selected with the $\sltmu$ algorithm by
constructing a global~$\chi^2$ quantity, $L$, based on a comparison
of the measured matching variables with their expectations.  The
first step in constructing $L$ is taking a sum, $Q$, of
individual~$\chi^2$ variables:
\begin{equation}
Q=\sum_{i=1}^{n} \frac{(X_i - \mu_i)^2}{\sigma_i^2}, \label{eq:Q}
\end{equation}
where $\mu_i$ and $\sigma_i$ are, respectively, the expected mean
and width of the distribution of the matching variable~$X_i$.  The
sum is taken over $n$~selected variables as described below.  We
construct $L$, by normalizing $Q$ according to
\begin{equation}
L=\frac{(Q - n)}{\sqrt{\mathrm{var}(Q)}}, \label{eq:L}
\end{equation}
where the variance, var$(Q)$, is calculated using the full
covariance matrix for the selected variables.  The normalization is
chosen to make $L$ independent of the number of variables~$n$.

The selected variables are the full set of matching variables,
$\Delta x$, $\Delta z$, $\Delta\phi_L$ in the CMU, CMP and CMX with
the following two exceptions:  The CMP chambers do not provide a
measurement of the longitudinal coordinate~$z$, and matching in
$\phi_L$ is not included for stubs in the muon chambers that have
only three hits. Because of their significantly poorer resolution,
track segments reconstructed only in the CMU or only in the CMP
chambers with only three hits are rejected~(if the $\sltmu$
candidate has stubs in both CMU and CMP, then a stub with only three
hits is allowed). These two exceptions are a new feature of the
algorithm, since the previous publication, that reduce backgrounds
from hadronic punch-through with a negligible effect on the
efficiency. Note that a muon that traverses both the CMU and the CMP
chambers yields two sets of matching measurements in $x$ and
$\phi_L$ and one $z$ matching measurement, and is referred to as a
CMUP muon.  All available matching variables are used in the
calculation of $L$ for a given muon candidate.

As described in Reference~\cite{SLTPRD}, the expected means and
widths in equation~\ref{eq:Q} are parameterized as a function of the
$\Pt$ of the muon using $J/\psi$ and $W$ and $Z$ bosons in the data.
We use the same parametrization described there.  The efficiency has
been remeasured, from the data, using the full dataset for this
analysis.

Using $\jpsi$ events only, we measure the efficiency as a function
of the quantity $L$ defined in equation~\ref{eq:L} (the efficiency
measurement is described in detail in Section~\ref{sec:SLTEff}).
%The result is shown in Figure~\ref{fig:sltlkcut}.
The efficiency plateaus at a value of $\mid L\mid~\le~3.5$, and we
therefore use this requirement to define an $\sltmu$ tag.

Beginning with the $W+$jets candidate dataset, selected as described
in Section~\ref{sec:EvSel} above, we require that at least one jet
in each event has an $\sltmu$ tag.  A jet is determined to have an
$\sltmu$ tag if a candidate muon with $\mid L\mid\le3.5$ is found
within a cone of $\Delta R<0.6$ centered on the jet axis.  When the
primary lepton is a muon, the event is rejected when the $\sltmu$
has opposite charge to the primary muon and together with that muon
has an invariant mass between 8 and 11 $\GeVcc$ or between 70 and
110 $\GeVcc$.  This rejects events in which an $\Upsilon$ or
$Z$~boson decays to a pair of muons, one of which becomes the
primary lepton while the other ends up in a jet and is tagged by the
$\sltmu$ algorithm.  Whether the primary is an electron or a muon,
events where the invariant mass is less than 5~$\GeVcc$ are also
removed to prevent sequential double-semileptonic $b\rightarrow
c\rightarrow s$~decays~(where the primary lepton and the $\sltmu$
tag are from these semileptonic decays, rather than the primary
lepton being from the decay of a $W$~boson) from entering the
sample, as well as events with a $\jpsi$~decay.  We further reject
events as candidate radiative Drell-Yan and $Z$ bosons if the tagged
jet has an electromagnetic energy fraction above 0.8 and only one
track with $\Pt>1.0\GeVc$ within a cone of $\Delta R=0.4$ about the
jet axis.

Three levels of selection are defined in this analysis.  Events that
pass the kinematic cuts and the dilepton and radiative-$Z$ vetoes,
but do not necessarily have an $\sltmu$-taggable track in them,
comprise the ``pretag'' sample.  Pretag events that have an
$\sltmu$-taggable track~($\Pt >3~\GeVc$, passing quality cuts,
pointing to the muon chambers) within $\Delta R<0.6$ of a jet with
$\Et>20$~GeV are called taggable events. Finally, the subset of
$\sltmu$-taggable events that have at least one $\sltmu$-tagged jet
are called tagged events.

\subsection{\label{sec:Dataset} Selected Event Samples}
Table~\ref{tab:Data} shows the number of pretagged, taggable and
tagged events in the electron and muon channels in this dataset as a
function of jet multiplicity.
\begin{table}[h]
\begin{center}
\begin{small}
\begin{tabular}{l c c c c c }
\hline \hline
        & 1 jet & 2 jets& 3 jets& $\ge$4 jets &  $\ge$3 jets \\ \hline
\multicolumn{6}{c}{Electrons} \\ \hline
      Pretag    & 79348 & 13068 & 1615  & 660   & 2275  \\
$\sltmu$ Taggable    & 43005 & 10479 & 1518  & 648   & 2166  \\
  $\sltmu$ Tagged    & 519   & 224   & 85    & 64    & 149   \\
\hline \multicolumn{6}{c}{CMUP Muons} \\ \hline
      Pretag    & 38165 & 6320  & 719   & 325   & 1044  \\
$\sltmu$ Taggable    & 20162 & 4921  & 673   & 312   & 985   \\
  $\sltmu$ Tagged    & 224   & 105   & 41    & 34    & 75    \\
\hline \multicolumn{6}{c}{CMX Muons} \\ \hline
      Pretag    & 23503 & 3672  & 422   & 162   & 584   \\
$\sltmu$ Taggable    & 12428 & 2864  & 396   & 160   & 556   \\
  $\sltmu$ Tagged    & 149   & 55    & 16    & 8 & 24    \\
\hline \multicolumn{6}{c}{Electrons+Muons} \\ \hline
      Pretag    & 141016& 23060 & 2756  & 1147  & 3903  \\
$\sltmu$ Taggable    & 75595 & 18264 & 2587  & 1120  & 3707  \\
  $\sltmu$ Tagged    & 892   & 384   & 142   & 106   & 248   \\
\hline \hline
\end{tabular}
\end{small}
\caption[Summary of event counts for 2~fb$^{-1}$ of CDF Run~II
data.]{Summary of event counts for 2~fb$^{-1}$ of CDF Run~II data
for the event selection described in Sections~\ref{sec:EvSel}
and~\ref{sec:SLT}.} \label{tab:Data}
\end{center}
\end{table}

%\subsection{\label{sec:MCdata}Monte Carlo Datasets}
\section{\label{sec:MCdata}Monte Carlo Datasets} The detector
acceptance of $\ttbar$ events is modeled using
$\Pythia$~v6.216~\cite{Pythia} and $\Herwig$~v.6.510~\cite{Herwig}.
This analysis uses the former for the final cross section estimate
and the latter to estimate the systematics resulting in the modeling
of $\ttbar$ production and decay. The $\Pythia$ event generator has
been tuned using jet data to better model the effects of multiple
interactions and remnants from the break-up of the proton and
antiproton. The generators are used with the {\sc CTEQ5L} parton
distribution functions~\cite{cteq5l}. Decays of $b$- and $c$-hadrons
are modeled using $\Evtgen$~\cite{evtgen}.

Events with a $W$~boson produced in association with multiple jets
are modeled using $\Alpgen$~v2.1~\cite{ALPGEN}, with parton
showering provided by $\Pythia$~v6.326 and HF hadron decays handled
by $\Evtgen$.  $\Alpgen$ calculates exact matrix elements at leading
order for a large set of parton level processes in QCD and
electroweak interactions.   The showering in $\Pythia$ may result in
multiple $\Alpgen$ samples covering the same phase space.  These
overlaps are removed using a jet-parton matching algorithm along
with a jet-based heavy flavor overlap removal
algorithm~\cite{Sherman}.

Estimates of backgrounds from diboson production~($WW$, $WZ$, and
$ZZ$) and Drell-Yan/$Z\rightarrow\tau\tau$ are derived using
$\Pythia$.  Drell-Yan to $\mu\mu$ events are modeled using $\Alpgen$
with $\Pythia$ showering while single-top production is modeled with
$\Madevent$~\cite{MadEvent}, also with $\Pythia$ showering.

The CDF II detector simulation reproduces the response of the
detector to particles produced in $\ppbar$ collisions.  The detector
geometry database used in the simulation is the same as that used
for reconstruction of the collision data.  Details of the CDF II
simulation can be found in~\cite{sim}.

\section{\label{sec:AccEff}Efficiency for Identifying ${\boldmath \ttbar}$ Events }
The efficiency for identifying $\ttbar$~events in this analysis is
factorized into the geometric times kinematic acceptance and the
$\sltmu$ tagging efficiency.  The acceptance is evaluated assuming a
top mass of $175~\GeVcc$ and includes the branching fraction to
leptons, which is assumed to have the SM value. The tagging
efficiency is the efficiency for $\sltmu$-tagging at least one jet
in events that pass the geometric and kinematic selection. Each
piece is described below.

\subsection{Geometric and Kinematic Acceptance}
\label{sec:acceptance} The acceptance of $\ttbar$ events in this
analysis is measured in $\Pythia$ and then corrected, using
measurements from the data, for effects that are not sufficiently
well modeled in the simulation: the lepton trigger efficiencies, the
fraction of the $\ppbar$ luminous region well-contained in the CDF
detector~(i.e. the $z$-vertex cut efficiency), and track
reconstruction and lepton identification efficiencies. The
efficiency of the $z$-vertex cut, $\mid z_0\mid<60$~cm, is measured
from minimum-bias triggered events to be $(96.3\pm0.2)$\%. The
correction factor for the difference between the track
reconstruction efficiencies in data and simulation is
$1.014\pm0.002$. Events in the Monte Carlo are not required to pass
any trigger, so the acceptance is multiplied by lepton trigger
efficiency. The lepton trigger and identification efficiencies,
measured using the unbiased leg of $Z$ boson decays to electrons and
muons, and the correction factors for each of the primary lepton
types are shown in Tab.~\ref{tab:lepid}.

%The efficiencies for triggering on and identifying primary leptons
%are each measured in $Z\rightarrow\mu\mu$ and $Z\rightarrow ee$
%events in which one leg is required to satisfy the trigger or
%selection requirements and the other leg is used for the efficiency
%measurement.  The ratio of the lepton-identification efficiency in
%data to that in simulation is applied as a correction factor to the
%acceptance calculated from simulation.

\begin{table*}[htbp]
\sans
\begin{center}
\begin{tabular}{l c c c}
\hline \hline Quantity & Electron & CMUP Muon & CMX Muon \\ \hline
Trigger efficiency      & 0.966$\pm$0.005   & 0.917$\pm$0.005   &
0.925$\pm$0.007   \\ \hline
Lepton ID efficiency (data) & 0.789$\pm$0.004   & 0.829$\pm$0.006   & 0.893$\pm$0.006   \\
Lepton ID efficiency (MC)   & 0.806$\pm$0.001   & 0.896$\pm$0.001
& 0.916$\pm$0.002   \\ \hline
Lepton identification correction& 0.978$\pm$0.005   & 0.926$\pm$0.007   & 0.975$\pm$0.007   \\
\hline \hline
\end{tabular}
\caption{Summary of lepton trigger and identification efficiencies.}
\label{tab:lepid}
\end{center}
\end{table*}

The raw acceptance is defined as the number of pretag events divided
by the total number of $\ttbar$ events in the $\Pythia$ sample.  The
acceptance,  after correcting for the differences between data and
simulation, is shown in Tab.~\ref{tab:acc} as a function of the
number of identified jets.
\begin{table*}[htbp]
 \sans
\begin{center}
%\begin{adjustwidth}{-5em}{-5em}
%\begin{ruledtabular}
\begin{tabular}{l c c c c c}
\hline \hline
    & 1 jet & 2 jets & 3 jets & $\geq$ 4 jets & $\geq$ 3 jets \\ \hline
%& \multicolumn{4}{c}{electrons} \\
Electron (\%)& 0.163$\pm$0.002 & 0.858$\pm$0.004 & 1.63$\pm$0.01 & 2.08$\pm$0.01 & 3.71$\pm$0.01\\
% muons below are CMUP+CMX
CMUP Muon (\%)  & 0.088$\pm$0.001 & 0.472$\pm$0.003 & 0.909$\pm$0.004 & 1.142$\pm$0.005 & 2.05$\pm$0.01 \\
CMX Muon (\%)   & 0.042$\pm$0.001 & 0.219$\pm$0.002 &
0.414$\pm$0.003 & 0.532$\pm$0.003 & 0.946$\pm$0.004 \\ \hline
Combined (\%)   & 0.292$\pm$0.002 & 1.544$\pm$0.005 & 2.946$\pm$0.008 & 3.743$\pm$0.009 & 6.69$\pm$0.01 \\
\hline \hline
\end{tabular}
%\end{ruledtabular}
%\end{adjustwidth}
\caption{Acceptance for $\ttbar$ events as a function of jet
multiplicity from $\Pythia$ Monte Carlo sample, after data/MC
corrections described in the text. In the combined acceptance we
account for the fact that the CMX detector was not operating early
in Run~II. The uncertainties listed are statistical only. }
\label{tab:acc}
\end{center}
\end{table*}

\subsection{\label{sec:SLTEff}Efficiency of the $\sltmu$ Algorithm}
The efficiency of the $\sltmu$ algorithm is measured from the data
using samples of $\jpsi$ and $Z$ decays triggered on a single muon.
The tagger is applied to the non-trigger muon~(probe leg).  If both
legs pass the trigger, only one of them is used.  To reduce
background in the $Z$ sample, the leg that is not used to measure
efficiency is required to be isolated and to be consistent with
being a minimum ionizing particle in the calorimeter.  We correct
for the remaining background using the invariant mass regions
outside the $Z$ mass window (``sidebands").

The efficiency of the $\sltmu$ is defined as:
\begin{equation}
\epsilon =
\frac{\mathrm{Number~of~tagged~muons}}{\mathrm{Number~of~taggable~muon
~tracks~with~a~stub}}.
\end{equation}

\begin{figure}[htbp]
\begin{center}
\includegraphics[width=3.375in]{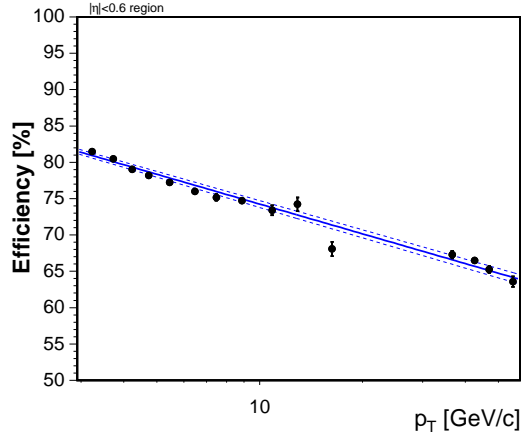}
\end{center}
\caption{The $\sltmu$ efficiency for CMU/CMP as a function of $\Pt$
measured from $\jpsi$ and $Z$ data for $\mid L\mid<3.5$. The solid
line is the fit to the data and the dashed lines indicate the
uncertainty on the fit.} \label{fig:eff_central}
\end{figure}

\begin{figure}[htbp]
\begin{center}
\includegraphics[width=3.375in]{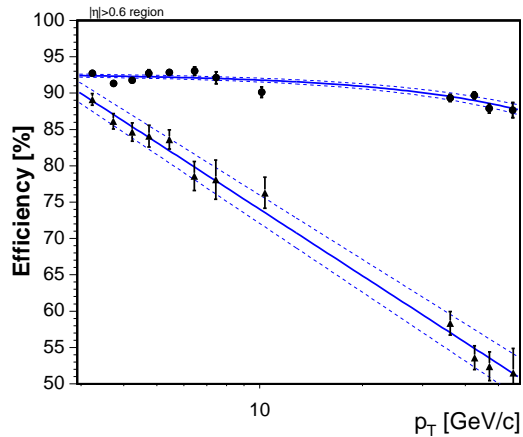}
\end{center}
\caption{The $\sltmu$ efficiency for CMX Arches (circles) and
Miniskirt/Keystone (triangles) as a function of $\Pt$ measured from
$\jpsi$ and $Z$ data for $\mid L\mid<3.5$. The solid curves are the
fits to the data and the dashed lines indicate the uncertainties on
the fits.} \label{fig:eff_cmx}
\end{figure}

The requirement in the denominator that the taggable muon track has
a stub in the requisite muon chambers decouples the muon
reconstruction efficiency, which is accounted for separately, from
the efficiency of the tagger. Figures~\ref{fig:eff_central}
and~\ref{fig:eff_cmx} show the efficiency for tagging muons with
$\mid L\mid~\le~3.5$ as a function of muon $\Pt$ from both $\jpsi$
and $Z$ data.  The decrease in efficiency with increasing $\Pt$ is
due, primarily, to non-Gaussian tails in the resolution functions.
These efficiency data are fit to functional forms~\cite{SLTPRD}
shown as the curves in the data. The dotted curves are those
obtained by varying the fit parameters by $\pm 1\sigma$. Although
the efficiency measurement is dominated by isolated muons, we do not
expect that it will depend on the isolation of the muon because the
muon chambers are well shielded from the inner detector. We have
checked this assumption by measuring the efficiency as a function of
the number of nearby tracks and found no dependence.

%%%%%%%%%%%%%%%%%%
%%%%%%%%%%%%%%%%%%
%%%%%%%%%%%%%%%%%%
%%%%%%%%%%%%%%%%%%

The efficiency for $\sltmu$-tagging a $\ttbar$ event is measured for
Monte Carlo events that pass the geometric and kinematic selection.
We model the $\sltmu$ tagging in these events by tagging muons from
semileptonic HF decay with a probability given by the efficiencies
in Fig.s~\ref{fig:eff_central} and~\ref{fig:eff_cmx}.  Events
without a `real' muon tag of this type can still be $\sltmu$-tagged
through a mistag.  Mistags in $\ttbar$ events are included by
applying the ``mistag matrix" described in the following section,
and are included as part of the signal efficiency. The $\sltmu$
tagging efficiency in $\ttbar$ events is given in
Tab.~\ref{tab:ttbartageff}.

\begin{table*}[htbp]
 \sans
\begin{center}
\begin{tabular}{l c c c c c}
\hline \hline
    & 1 jet & 2 jets & 3 jets & $\geq$ 4 jets & $\geq$ 3 jets \\ \hline
Electron (\%)   & 7.0$\pm$0.3 & 11.4$\pm$0.2 & 12.9$\pm$0.1 & 14.9$\pm$0.1 & 14.0$\pm$0.1\\
CMUP Muon (\%)  & 5.6$\pm$0.3 & 10.7$\pm$0.2 & 11.8$\pm$0.1 & 14.1$\pm$0.1 & 13.1$\pm$0.1 \\
CMX Muon (\%)   & 6.7$\pm$0.5 & 11.2$\pm$0.3 & 12.3$\pm$0.2 & 14.2$\pm$0.2 & 13.4$\pm$0.2 \\
Average  (\%)   & 6.5$\pm$0.2 & 11.2$\pm$0.1 & 12.5$\pm$0.1 & 14.6$\pm$0.1 & 13.6$\pm$0.1 \\
\hline \hline
\end{tabular}
\caption[$\ttbar$ event tagging efficiency for SLT muons as a
function of jet multiplicity.]{$\ttbar$ event tagging efficiency for
SLT muons as a function of jet multiplicity from $\Pythia$ Monte
Carlo sample. The lepton category refers to the primary lepton.  The
average tagging efficiency is determined by weighting each channel
by the acceptance and luminosity for each channel.  The listed
uncertainties are statistical only.} \label{tab:ttbartageff}
\end{center}
\end{table*}

\section{\label{sec:FakeMatrix} Predicting the Number of tags from light-quark
jets}

As a prelude to the evaluation of the backgrounds to the $\ttbar$
signal we describe a new method, developed for this analysis, for
predicting the number of $\sltmu$ tags that come from light quark
jets. We refer to these as ``mistags", and they result from a
combination of hadronic punch-through of the calorimeter and muon
steel, and hadronic decays-in-flight.

To predict the number of mistags in our sample we use a track-based
mistag probability that is a function of track $\Pt$ and $\eta$. We
use reconstructed $\dstar$ and $\lambdaz$ to identify a clean sample
of pions, kaons and protons and measure the probability per taggable
track, in 8 bins of $\Pt$ and 9 bins of $\eta$,  for each to satisfy
the $\sltmu$ $\mid L\mid<3.5$ requirement.  Details of the
reconstruction technique, the measurement of the tagging
probabilities and the assembly and testing of the two-dimensional
($8\times9$) ``mistag matrix" are described in what follows.

\subsection{Data samples}\label{sec:dstlam}
To identify kaons and pions we reconstruct $\dstar^+ \rightarrow
\dzero \pi^+\rightarrow K^-\pi^+\pi^+$ decays, and their charge
conjugates.
%`$\dstar$' includes both $\dstarp$ and $\dstarn$.
 This dataset is collected using a
two-track trigger that requires two oppositely charged tracks with
$\Pt \geq2~\GeVc$. The tracks are also required to have a scalar sum
$P_{T1}+P_{T2}\geq5.5~\GeVc$, an opening angle between them of
$2^{\circ}\leq|{\Delta\phi}|\leq 90^{\circ}$, and originate from a
displaced vertex.

A sample of protons is obtained by reconstructing
$\Lambda\rightarrow p\pi^-$ decays. These events are collected using
another two-track trigger similar to the one described above, but
with an opening angle requirement of
$20^{\circ}\leq|{\Delta\phi}|\leq135^{\circ}$ and the invariant mass
of the track pair~(assumed to be pions) required to be $4~\GeVcc\leq
M(\pi,\pi)\leq 7~\GeVcc$.

\subsection{Event Reconstruction} \label{sec:reco} We apply the
following track quality criteria in the reconstruction of both
$\dstar$~\cite{Chen:2003qe} and $\lambdaz$~\cite{lambda} decays:
\begin{itemize}
\item the number of COT axial superlayers with $\geq 5$ hits is $\geq 3$;
\item the number of COT stereo superlayers with $\geq 5$ hits is $\geq 2$;
\item the track has $|z_{0}|\leq 60$~cm.
\end{itemize}

\begin{figure*}
\begin{center}
$\begin{array}{c@{\hspace{0.25in}}c}
    \includegraphics[width=2.7in]{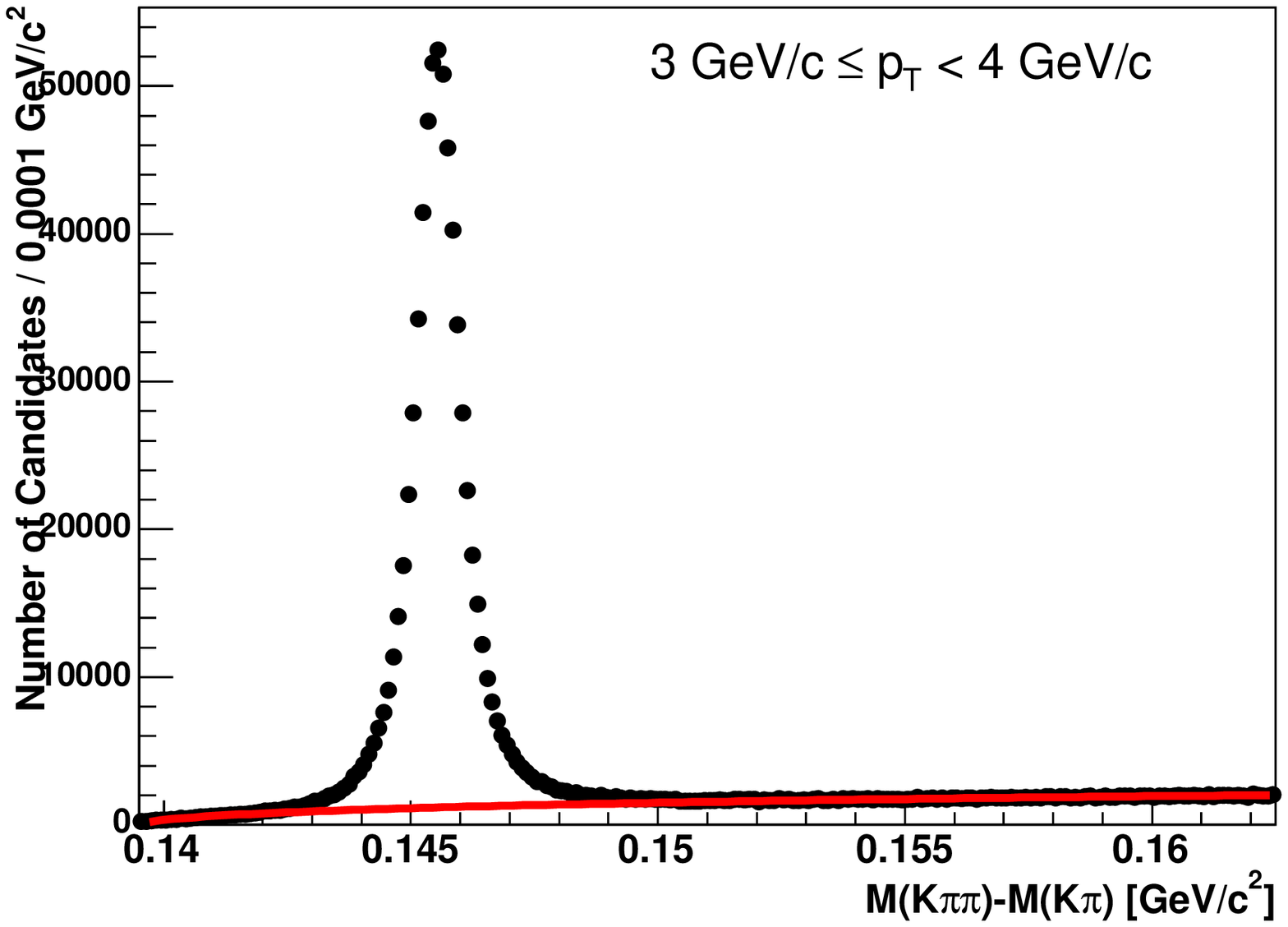} &
        \includegraphics[width=2.7in]{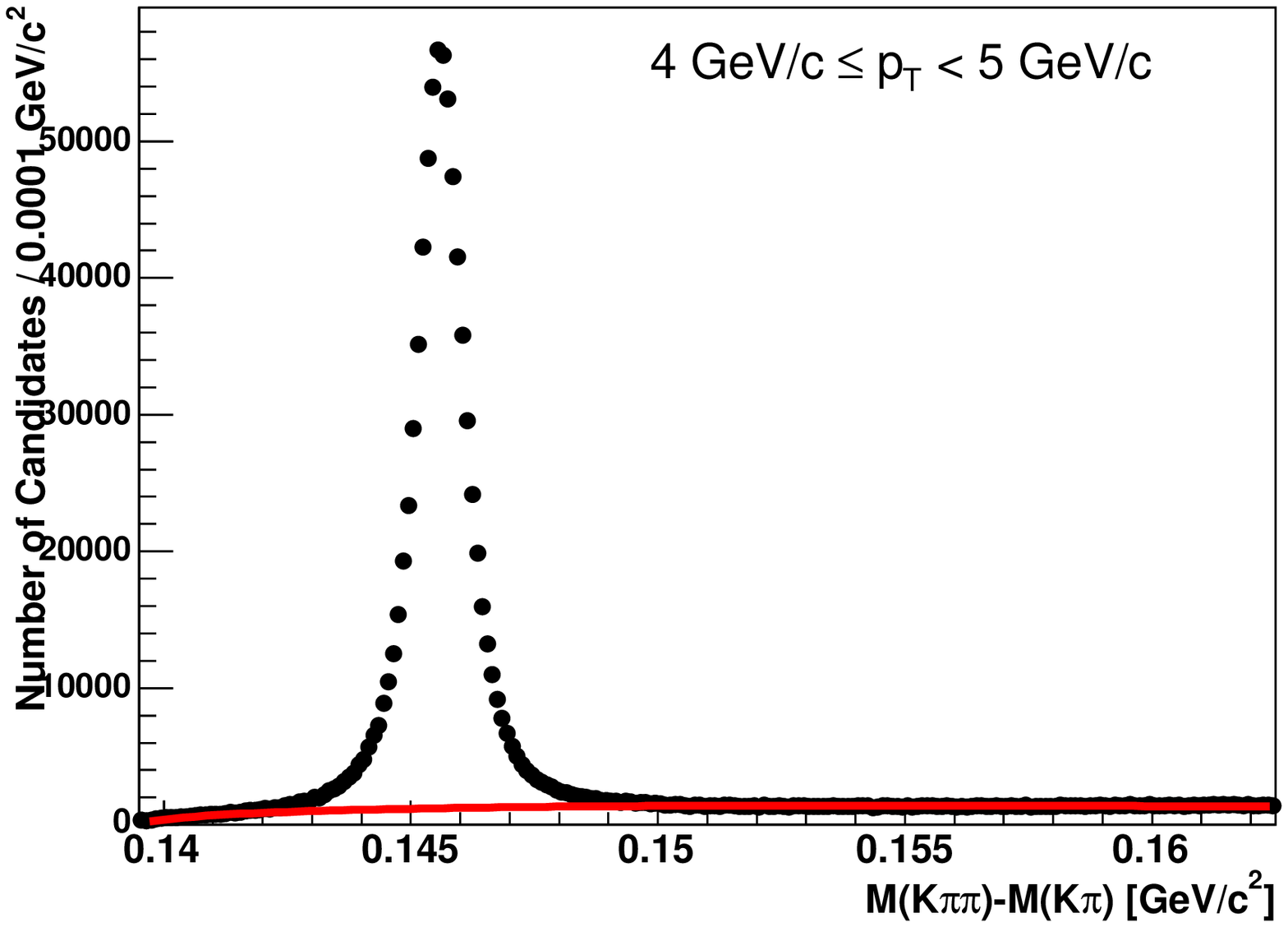} \\
    \includegraphics[width=2.7in]{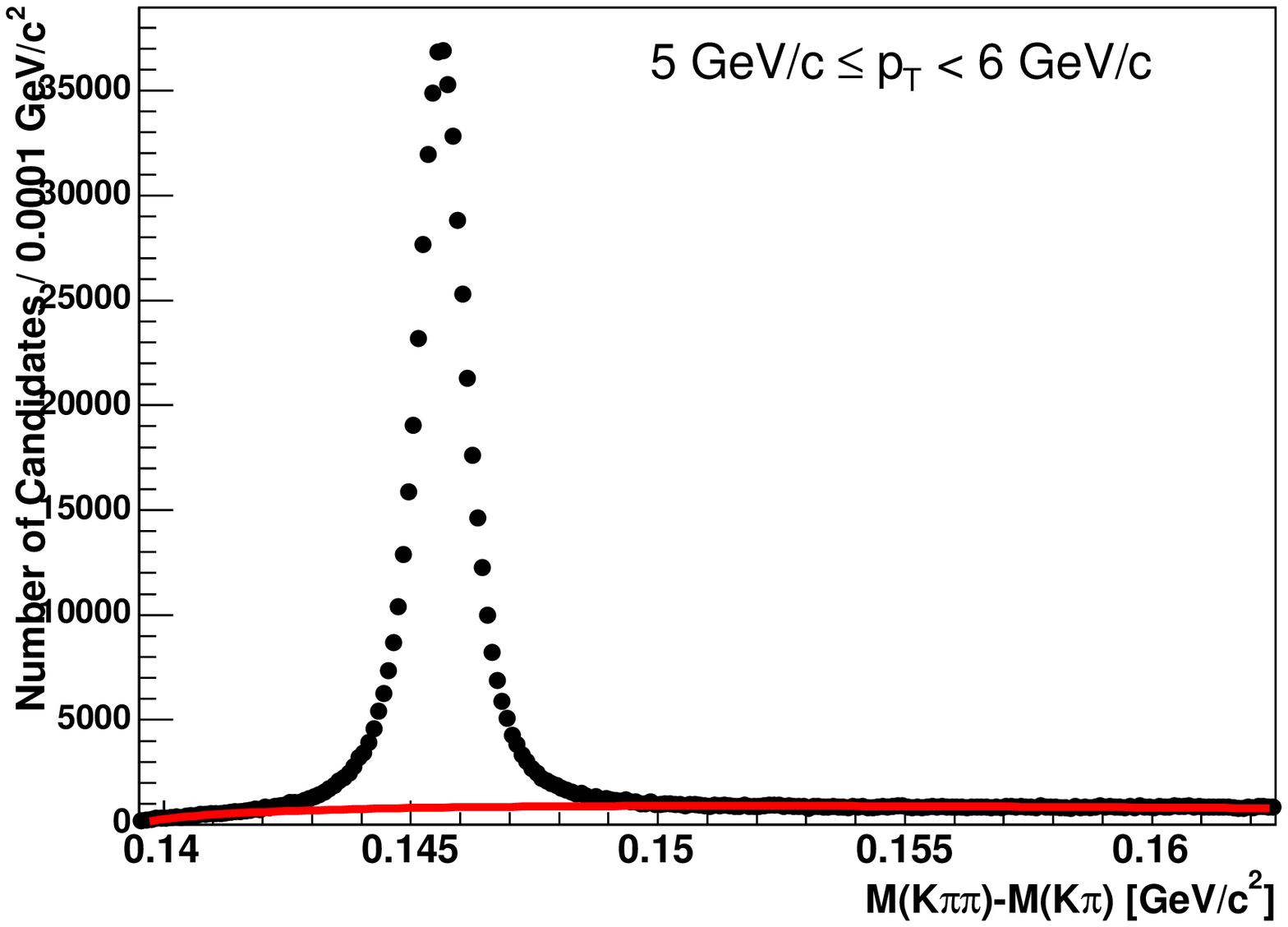} &
        \includegraphics[width=2.7in]{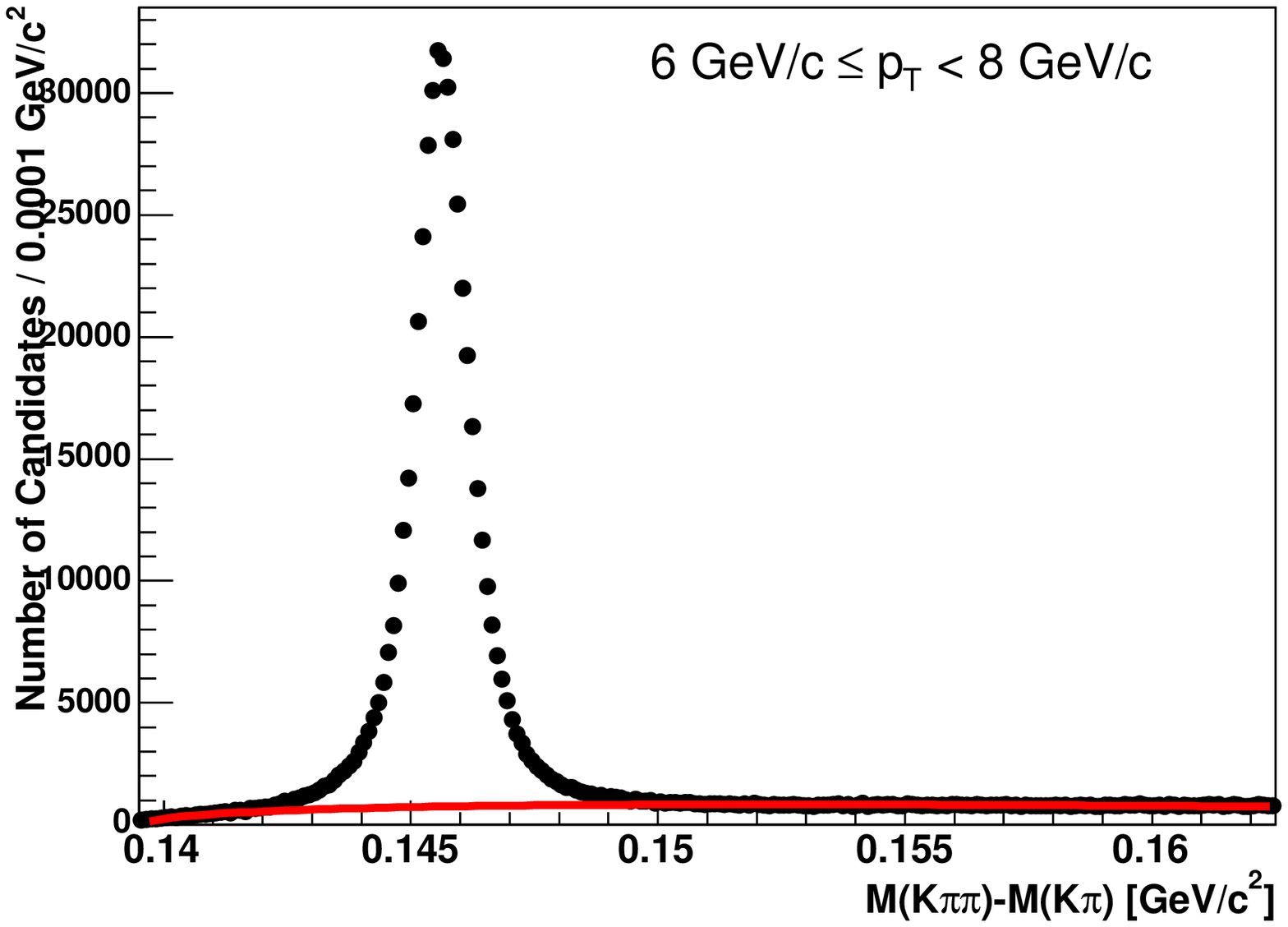} \\
    \includegraphics[width=2.7in]{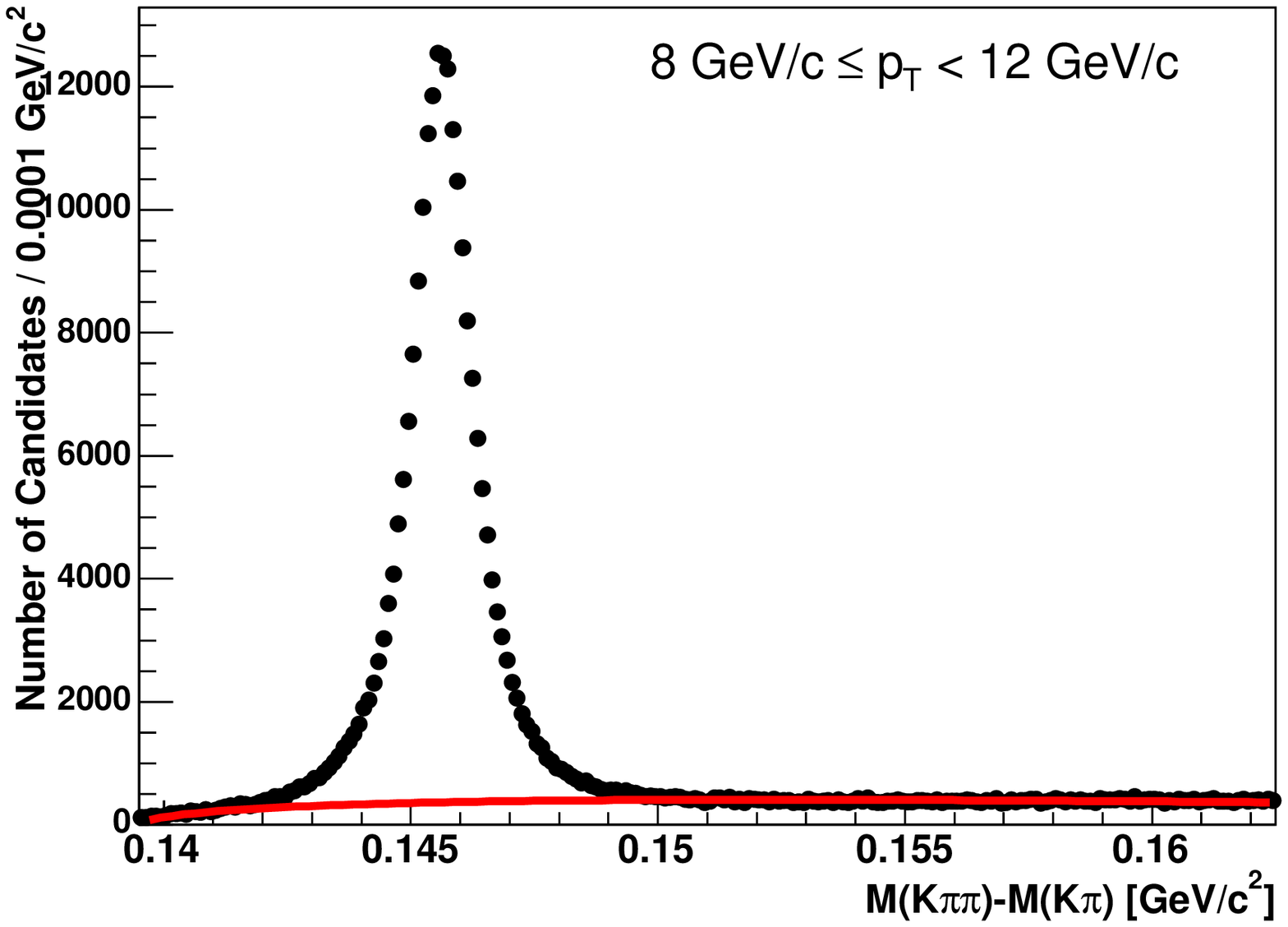} &
        \includegraphics[width=2.7in]{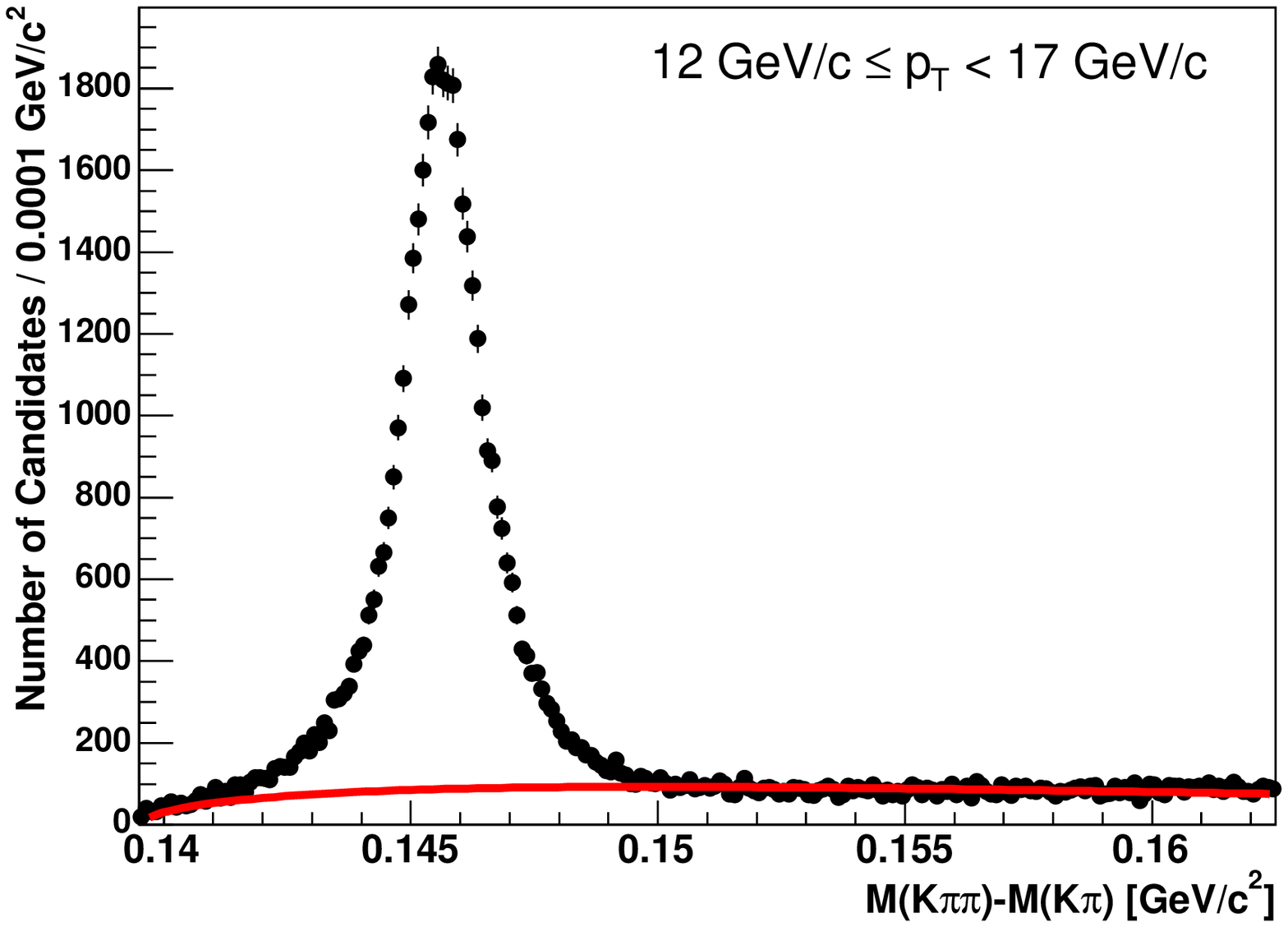} \\
    \includegraphics[width=2.7in]{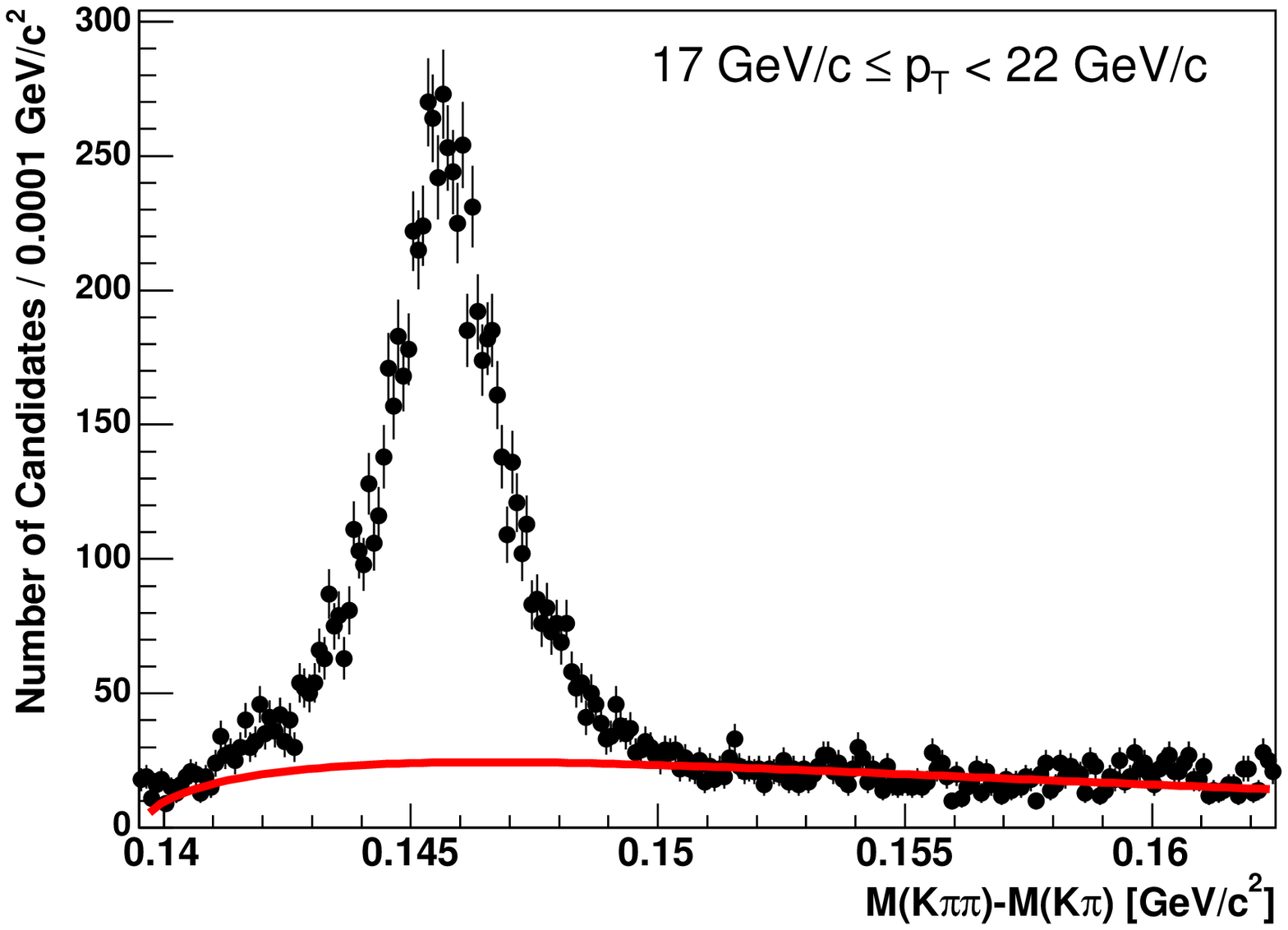} &
        \includegraphics[width=2.7in]{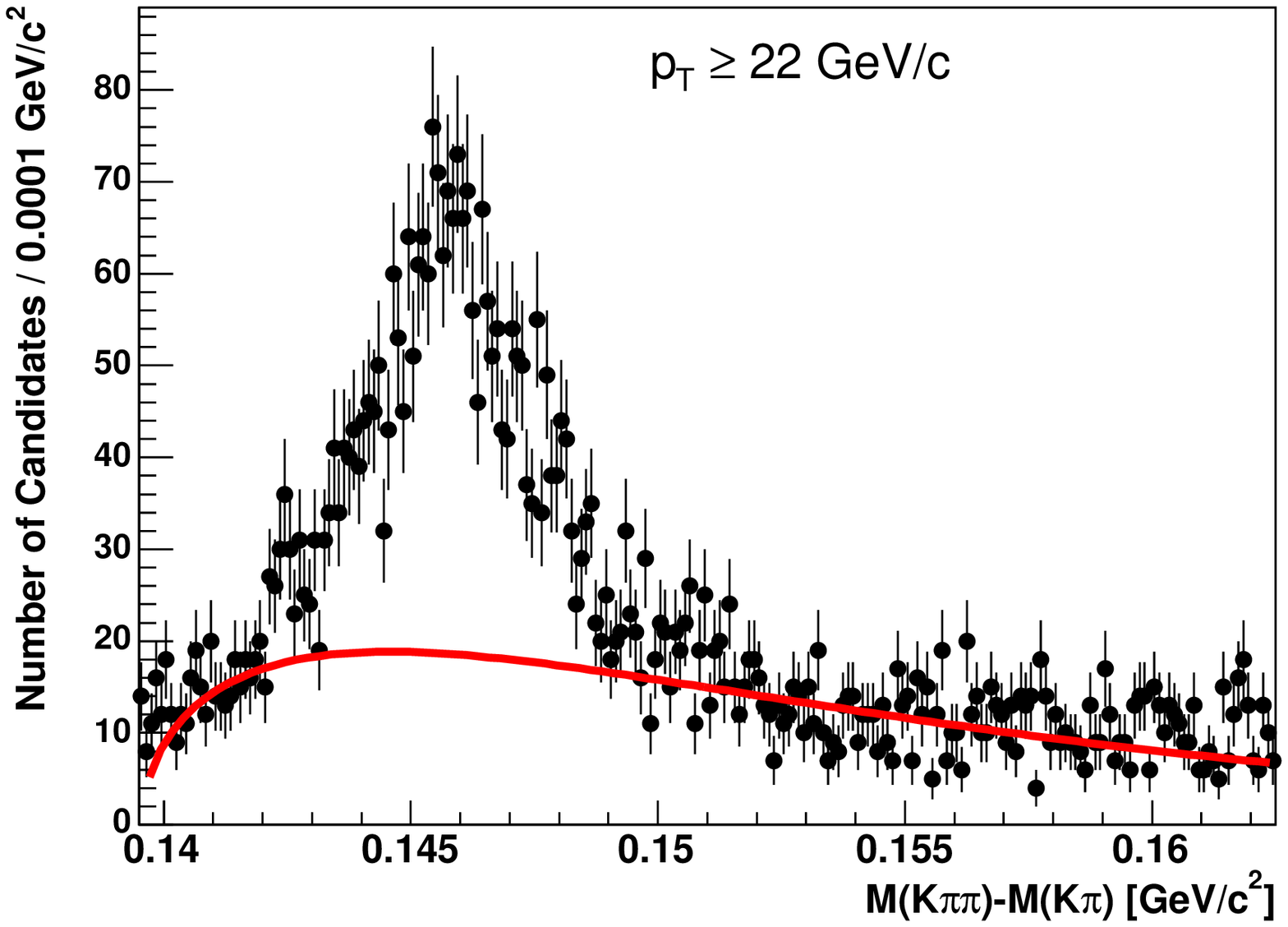} \\
\end{array}$
\end{center}
\caption{The $m(K\pi\pi)-m(K\pi)$ distribution for
$\dstar^{\pm}\rightarrow\dzero\pi^{\pm},\dzero\rightarrow
K^{\mp}\pi^{\pm}$ candidates in different $\sltmu$-track-$\Pt$ bins.
The line in each plot represents the fit to the sideband regions.}
\label{fig:diffmass}
\end{figure*}

The $\dstar$ reconstruction then proceeds through the examination of
the mass difference $\Delta m = m(K\pi\pi)-m(K\pi)$ with the
following criteria:
\begin{itemize}
\item the kaon must have opposite charge to each of the two pions;
\item $|\Delta z_{0}|\leq 5$~cm between any two tracks;
\item the soft pion from the $\dstar\rightarrow\dzero\pi$ decay must have $\Pt\geq 0.5~\GeVc$;
\item the kaon and pion from the $\dzero$ decay must each have $\Pt \geq 2~\GeVc$;
\item the kaon and pion tracks must have impact parameter, $|d_{0}|\leq0.2$~cm;
\item $|m(K\pi)-m(\dzero)|\leq 0.03~\GeVcc$;
\item At least one of the tracks (K or $\pi$) from the \dzero~must be $\sltmu$ taggable (including having $\Pt \geq 3~\GeVc$).
\end{itemize}
As shown in Fig.~\ref{fig:diffmass}, a clean $\dstar$ signal is
obtained for the right-sign $\Delta m$ distribution.

\begin{figure*}
\begin{center}
$\begin{array}{c@{\hspace{0.25in}}c}
    \includegraphics[width=2.7in]{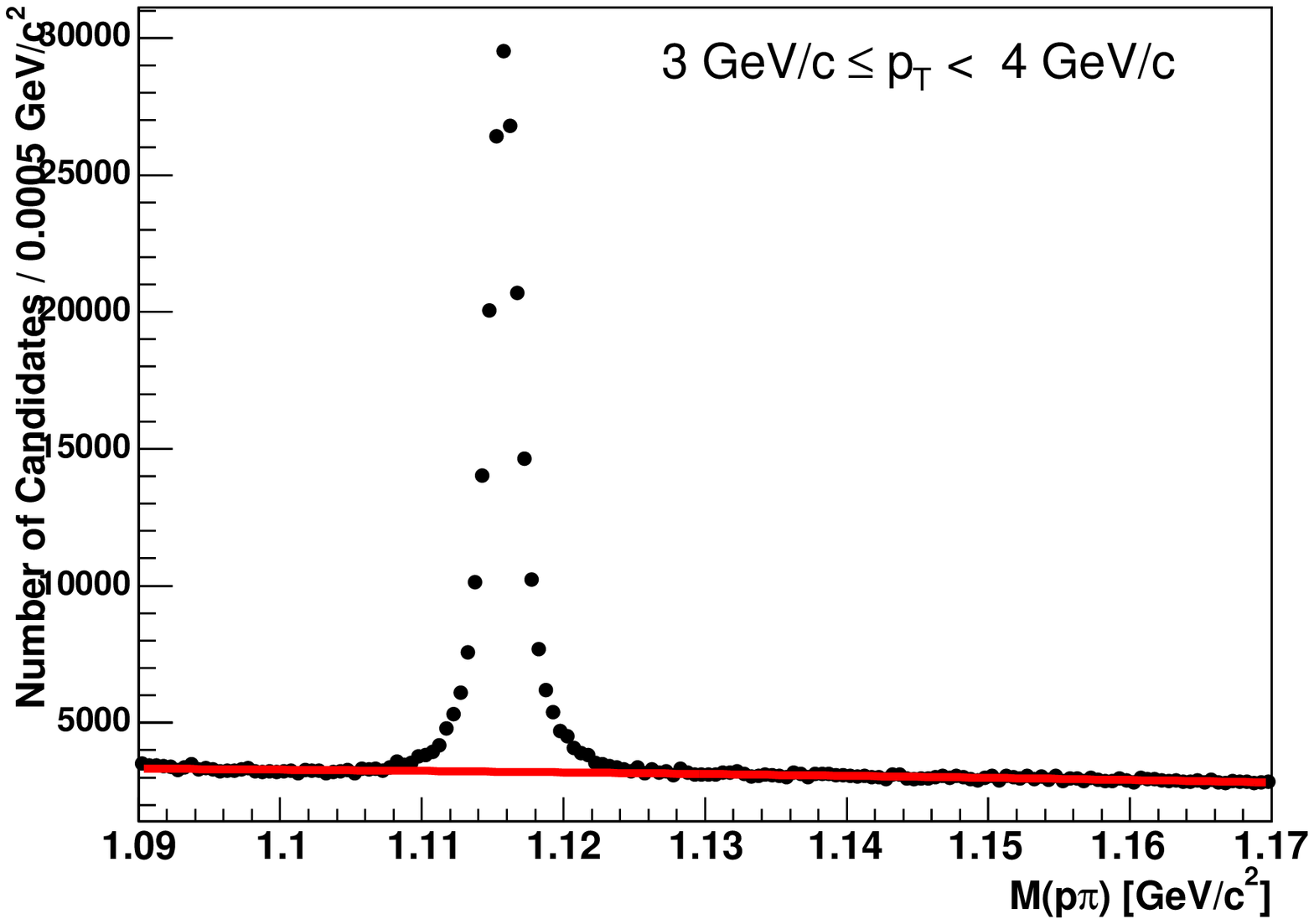} &
        \includegraphics[width=2.7in]{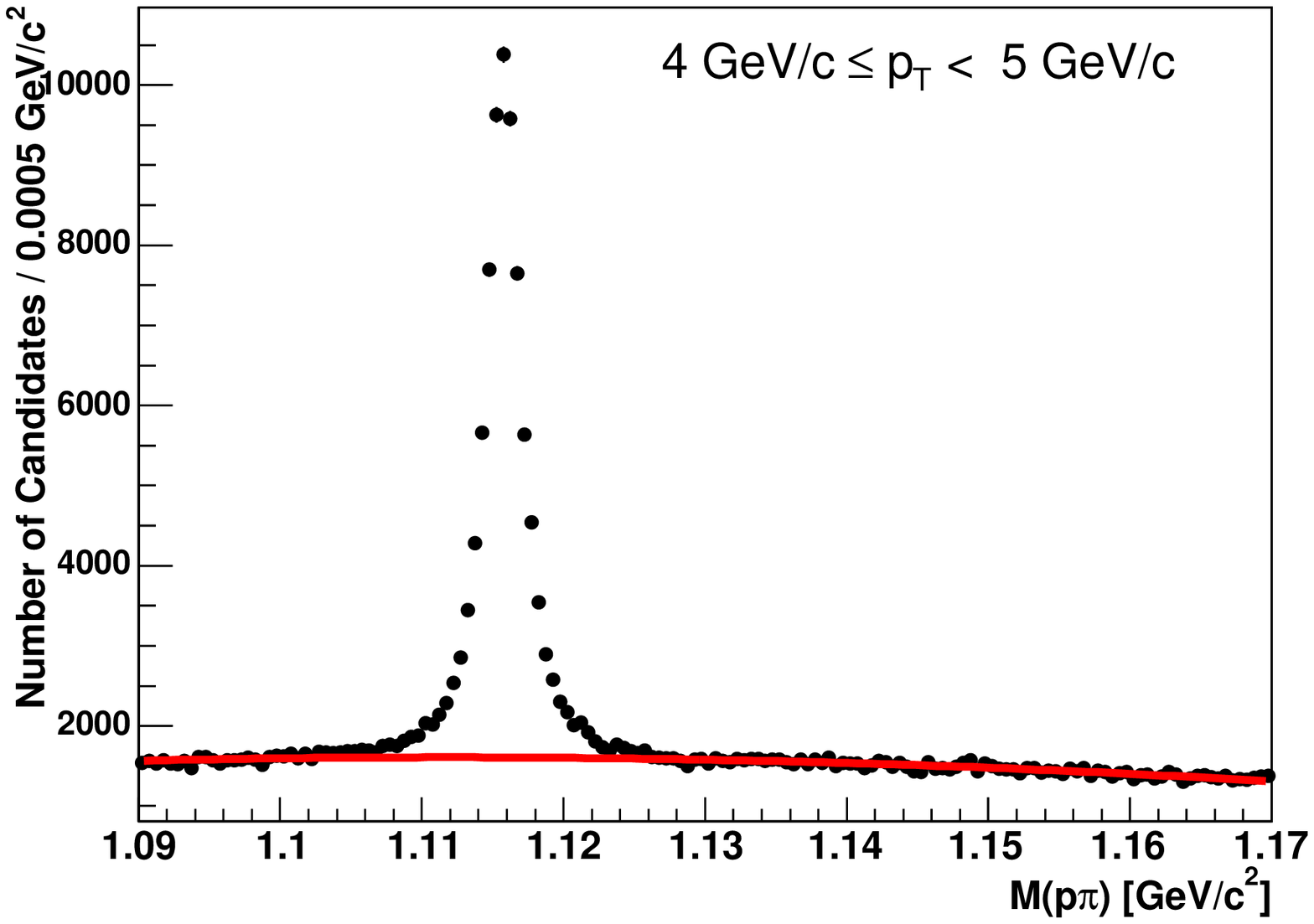} \\
    \includegraphics[width=2.7in]{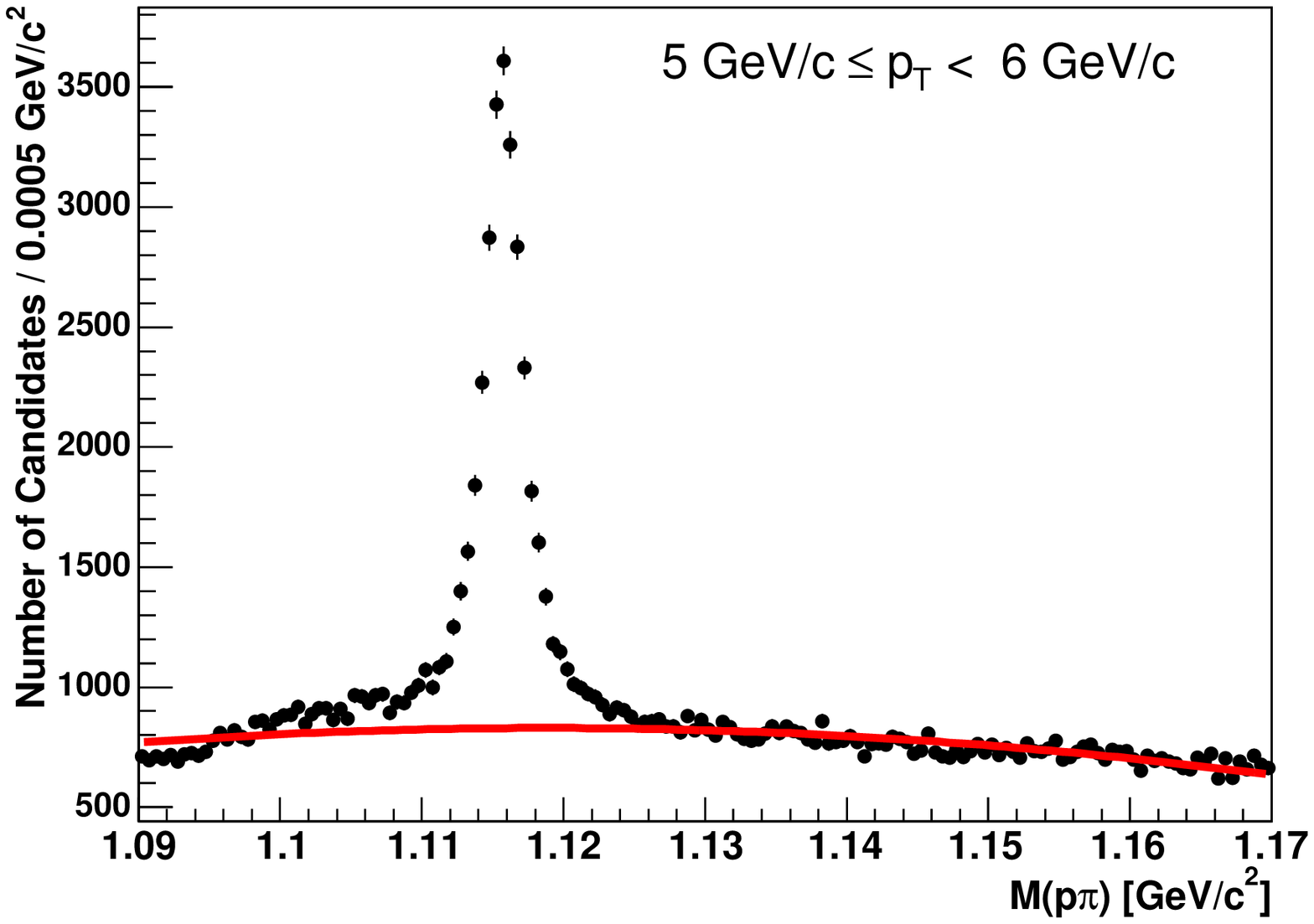} &
        \includegraphics[width=2.7in]{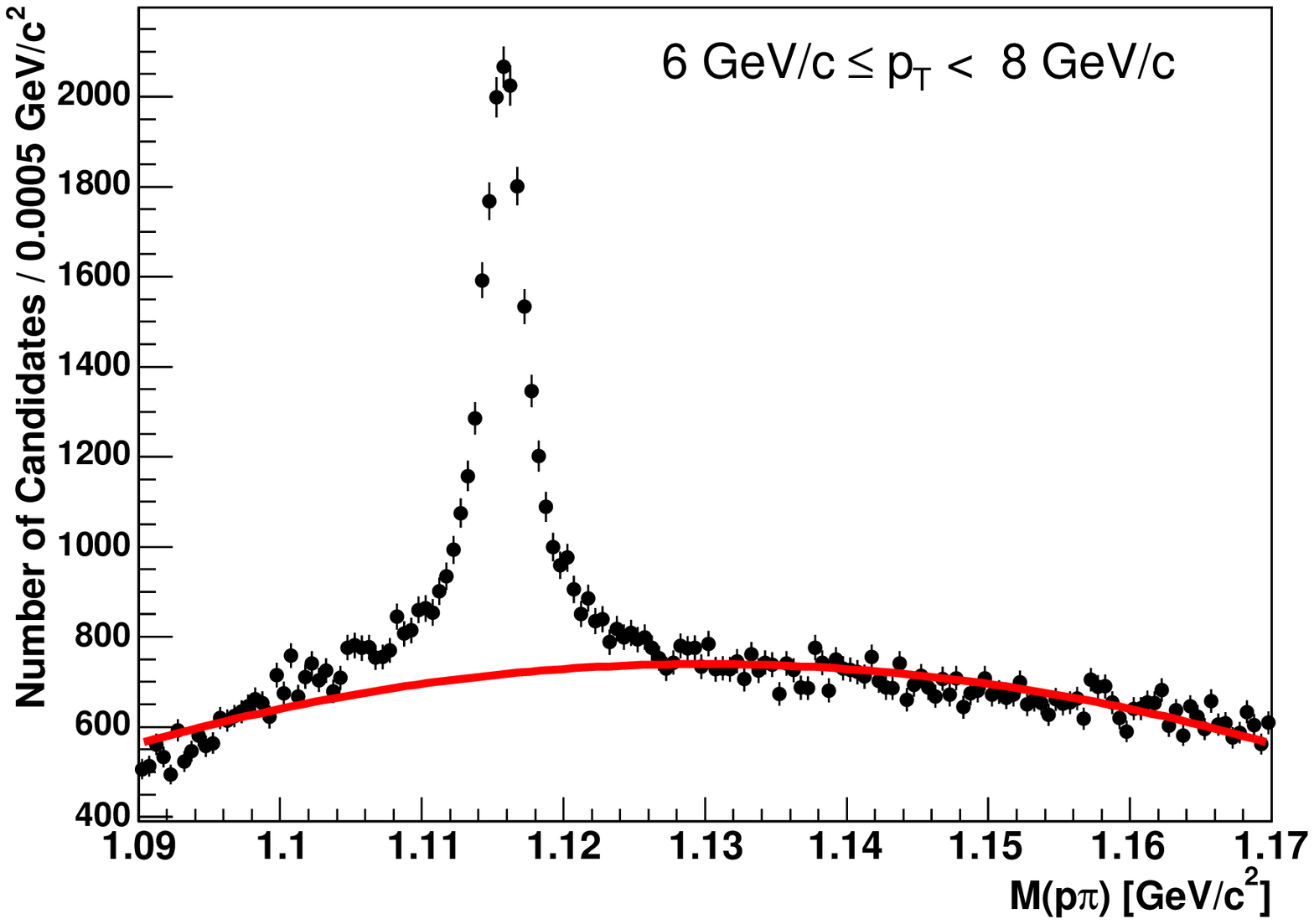} \\
    \includegraphics[width=2.7in]{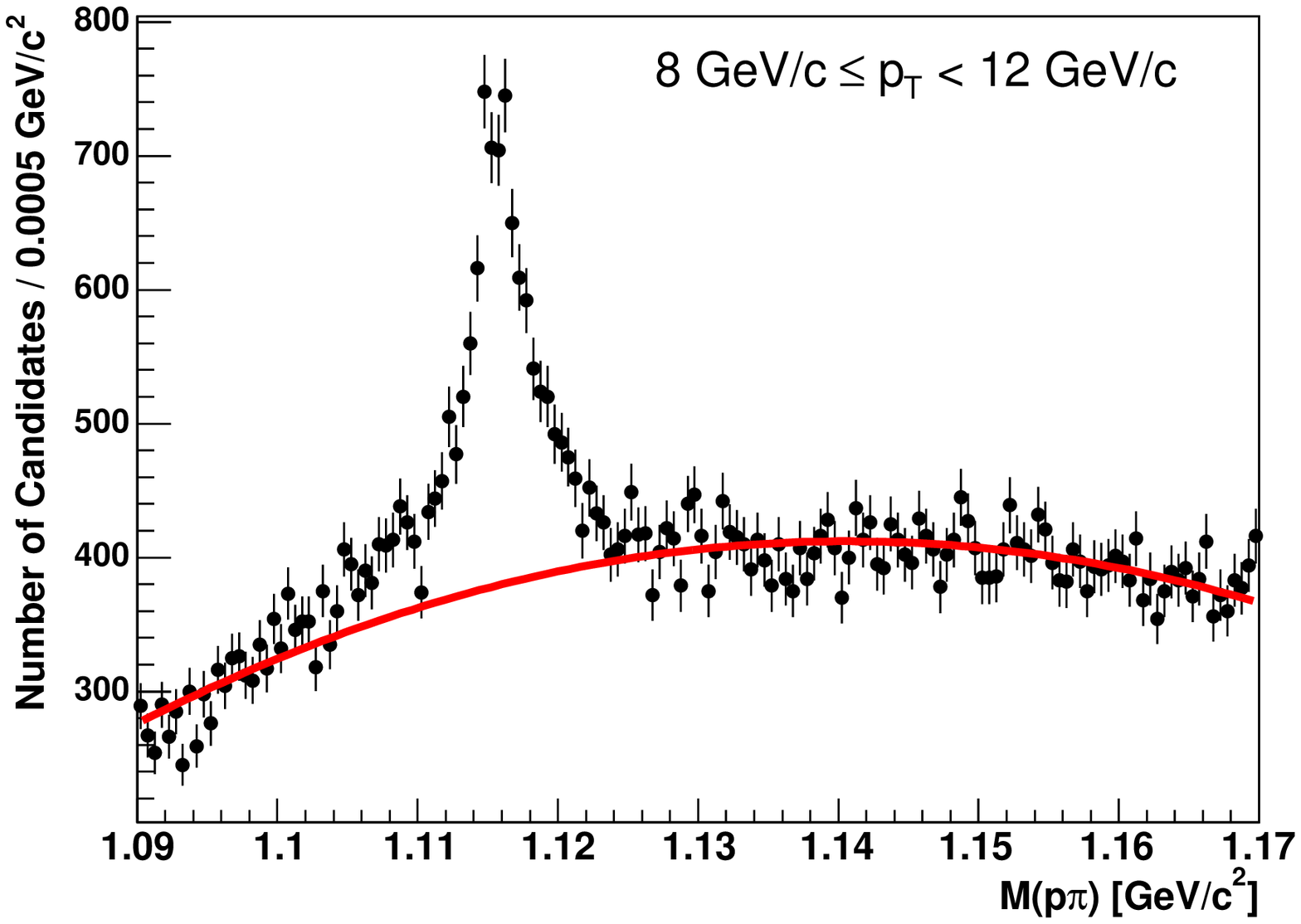} &
        \includegraphics[width=2.7in]{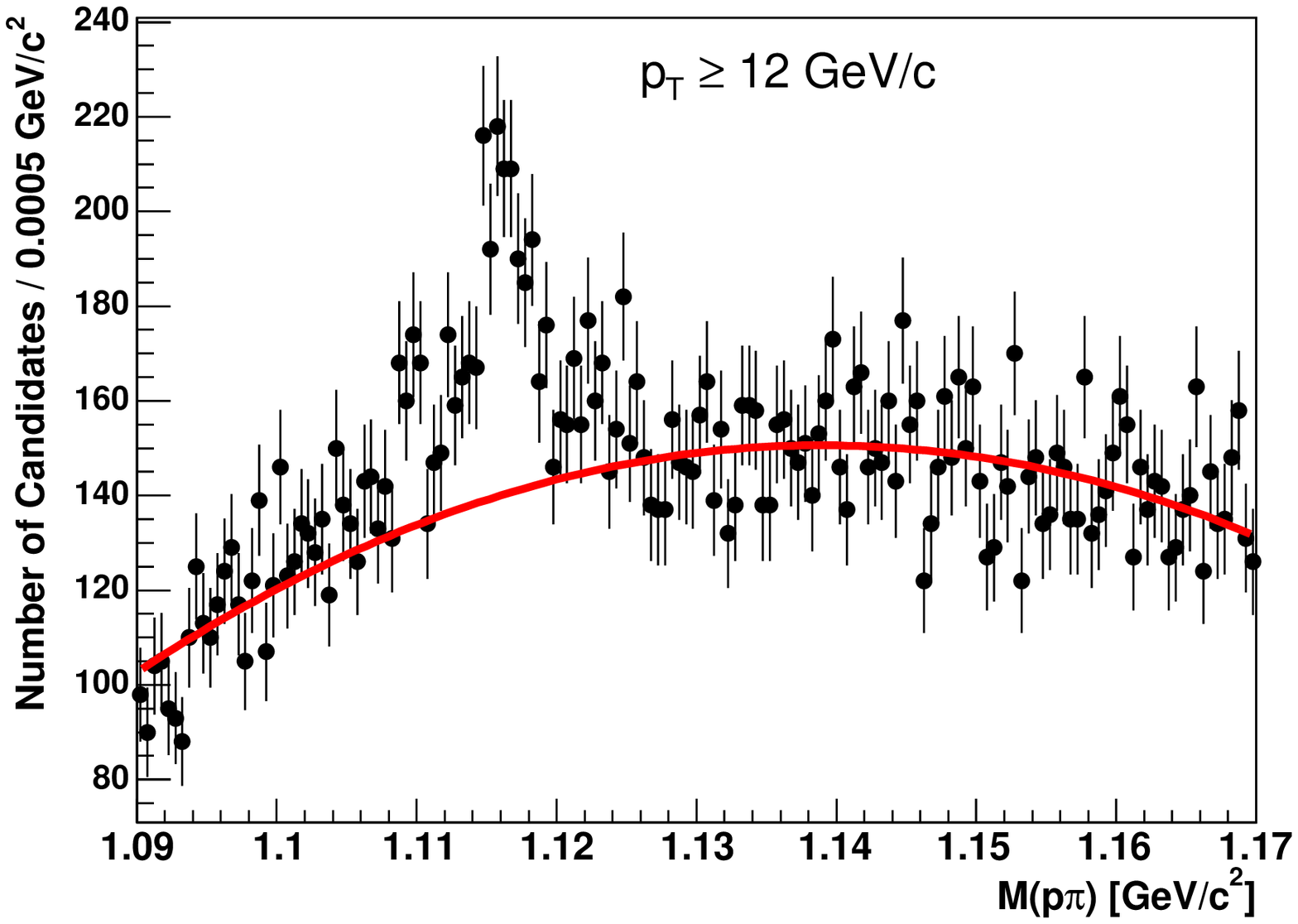} \\
\end{array}$
\end{center}
\caption{The $m(p\pi)$ distribution for $\lambdaz\rightarrow p\pi$
candidates in different $\sltmu$-track-$\Pt$ bins.  The line in each
plot represents the fit to the sideband regions.}
\label{fig:lambdamass}
\end{figure*}

The reconstruction of $\Lambda$ decays requires the following
criteria:

\begin{itemize}
\item the pion and proton must have opposite charge;
\item $|\Delta z_{0}|\leq 2$~cm between the two tracks;
\item the $\chi^2$ of the vertex fit must be $\leq 10$;
\item the vertex must have $L_{xy}\geq 0.5$~cm, where $L_{xy}$ is
defined as the projection onto the net momentum direction, in the
$r-\phi$ plane, of the vector pointing from the primary to the
secondary vertex.;
\item the proton $\Pt$ is greater than the pion $\Pt$;
\item the pion must have $\Pt\geq 0.4~\GeVc$;
\item the proton must have $|d_{0}|\leq 0.2$~cm;
\item the proton must be $\sltmu$ taggable (including having $\Pt \geq 3~\GeVc$).
\end{itemize}
Figure~\ref{fig:lambdamass} shows the invariant mass distribution in
the $p\pi$ mass hypothesis.

\begin{figure}
\begin{center}
\includegraphics[width=3.375in]{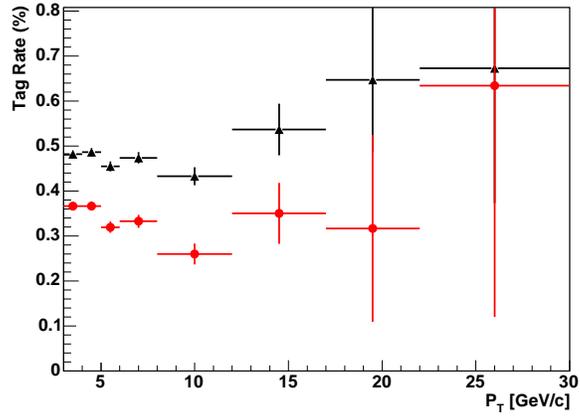} \\
\mbox{\bf (a)} \\
\includegraphics[width=3.375in]{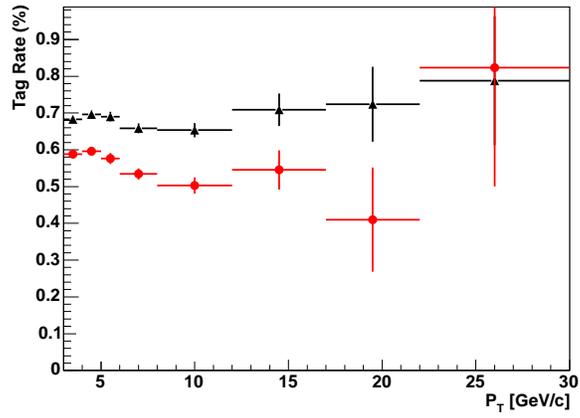}  \\
\mbox{\bf (b)} \\
\includegraphics[width=3.375in]{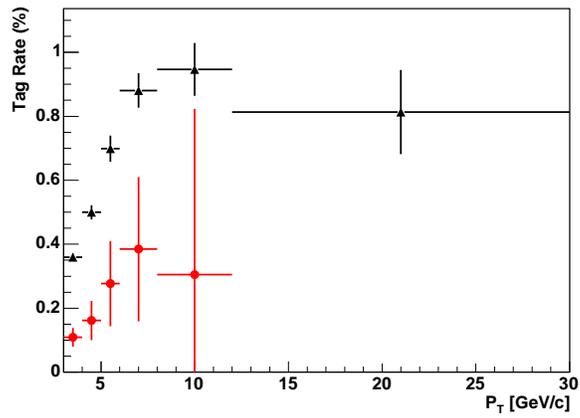} \\
\mbox{\bf (c)} \\
\end{center}
\caption{The measured~(triangles) and sideband-subtracted~(circles)
tag probabilities as a function of track $\Pt$ for (a)~pions,
(b)~kaons, and (c)~protons.  The uncertainties shown are statistical
only.} \label{fig:sidesub}
\end{figure}

We define a signal region for $\dstar$ and $\lambdaz$ decays as well
as sideband regions for each.  We measure the sideband-subtracted
tagging probability for $K$, $\pi$ and $p$ tracks using events in
the signal, corrected for the enhanced probabilities in the
sidebands. The sideband regions have a higher $\sltmu$ per track tag
probability because they are enriched in HF as a result of the
two-track trigger described above.  The signal and sideband regions
are given in Tab.~\ref{tab:windows} and the sideband subtraction is
done using the fits~\cite{UlyThesis} shown in
Fig.s~\ref{fig:diffmass} and~\ref{fig:lambdamass}.  The tag
probabilities before and after sideband subtraction are shown as a
function of $\Pt$ in Fig.~\ref{fig:sidesub}.  We note that there are
systematic uncertainties due to the choice of fit functions and in
particular the quality of the fits in the sideband regions. These
systematics, and all others associated with the construction of the
mistag matrix, are evaluated by testing the predictive power of the
matrix on a variety of independent data samples, as described in
Section~\ref{sec:MisSys}.

\begin{table}%[p]
\begin{center}
\begin{tabular}{l c}
\hline \hline Region          & Mass Window ($\MeVcc$) \\ \hline
$\dstar$ Signal     & 142.421 $<\Delta m<$ 148.421 \\
$\dstar$ Sidebands  & 139.6 $<\Delta m< $ 141 or \\
            & 152 $<\Delta m<$ 162.5 \\ \hline
$\lambdaz$ Signal   & 1109.683 $<m<$ 1121.683 \\
$\lambdaz$ Sidebands    & 1090 $<m<$ 1105.683 or \\
            & 1125.683 $<m<$ 1170 \\
\hline \hline
\end{tabular}
\end{center}
\caption{Mass windows used in determining the status of a $\dstar$
or $\lambdaz$ candidate.} \label{tab:windows}
\end{table}

The mistag matrix is designed to predict $\sltmu$ tags that arise
from both hadronic punch-through and decays-in-flight.  When a pion
or a kaon from a $\dstar$ decays in flight, the track may be poorly
reconstructed causing the reconstructed mass to fall outside of the
signal region defined in Tab.~\ref{tab:windows}. We measure the size
of this effect using $\dstar$ decays in a Monte Carlo sample and
make a correction.  The correction factor is calculated in three
bins in $\Pt$ (limited by the sample size of the Monte Carlo) and
shown in Tab.~\ref{tab:difcorr}.  Full details of the calculation of
the correction factor are given in~\cite{UlyThesis}.

\begin{table*}
\sans
\begin{center}
\begin{tabular}{l c c c c c}
\hline \hline
            & $\Pt$     & Frac. DIF       & After Reco.     & DIF Tag prob.   & Corr. Factor \\
            & [$\GeVc$] & [\%]            & [\%]            & [\%]           &             \\ \hline
            & 3--4      & 0.40 $\pm$ 0.05 & 0.20 $\pm$ 0.04 & 44.6 $\pm$ 2.0 & 1.25 $\pm$ 0.08 \\
   $\pi$    & 4--6      & 0.25 $\pm$ 0.04 & 0.16 $\pm$ 0.03 & 60.1 $\pm$ 2.3 & 1.16 $\pm$ 0.08 \\
            & $>6$      & 0.20 $\pm$ 0.04 & 0.16 $\pm$ 0.04 & 75.6 $\pm$ 2.6 & 1.09 $\pm$ 0.14 \\ \hline

            & 3--4      & 0.99 $\pm$ 0.09 & 0.77 $\pm$ 0.08 & 10.2 $\pm$ 1.3 & 1.04 $\pm$ 0.02 \\
     $K$    & 4--6      & 0.65 $\pm$ 0.06 & 0.51 $\pm$ 0.06 & 10.8 $\pm$ 1.3 & 1.02 $\pm$ 0.02 \\
            & $>6$      & 0.39 $\pm$ 0.05 & 0.23 $\pm$ 0.04 & 18.2 $\pm$ 1.8 & 1.05 $\pm$ 0.02 \\ \hline
\hline
\end{tabular}
\end{center}
\caption{Relevant numbers, from Monte Carlo simulation, in the
determination of the decay-in-flight correction for $\dstar$ decays.
These include the fraction of taggable tracks from $\dstar$s that
decay-in-flight (DIF), the same fraction after all reconstruction
requirements that fall inside the $\dstar$ signal window (shown in
Tab.~\ref{tab:windows}) and the probability for tagging a
decay-in-flight.  The last column gives the correction factor that
is applied to the measured tag probability to account for the bias
against decays-in-flight.} \label{tab:difcorr}
\end{table*}

\subsection{The Mistag Matrix}
\label{sec:matrix}

At this point we have $\sltmu$ tag probabilities for tracks from
$\pi$, $K$ and $p$, corrected for backgrounds (sideband-subtracted)
and for a bias against $\pi$ and $K$ decays-in-flight.  What remains
is to assemble these separate $\sltmu$ tag probabilities into a full
mistag matrix that can be used to predict the number of tags in
light-flavor jets in $W+$jets events.

\begin{figure}
\begin{center}
\includegraphics[width=3.375in]{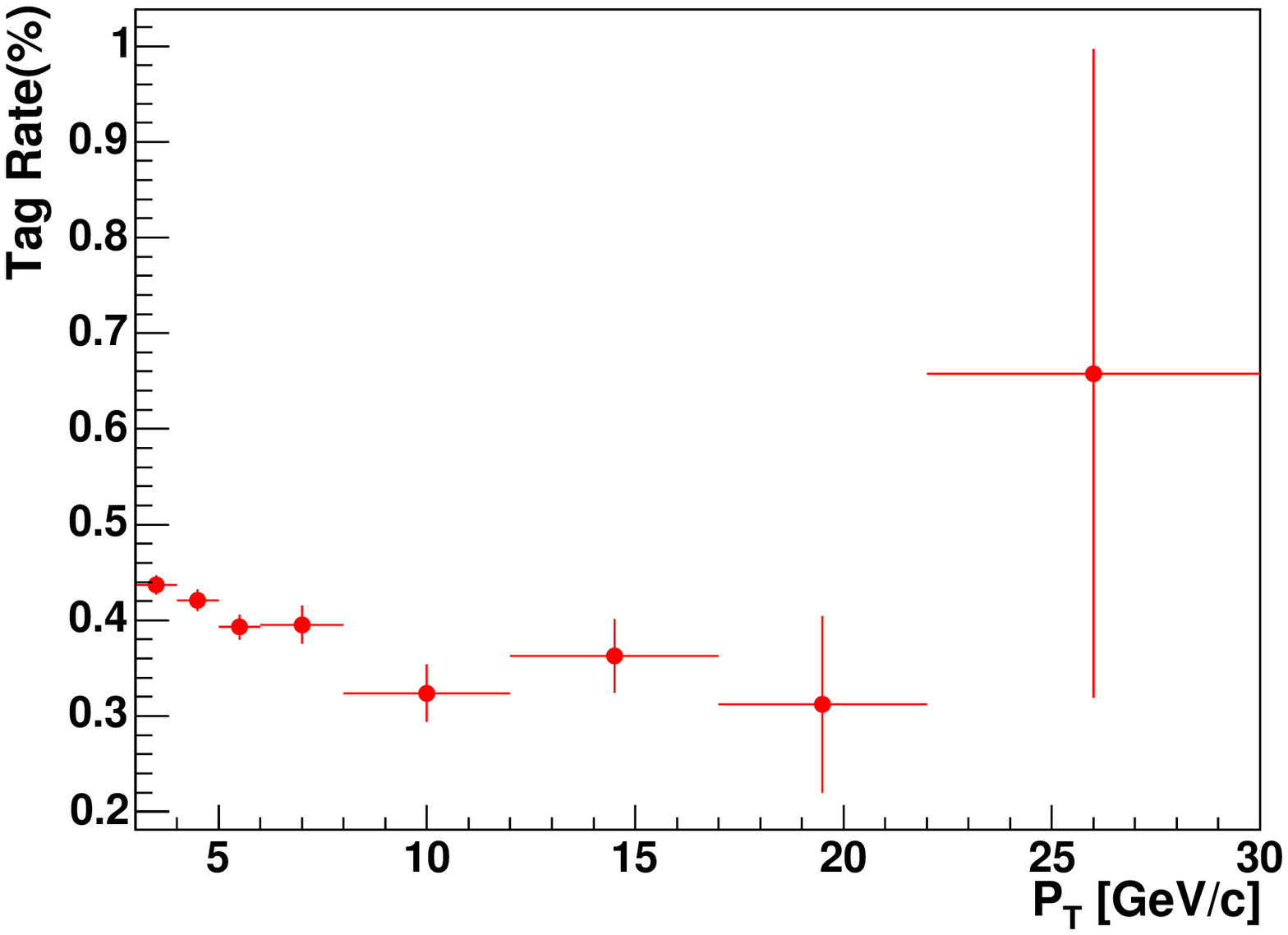}   \\
\includegraphics[width=3.375in]{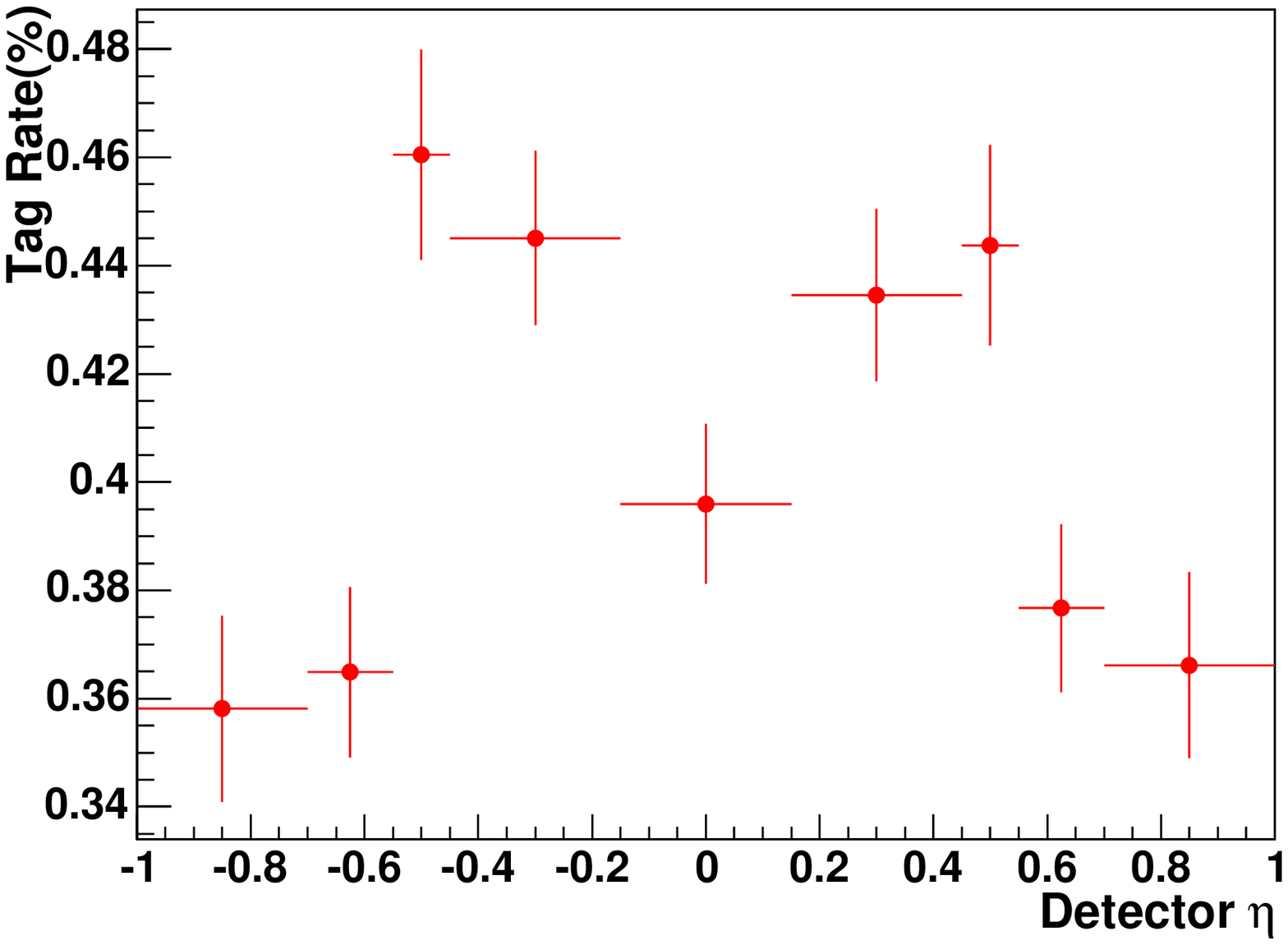} \\
\end{center}
\caption{The mistag probability per track as a function of track
$\Pt$ and detector $\eta$.  The histogram binning matches that of
the matrix.  The uncertainties shown are statistical only.  The
structure in the $\eta$ distribution is an artifact of the profile
of the absorber and geometrical coverage of the muon system.}
\label{fig:finalmat}
\end{figure}

To assemble the final mistag matrix, we take a weighted sum of the
individual $\pi$, $K$ and $p$ matrices as follows:
\begin{equation}
M_{ij}=W_{\pi}\cdot M^{\pi}_{ij} + W_{K}\cdot M^{K}_{ij} +
W_{p}\cdot M^{p}_{ij}. \label{eqn:finalmat}
\end{equation}
where $M_{ij}$ is the entry in the $i^{th}~\Pt$ and $j^{th}~\eta$
bin of the final matrix and $M^{\pi}_{ij}$, $M^{K}_{ij}$ and
$M^{p}_{ij}$ are the corresponding entries in the $\pi$, $K$ and $p$
matrices.  The weights $W_{\pi}=71.9$\%, $W_K=15.6$\% and
$W_p=12.5$\%, are taken from the taggable-track particle content of
light-quark jets in $\Alpgen$ $W+$jets Monte Carlo.
Figure~\ref{fig:finalmat} shows the final tag probability for the
eight $\Pt$ bins (integrated over $\eta$) and the nine $\eta$ bins
(integrated over $\Pt$).  The features in the $\eta$ distribution
are due to the profile of absorber in front of, and the coverage of,
the muon system. The average tag probability per track in the matrix
is ($0.41\pm0.01$)\% per track.

The uncertainties associated with the probabilities in the mistag
matrix include uncertainties from the sideband subtraction, the
decay-in-flight correction and the weighting of $\pi$, $K$ and $p$
probabilities, to name just a few.  To evaluate the overall
systematic uncertainty on the number of light-quark tags predicted
by the matrix, one possibility would be to carefully evaluate the
size of each of these uncertainties.  However, there is no
straightforward way to do this.  Instead, as described in detail in
Section~\ref{sec:MisSys}, we directly test the predictive power of
the matrix, using event samples acquired with jet triggers, and use
the control samples of $W+1$ and 2 jet events to further validate
the technique and establish that we have not underestimated the size
of the systematic uncertainty.

\section{Background Evaluation} \label{ch:Backgrounds}

The dominant background contribution to the $\ttbar$ signal in this
analysis comes from mistags in $W$+jets events.
%In this context,
%mistags are particles identified as $\sltmu$ tags whose origin is
%not semileptonic HF decay to muons. This includes non-muons which
%are tagged, such as a pion faking a muon, as well as muons from pion
%or kaon decays-in-flight.
Another smaller, yet still significant background comes from
$W$~bosons produced in association with heavy flavor~($W\bbbar$,
$W\ccbar$ and $Wc$).  The estimate of the mistag background is
described in Section~\ref{sec:fakeBkg}, while the $W$+HF background
estimate is described in Section~\ref{sec:hfBkg}.

Other backgrounds that can produce a $W$~boson and an $\sltmu$ tag
that are not accounted for by the mistag matrix include dibosons
($WW$, $ZZ$, $WZ$), $Z\rightarrow\tptm$, single~top, QCD multijet
backgrounds including $\bbbar$, and residual Drell-Yan~($\mu\mu$)
events not removed by the dimuon removal. QCD and Drell-Yan
backgrounds are measured using the data, as described in detail in
Sections~\ref{sec:qcdBkg} and~\ref{sec:dyBkg} below.  The remaining
backgrounds are estimated from Monte Carlo as described in
Section~\ref{sec:mcBkg}. We treat QCD independently of the
calculation of mistags in $W$+jets events because events that enter
our sample by mimicking the signature of a $W$~boson can have a
significantly larger tag rate than true $W$ events. The enhanced tag
rate arises due to the contribution of $\bbbar$ events to the QCD
background and because of the correlation between the tag rate and
measured $\met$ in events in which the $\met$ arises from jet
mismeasurement or semileptonic HF decay rather than from a neutrino
in a $W$ boson decay. In order to avoid double counting we correct
the estimate of tags in $W$+jets events by $(1-F_{QCD})$, where
$F_{QCD}$ is the QCD multijet fraction in the $W$+jets candidate
sample.

\subsection{Mistags} \label{sec:fakeBkg} The background due to
mistags is evaluated using the track-based mistag matrix described
in Section~\ref{sec:FakeMatrix}. To predict the number of events
from $W$+jets with at least 1 mistag, we apply the mistag matrix to
all pretag events according to:
\begin{equation}
N_{\rm raw}^{\rm Wjtag}= \sum_{\mathrm{events}}
\left[1-\prod_{i=1}^{N_{\rm trk}}\left(1-{\cal
P}({\Pt}_i,\eta_i)\right)\right ] , \label{eq:Fake}
\end{equation}
where the sum runs over each event in the pretag sample, and the
product is over each taggable track in the event. ${\cal
P}({\Pt}_i,\eta_i)$ is the probability from the mistag matrix for
tagging the $i^{th}$ track with parameters ${\Pt}_i$ and $\eta_i$.
Note that the sum over the events in equation~\ref{eq:Fake} includes
any $\ttbar$ events that are in the pretag sample. We correct for
the resulting overestimate of the background at the final stage of
the cross section calculation (see Section~\ref{sec:xsCalc}).  It
also includes $W+$HF events, diboson events, etc. Therefore mistags
from these backgrounds are included here.  Tags from muons resulting
from the decay of HF hadrons or $W$~or $Z$~bosons in these
backgrounds are calculated separately using Monte Carlo simulations,
as described in Sections~\ref{sec:hfBkg} and~\ref{sec:mcBkg}.  To
avoid any double counting the Monte Carlo estimates of the
contributions from these backgrounds do not include any mistags.

%No correction is made for the inclusion of these other backgrounds.
%Instead, the mistag matrix is used to predict the mistags from those
%backgrounds, and the estimates of the contributions from those
%backgrounds are limited to the {\it real} tags in those events~(i.e.
%tags from muons resulting from the decay of HF hadrons or $W$~or
%$Z$~bosons).

A fraction, $F_{QCD}$, of the events in the signal region are QCD
events for which the background is estimated separately. Therefore,
we correct the prediction of equation~\ref{eq:Fake} according to
\begin{equation}
N_{\rm corr}^{Wjtag}=(1-F_{QCD})\cdot N_{\rm raw}^{\rm Wjtag}.
\label{eq:fakeqcd}
\end{equation}

The background estimate from the application of the mistag matrix is
shown in Tab.~\ref{tab:FakeBkg}.  We list here both the raw
prediction and that corrected by $(1-F_{QCD})$. The calculation of
$F_{QCD}$ is described in Section~\ref{sec:Fqcd}.

\begin{table}[htbp]
\sans
\begin{center}
\begin{tabular}{l c c c c c}
\hline \hline
                & 1 jet     & 2 jet & 3 jets & $\ge$ 4 jets &  $\ge$ 3 jets \\ \hline
$N_{\rm raw}^{\rm Wjtag}$   & 641$\pm$32 & 238$\pm$12 & 55.0$\pm$2.8 & 32.7$\pm$1.6 & 87.5$\pm$4.4 \\
$N_{\rm corr}^{\rm Wjtag}$  & 622$\pm$31 & 226$\pm$12 & 53.0$\pm$2.7 & 31.4$\pm$1.6 & 84.5$\pm$4.3 \\
\hline \hline
\end{tabular}
\caption{Summary of background estimate from mistags in $W$+jets
events.  These numbers include a contribution from $\ttbar$ events
in the $W$+jets sample that is removed in the final cross section
calculation, as described in Section~\ref{sec:xsCalc}}.
\label{tab:FakeBkg}
\end{center}
\end{table}

\subsection{$W$+Heavy Flavor} \label{sec:hfBkg} The evaluation of
background tags from the semileptonic decays of HF quarks in
$W\bbbar$, $W\ccbar$ and $Wc$ events is done using the $\Alpgen$
Monte Carlo program. We determine the fraction of $W$+jets events
that contain heavy flavor at the pretag level, $F_{HF}$, and the
tagging efficiency, $\epsilon_{HF}$, for these events and then
normalize the total to the number of $W+$jets events seen in the
data.  The final prediction of the number of tags from
$W$+heavy-flavor events is:
\begin{equation}
N_{HF}=(1-F_{QCD}-F_{other})\cdot N_{pretag}\cdot F_{HF}\cdot
\epsilon_{HF}. \label{eq:NHF}
\end{equation}
where $F_{QCD}$ is the fraction of QCD events in the pretag sample
and $F_{other}$ is the fraction of other, non-$W+$jets backgrounds.
As with the mistag prediction, correction for $\ttbar$ in the pretag
sample is done as part of the final cross section calculation.

This procedure is used because the theory cross sections for the
$W\bbbar$, $W\ccbar$ and $Wc$ processes have large uncertainties,
whereas the uncertainties on the fraction of events with
heavy-flavor jets are smaller. This procedure follows that used
in~\cite{SecVtxPRD}.

\subsubsection{Heavy-Flavor Fractions \& Tagging Efficiency}
\label{sec:hffrac} The HF fractions of events in the $W$+jets sample
are determined by measuring the fractions in Monte Carlo and then
scaling those fractions by a multiplicative factor of $1.15\pm0.35$,
determined by comparing measured HF fractions in inclusive jet data
with those predicted by $\Alpgen$.

The $\Alpgen$ HF fractions, broken down according the number of $b$-
or $c$-jets, are shown in Tab.~\ref{tab:hffrac}.  In addition to the
uncertainty on the HF-fraction scaling, an additional uncertainty on
the $\Alpgen$ fractions is determined by varying the $\Alpgen$
generator parameters such as $Q^2$ and the quark masses.

The Monte Carlo is also employed to determine the efficiency for
tagging a muon from a semileptonic heavy-flavor decay in
$W$+heavy-flavor events.    As with the $\ttbar$ tagging efficiency
described in Section~\ref{sec:AccEff}, tags are assigned based on
the $\sltmu$ tagging efficiency measured in the data
(Fig.~\ref{fig:eff_central} and~\ref{fig:eff_cmx}). The results are
shown in Tab.~\ref{tab:hffrac}.  Note that we do not include here
the additional efficiency that arises from mistags in real HF jets,
because this is included in the mistag evaluation given in
Tab.~\ref{tab:FakeBkg}.

\begin{table*}[htbp]
\sans
\begin{center}
\begin{tabular}{l c c c c c}
\hline \hline     & 1 jet         & 2 jets    & 3 jets    & $\geq4$
jets  & $\geq3$ jets  \\ \hline \multicolumn{6}{c}{Category 2 $b$}
\\ \hline
$F_{HF}~(\%)$    & & 0.9$\pm$0.3   & 1.8$\pm$0.7   & 2.8$\pm$1.1   & 2.0$\pm$0.8   \\
$\epsilon_{HF}$ & & 7.8$\pm$0.2   & 8.4$\pm$0.2   & 8.5$\pm$0.3   &
8.4$\pm$0.2   \\  \hline \multicolumn{6}{c}{Category 1 $b$}\\
\hline
$F_{HF}~(\%)$ & 0.7$\pm$0.3       & 1.4$\pm$0.5   & 2.6$\pm$1.0   & 3.0$\pm$1.1   & 2.7$\pm$1.0   \\
$\epsilon_{HF}$ & 3.54$\pm$0.05 & 4.30$\pm$0.06 & 5.5$\pm$0.1   &
5.8$\pm$0.2   & 5.53$\pm$0.09 \\  \hline \multicolumn{6}{c}{Category
2 $c$} \\  \hline
$F_{HF}~(\%)$ &           & 1.3$\pm$0.5   & 2.8$\pm$1.1   & 4.5$\pm$1.7   & 3.1$\pm$1.2   \\
$\epsilon_{HF}$ &       & 3.1$\pm$0.1   & 3.6$\pm$0.1   &
3.5$\pm$0.2 &
3.6$\pm$0.1   \\  \hline \multicolumn{6}{c}{Category 1 $c$} \\
\hline $F_{HF}~(\%)$ & 5.5$\pm$2.1 & 8.9$\pm$3.4   & 11.0$\pm$4.1 &
11.5$\pm$4.4 & 11.1$\pm$4.2  \\
$\epsilon_{HF}$ & 1.52$\pm$0.02 & 1.70$\pm$0.03 & 2.04$\pm$0.07 &
2.05$\pm$0.06 & 2.04$\pm$0.06 \\ \hline $W\bbbar$+$W\ccbar$+$Wc$
Background & 145$\pm$55   & 66.6$\pm$25.2 & 15.3$\pm$5.8 &
8.5$\pm$3.2  & 23.0$\pm$8.7
%& 145.2$\pm$38.4   & 65.72$\pm$17.47 & 14.98$\pm$3.98 &
%8.37$\pm$2.23  & 22.61$\pm$6.00
\\ \hline \hline \hline
\end{tabular}
\caption{The heavy-flavor fractions, $F_{HF}$, tagging efficiencies,
$\epsilon_{HF}$, and $W+$heavy-flavor background evaluated using
$\Alpgen$ Monte Carlo. The fractions are scaled by 1.15 as described
in the text. The uncertainty on the heavy-flavor fractions includes
that from the scaling factor and from variation of the $\Alpgen$
parameters.} \label{tab:hffrac}
\end{center}
\end{table*}

Armed with these HF fractions and tagging efficiencies, the number
of tagged events from $W+$HF is evaluated according to
equation~\ref{eq:NHF} above.  The results are given in the last line
of Tab.~\ref{tab:hffrac}.

\subsection{QCD Background} \label{sec:qcdBkg}

The background due to tags in QCD events that enter the signal
sample is estimated by calculating the fraction of QCD events in the
$W$+jets data and applying the standard mistag matrix times a
multiplicative factor.  The multiplicative factor is required
because the tagging rate of QCD events that enter the pretag sample
is higher than the corresponding tagging rate for $W$+jets events.

\subsubsection{The QCD Fraction} \label{sec:Fqcd}
The fraction of QCD events before $\sltmu$ tagging is determined
using the isolation, $I$ (see Section~\ref{sec:EvSel}), and $\met$
of events with high-$\Pt$ leptons and jets. Under the assumption
that $I$ and $\met$ are uncorrelated for QCD events, the number of
QCD events in the $\ttbar$ signal region can be found by
extrapolation from the non-signal regions
\begin{equation}
N_{D}^{QCD}=\frac{N_C}{N_A} N_B, \label{eq:nqcdD}
\end{equation}
Where region $D$, the signal region, and regions $A,B$ and $C$ are
defined according to
\begin{eqnarray}
\mathrm{Region~A}: \met < 20~\GeV; & I > 0.2 \nonumber\\
\mathrm{Region~B}: \met < 20~\GeV; & I < 0.1 \nonumber\\
\mathrm{Region~C}: \met > 30~\GeV; & I > 0.2 \nonumber\\
\mathrm{Region~D}: \met > 30~\GeV; & I < 0.1 \nonumber .
\end{eqnarray}

The event counts used in equation~\ref{eq:nqcdD} are corrected for
Monte Carlo predictions of the number of $W+$jets and $\ttbar$
events in Regions $A,B$ and $C$.  The QCD fraction, $F_{QCD}$ is
then given by $N_{D}^{QCD}$ divided by the total number of events in
Region D.
%$F_{QCD}=N_{D}^{QCD}/N_D^{tot}$, where $N_D^{tot}$
%is the total number of events in Region $D$.
The QCD fractions are given in Tab.~\ref{tab:fqcd}.

\begin{figure}
\begin{center}
\includegraphics[width=3.375in]{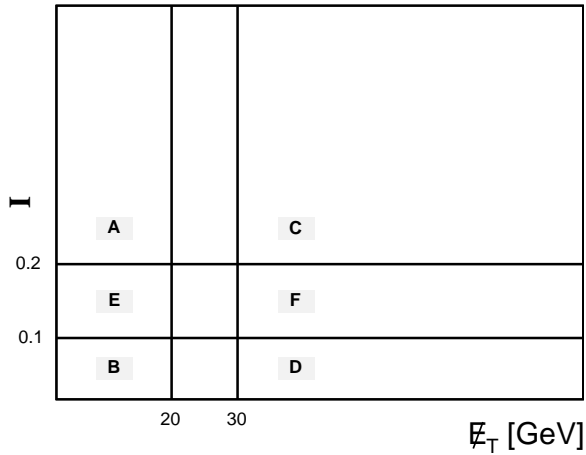}
\end{center}
\caption{A diagram illustrating the regions defined in the $\met$
vs. I plane.} \label{fig:MetIso}
\end{figure}

To evaluate the accuracy of the $\met$-$I$ prediction, two
complementary regions in the plane are defined as:
\begin{eqnarray}
\mathrm{Region~E}: \met < 20~\GeV; & 0.1 < I < 0.2 \nonumber\\
\mathrm{Region~F}: \met > 30~\GeV; & 0.1 < I < 0.2 \nonumber .
\end{eqnarray}
The different regions in the $\met$-$I$ plane are shown in
Fig.~\ref{fig:MetIso}. Region~F is outside the signal region and,
once contamination from $W$+jets and $\ttbar$ is removed, should
have a QCD fraction, $F_{QCD}^{F}$, of approximately 1.0.
$F_{QCD}^{F}$ is given by:
\begin{equation}
F_{QCD}^{F}=\frac{N_C\cdot N_E}{N_A\cdot N_F}. \label{eq:fqcdF}
\end{equation}
We use the difference of $F_{QCD}^{F}$ from 1.0 to estimate a
systematic uncertainty on the $\met$ vs. $I$ technique.  The results
are given in Tab.~\ref{tab:fqcd}.  Given the deviation from 1.0 in
the $\geq 3$ jets data, we assign an 11\%~(120\%) systematic
uncertainty to $F_{QCD}$ for electrons~(muons).
\begin{table*}[htbp]
%\begin{adjustwidth}{-4em}{-4em}
\sans
\begin{center}
\renewcommand{\arraystretch}{1.25}
\begin{small}
\begin{tabular}{l c c c c c}
\hline \hline
        & 1 jet         & 2 jets        & 3 jets          & $\ge$ 4 jets    &  $\ge$ 3 jets   \\ \hline
\multicolumn{6}{c}{Electron channel} \\ \hline
F$_{QCD}$ & 0.0423$\pm$0.0009 & 0.070$\pm$0.002 & 0.049$\pm$0.003 & 0.056$\pm$0.006 & 0.051$\pm$0.003 \\
Region F               & 0.95$\pm$0.04     & 0.97$\pm$0.06   & 0.84$\pm$0.10   & 1.06$\pm$0.24   & 0.89$\pm$0.09   \\
\hline \multicolumn{6}{c}{Muon channel} \\ \hline
F$_{QCD}$ & 0.0118$\pm$0.0004 & 0.020$\pm$0.001 & 0.013$\pm$0.004 & 0.007$\pm$0.004 & 0.011$\pm$0.003 \\
Region F               & 0.58$\pm$0.05     & 0.65$\pm$0.07   & 0.31$\pm$0.09   & 2.27$\pm$4.25   & 0.45$\pm$0.13   \\
\hline \hline
\end{tabular}
\renewcommand{\arraystretch}{1.00}
\end{small}
\caption{The fractions, $F_{QCD}$, of lepton-plus-jets events due to
QCD multijet processes before $\sltmu$ tagging.  The uncertainties
on the $F_{QCD}$ values are statistical only.  Also shown is the
measured QCD fraction in Region~F, used to assign a systematic
uncertainty on the $F_{QCD}$ prediction.} \label{tab:fqcd}
\end{center}
%\end{adjustwidth}
\end{table*}

\subsubsection{The Tag Rate of QCD Events} \label{sec:kqcd}
\begin{figure}
\begin{center}
\includegraphics[width=3.375in]{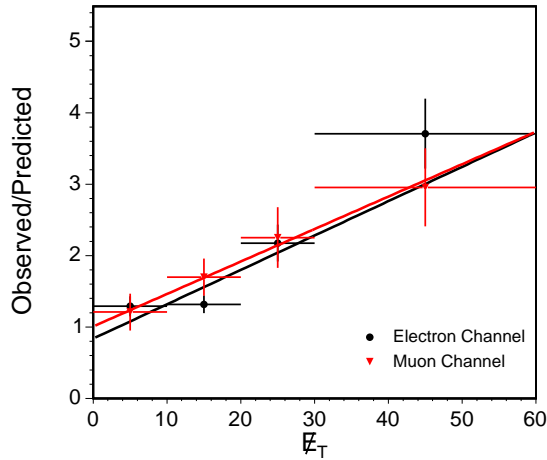}
\end{center}
\caption[The ratio of observed to predicted tags as a function of
$\met$.]{The ratio of observed to predicted tags as a function of
$\met$ in data with a non-isolated primary lepton~($I\geq 0.2$).}
\label{fig:kmet}
\end{figure}

The tag rate in QCD events that populate our signal region is
enhanced relative to the rate predicted by the mistag matrix. There
are two sources for this enhancement. First, much of the $\met$ in
QCD events is due to mismeasurement of jet energies, which is
correlated with the tag rate~(see Section~\ref{sec:MisSys}).  As
seen in Fig.~\ref{fig:kmet}, the ratio of observed to predicted tags
increases with $\met$. Second, QCD includes $\bbbar$~and
$\ccbar$~events in which the high-$\Pt$ lepton comes from the
semileptonic decay of one of the $b$~or $c$~quarks; if the other
also decays semileptonically, it may be tagged by the $\sltmu$.

In the $\met$-$I$ plane, the region closest kinematically to the
signal region (Region~D) is the high-isolation and high-$\met$
region, Region~C. Region~C has the same $\met$ requirement that
Region~D does, and likewise it requires a high-$\Pt$ lepton in the
event. Therefore, the tag rate measured in Region~C is a good
representation of that of QCD events in Region~D.

The tag rate of QCD events is measured as an enhancement factor,
$k$, times the $W$+jets mistag probability.  We calculate $k$ as the
ratio of observed to predicted (by the mistag matrix) $\sltmu$ tags
in Region~C. The results are shown in Tab.~\ref{tab:QCDbkg}.

\subsubsection{The QCD Background Estimate} Having determined
$F_{QCD}$ and $k$, the QCD background is given by:
\begin{equation}
%N_{QCD}={\cal P}(\Pt ,\eta)\cdot N_D\cdot k\cdot F_{QCD}
N_{QCD}= N_{\rm raw}^{\rm Wjtag} \cdot k\cdot F_{QCD}
\label{eq:NQCD}
\end{equation}
where $N_{\rm raw}^{\rm Wjtag}$ comes from equation~\ref{eq:Fake}.
%represents the application of the mistag matrix to the events in
%Region D (the signal region: $\met >30~\GeV$, $I <0.1$ ).

\begin{table*}[htbp]
\sans
%\begin{adjustwidth}{-4em}{-4em}
\begin{center}
\renewcommand{\arraystretch}{1.25}
\begin{small}
\begin{tabular}{l c c c c c}
\hline \hline
        & 1 jet         & 2 jets        & 3 jets          & $\ge$ 4 jets    &  $\ge$ 3 jets   \\ \hline
\multicolumn{6}{c}{Electron channel} \\ \hline
$F_{QCD}$ & 0.042$\pm$0.005   & 0.070$\pm$0.008   & \multicolumn{3}{c}{$0.051\pm 0.006$} \\ %0.049$\pm$0.003 & 0.056$\pm$0.006 & 0.051$\pm$0.003 \\
        k & 5.3$\pm$0.4       & 3.9$\pm$0.4       & \multicolumn{3}{c}{$3.7\pm 0.5$} \\ %3.7$\pm$0.6     & 3.8$\pm$0.9     & 3.7$\pm$0.5     \\ \hline
$N_{QCD}$ & 82.6$\pm$11.9   & 38.1$\pm$6.1    & 6.3$\pm$1.2   & 3.6$\pm$0.7   & 9.9$\pm$1.9   \\
\hline \multicolumn{6}{c}{Muon channel} \\ \hline
F$_{QCD}$ & 0.0118$\pm$0.0004 & 0.0205$\pm$0.0009 & \multicolumn{3}{c}{$0.011\pm0.014$} \\ %0.013$\pm$0.004 & 0.007$\pm$0.004 & 0.011$\pm$0.003 \\
        k & 2.9$\pm$0.4     & 3.4$\pm$0.4     & \multicolumn{3}{c}{$3.0\pm 0.5$} \\ %2.6$\pm$0.6     & 3.6$\pm$1.0     & 3.0$\pm$0.5     \\ \hline
$N_{QCD}$ & 9.3$\pm$11.3  & 6.8$\pm$8.3   & 0.7$\pm$0.9   & 0.4$\pm$0.6   & 1.2$\pm$1.4   \\
\hline \multicolumn{6}{c}{Combined channels} \\ \hline
Total $N_{QCD}$ & 92$\pm$17   & 44.9$\pm$10.4 & 7.0$\pm$1.5   & 4.1$\pm$0.9   & 11.1$\pm$2.4  \\
\hline \hline
\end{tabular}
\renewcommand{\arraystretch}{1.00}
\end{small}
\caption{Summary of the QCD background estimate. The uncertainties
on the QCD fractions,~$F_{QCD}$, and on the number of QCD
events,~$N_{QCD}$, are systematic and statistical combined. The
$F_{QCD}$ and k values in the third column apply to 3, $\ge 4$ and
$\ge 3$ jets.} \label{tab:QCDbkg}
\end{center}
%\end{adjustwidth}
\end{table*}

\subsection{Drell-Yan${\boldmath\rightarrow\mu\mu}$}
\label{sec:dyBkg} Drell-Yan$\rightarrow\mu\mu$ events can survive
the $Z$, $\Upsilon$ and $\jpsi$ vetoes if one muon leg fails the
isolation requirement. We evaluate the number of residual Drell-Yan
events that remain in the signal sample after the dimuon vetoes by
measuring the number of $Z\rightarrow\mu\mu$ events in the data,
$N_{tag}^{Z}$, inside the $Z$ mass window, where one leg of the $Z$
is identified as an $\sltmu$ (these events are normally removed from
the signal sample). We then use an $\Alpgen$
$Z/\gamma^{*}\rightarrow\mu\mu$ Monte Carlo sample to estimate the
ratio, $R^{out/in}$, of events outside the $Z$-mass window to events
inside.  To increase the statistical precision the ratio is measured
with the $\met$ and $H_T$ and dilepton rejection cuts removed and
without requiring that the $\sltmu$ be inside a jet.  We use the
Monte Carlo to measure the ratio of efficiencies
$\epsilon^{out}/\epsilon^{in}$ of these requirements. With these
pieces, the number of residual Drell-Yan events, $N_{DY}$ is
\begin{equation}
N_{DY}=N_{tag}^{Z}\cdot R^\frac{out}{in}\cdot
\frac{\epsilon^{out}(\met,H_T,\mathrm{dilep,SLT\text{--}jet})}{\epsilon^{in}(\met,H_T,\mathrm{dilep,SLT\text{--}jet})},
\label{eq:dybkg}
\end{equation}
The results of the Drell-Yan calculation are given in
Tab.~\ref{tab:DYbkg}.
\begin{table*}[htbp]
\sans
\begin{center}
\begin{tabular}{l  c c c c c}\hline
                        &  1 jet   & 2 jets & 3 jets & $\geq$ 4 jets  & $\geq$ 3 jets \\ \hline
$N^Z_{tag}$ (data)  & 27       & 25     & 3      & 0          & 3         \\
$R^\frac{out}{in}$  & 0.325$\pm$0.003 & 0.315$\pm$0.005 & \multicolumn{3}{c} {0.312$\pm$0.006} \\
$\epsilon^{out}$ (\%)   & 0.71$\pm$0.04   & 1.62$\pm$0.10   & \multicolumn{3}{c} {2.85$\pm$0.24} \\
$\epsilon^{in}$ (\%)    & 1.03$\pm$0.05   & 3.10$\pm$0.12   & \multicolumn{3}{c} {3.24$\pm$0.18} \\
$R^\frac{out}{in}\cdot\frac{\epsilon^{out}}{\epsilon^{in}}$
            & 0.223$\pm$0.018 & 0.165$\pm$0.013 & \multicolumn{3}{c} {0.274$\pm$0.027} \\\hline
Drell-Yan Total     & 6.02$\pm$1.25   & 4.12$\pm$0.88   & 0.82$\pm$0.44 & 0.00$\pm$0.19 & 0.82$\pm$0.48\\
 \hline\hline
\end{tabular}
\caption[Drell-Yan background summary.]{Drell-Yan background
summary. Uncertainties are statistical only.  The values in the
third column apply to 3, $\ge 4$ and $\ge 3$ jets.}
\label{tab:DYbkg}
\end{center}
\end{table*}

\subsection{Monte Carlo Driven Backgrounds} \label{sec:mcBkg}
Backgrounds from dibosons~($WW$, $WZ$, $ZZ$),
$Z\rightarrow\tau^{+}\tau^{-}$, and single top are determined from
Monte~Carlo. For each of these backgrounds, the estimated number of
tags is calculated as
\begin{equation}
N_{i}=\sigma_{i}\cdot A_{i}\cdot\epsilon_{i}\cdot\int L \mathrm{dt}.
\label{eq:mcbkg}
\end{equation}
Here $\sigma_i$ is the theoretical cross section.  The acceptance,
$A_i$, and the $\sltmu$ tagging efficiency, $\epsilon_i$, are
calculated from the Monte Carlo.  As with the $W+$HF evaluation in
Section~\ref{sec:hfBkg}, the efficiency includes only that due to
tagging a muon from a semileptonic heavy-flavor decay.  We do not
include mistags in the efficiency evaluation because this is
included as part of the background determined by the mistag matrix.
The background evaluations are shown in Tab.~\ref{tab:MCBkgs}.
\begin{table*}[h]
\sans
\begin{center}
\begin{tabular}{l c c c c c}
\hline
    & 1 jet & 2 jet & 3 jets & $\ge$ 4 jets &  $\ge$ 3 jets \\ \hline \hline
\multicolumn{6}{c}{$WW$} \\ \hline
$\sigma_{theory}$ & \multicolumn{5}{c}{12.4$\pm$1.2~pb} \\
$A$ (\%)    & 2.44$\pm$0.01   & 2.62$\pm$0.01   & 0.403$\pm$0.004 & 0.121$\pm$0.002 & 0.524$\pm$0.005  \\
$\epsilon$ (\%) & 0.49$\pm$0.03   & 0.76$\pm$0.04   & 0.88$\pm$0.09
& 1.56$\pm$0.23   & 1.06$\pm$0.09    \\ \hline
$N_{WW}$    & 2.986$\pm$0.299 & 5.001$\pm$0.394 & 0.892$\pm$0.190 & 0.475$\pm$0.118 & 1.395$\pm$0.228  \\
\hline \multicolumn{6}{c}{$WZ$} \\ \hline
$\sigma_{theory}$ & \multicolumn{5}{c}{3.96$\pm$0.40~pb} \\
$A$ (\%)    & 1.085$\pm$0.007 & 1.317$\pm$0.007 & 0.233$\pm$0.003 & 0.070$\pm$0.002 & 0.302$\pm$0.004  \\
$\epsilon$ (\%) & 0.85$\pm$0.06   & 1.72$\pm$0.07   & 1.46$\pm$0.16
& 2.69$\pm$0.39   & 1.77$\pm$0.15    \\ \hline
$N_{WZ}$    & 0.740$\pm$0.075 & 1.821$\pm$0.128 & 0.274$\pm$0.044 & 0.151$\pm$0.036 & 0.432$\pm$0.058  \\
\hline \multicolumn{6}{c}{$ZZ$} \\ \hline
$\sigma_{theory}$ & \multicolumn{5}{c}{3.4$\pm$0.3~pb} \\
$A$ (\%)    & 0.104$\pm$0.002 & 0.097$\pm$0.002 & 0.060$\pm$0.001 & 0.012$\pm$0.001 & 0.042$\pm$0.001  \\
$\epsilon$ (\%) & 1.0$\pm$0.2     & 2.4$\pm$0.3     & 2.3$\pm$0.5
& 1.6$\pm$0.5     & 2.1$\pm$0.4      \\ \hline
$N_{ZZ}$    & 0.07$\pm$0.02   & 0.16$\pm$0.05   & 0.05$\pm$0.02   & 0.013$\pm$0.006 & 0.06$\pm$0.02  \\
\hline
\multicolumn{6}{c}{$\mathrm{Drell-Yan}\rightarrow\tau\tau$}\\
\hline
$\sigma_{theory}$ & \multicolumn{5}{c}{333$\pm$4.2~pb} \\
$A$ (\%)    & 0.112$\pm$0.001 & 0.054$\pm$0.001 & 0.0058$\pm$0.0004 & 0.0014$\pm$0.0002 & 0.0073$\pm$0.0004  \\
$\epsilon$ (\%) & 0.4$\pm$0.1     & 0.4$\pm$0.1     & 1.6$\pm$0.7
& 1.3$\pm$0.5       & 1.3$\pm$0.5  \\ \hline
$N_{\mathrm{Drell-Yan}\rightarrow\tau\tau}$ & 2.65$\pm$0.57 & 1.54$\pm$0.43 & 0.65$\pm$0.28 & 0.13$\pm$0.05 & 0.65$\pm$0.27  \\
\hline \multicolumn{6}{c}{$s$-channel Single Top} \\ \hline
$\sigma_{theory}$ & \multicolumn{5}{c}{0.88$\pm$0.11~pb} \\
$A$ (\%)    & 1.12$\pm$0.01 & 2.66$\pm$0.01 & 0.717$\pm$0.005 & 0.203$\pm$0.003 & 0.920$\pm$0.006  \\
$\epsilon$ (\%) & 5.0$\pm$0.1 & 9.7$\pm$0.1 & 10.2$\pm$0.2 &
11.1$\pm$0.4 & 10.4$\pm$0.2  \\ \hline
$N_{s-\mathrm{chan}}$ & 1.00$\pm$0.11 & 4.61$\pm$0.46 & 1.31$\pm$0.15 & 0.40$\pm$0.05 & 1.71$\pm$0.19  \\
\hline \multicolumn{6}{c}{$t$-channel Single Top} \\ \hline
$\sigma_{theory}$ & \multicolumn{5}{c}{1.98$\pm$0.08~pb} \\
$A$ (\%)    & 1.91$\pm$0.01 & 2.10$\pm$0.01 & 0.345$\pm$0.003 & 0.057$\pm$0.001 & 0.402$\pm$0.004  \\
$\epsilon$ (\%) & 4.38$\pm$0.09 & 5.19$\pm$0.09 & 5.96$\pm$0.24 &
7.37$\pm$0.64 & 6.16$\pm$0.22  \\ \hline
$N_{t-\mathrm{chan}}$ & 3.36$\pm$0.37 & 4.39$\pm$0.47 & 0.83$\pm$0.11 & 0.17$\pm$0.03 & 1.00$\pm$0.13   \\
\hline
\end{tabular}
\caption{Summary of Monte Carlo derived backgrounds.  The
theoretical cross
sections~\cite{Campbell:1999ah},~\cite{Acosta:2004uq},~\cite{Sullivan:2004ie}
are inclusive. The acceptance, A, includes the branching fraction to
events with $N$~jets, and the efficiency for finding an $\sltmu$ in
these events is $\epsilon$.} \label{tab:MCBkgs}
\end{center}
\end{table*}

\section{Systematic Uncertainties}\label{sec:sys}
Systematic uncertainties in this analysis come from Monte Carlo
modeling of the geometrical and kinematic acceptance, knowledge of
the $\sltmu$ tagging efficiency, the effect on the acceptance of the
uncertainty on the jet energy scale, uncertainties on the background
predictions, and the uncertainty on the luminosity.  The evaluation
of the size of each of these uncertainties is described below.

\subsection{Systematic Uncertainties on Acceptance and Efficiency}\label{sec:AccEffSys}
Monte Carlo modeling of geometrical and kinematic acceptance
includes effects of parton distribution functions (PDFs),
initial-state radiation (ISR), final-state radiation (FSR), and jet
energy scale. These are estimated by comparing different choices for
PDFs, varying ISR, FSR and the jet energy scale in the Monte Carlo
and comparing the $\Pythia$ generator with $\Herwig$.

The PDF uncertainty is evaluated from 3 contributions. The first is
obtained by varying the PDF according to the 20 CTEQ
eigenvectors~\cite{cteq5l}to account for the uncertainty on the PDF
fit. The second is the difference between the CTEQ5L PDF used for
the acceptance measurement with that obtained using
MRST98~\cite{mrst98} in the default configuration to account for the
type of PDF fit used. The third is evaluated comparing the default
MRST with two alternative choices of $\alpha_s$ to get an estimate
of the uncertainty due to the value of $\alpha_s$. The three
contributions in quadrature yield an acceptance uncertainty of
0.9\%.

The uncertainty due to the limited knowledge of ISR is constrained
by studies of radiation in Drell-Yan events in the data.  We vary
both ISR and FSR in the $\ttbar$ Monte Carlo within the allowed
range and add the deviations in quadrature.  The systematic
uncertainty due to this effect is 0.8\%.

The uncertainty on the acceptance due to the uncertainty in the jet
energy scale is measured by shifting the energies of the jets in
$\ttbar$ Monte Carlo by $\pm 1\sigma$ of the jet energy
scale~\cite{JetCorr}.  The resulting uncertainty on the acceptance
is 4.1\%.

The effects of generator modeling of the $\ttbar$ kinematics are
measured by comparing the acceptance from $\Pythia$ and $\Herwig$.
The result is a 2.4\% uncertainty.

As described in Section~\ref{sec:acceptance}, a scale factor is
applied to the $\ttbar$ Monte Carlo data set to correct for lepton
ID efficiency differences between data and Monte Carlo.  This scale
factor has an associated uncertainty that yields a 2.9\% uncertainty
on the total $\ttbar$ acceptance.

The systematic uncertainty on the $\sltmu$ tagging efficiency in
$\ttbar$ events is comprised of three parts. First, the uncertainty
due to the $\Pt$ dependence of the $\sltmu$ efficiency curves which
is evaluated by remeasuring the $\ttbar$ event tagging efficiency
with the $\pm 1\sigma$ curves shown as the dashed lines in
Fig.~\ref{fig:eff_central} and~\ref{fig:eff_cmx}.  Next, the tagging
efficiency measurement assumes that the efficiency for finding
tracks in jets in the COT is properly modeled in the simulation.
This assumption comes with a 5\% uncertainty, which was evaluated by
embedding MC tracks in data events. Finally, the statistical
uncertainty on the $\sltmu$ efficiency in $\ttbar$ events is
absorbed as a systematic uncertainty.  The three contributions
combine to give a systematic uncertainty of 5.1\%.

Adding all these contributions in quadrature gives a total
``Acceptance Modeling and Efficiency" systematic uncertainty of
7.7\%.

\subsection{Systematic Uncertainty of the Mistag Prediction} \label{sec:MisSys}
To measure the uncertainty on the predicted number of tags in
light-flavor jets, we test the predictive power of the mistag matrix
on large samples of events triggered on a single jet with an
(uncorrected) $\Et$ threshold of 20, 50, 70 or 100 GeV. Care must be
taken in several areas.  Because of the $\Et$ threshold on one jet
in the event, that `trigger jet' will have a bias against $\sltmu$
tags because particles that reach the muon chambers do not deposit
all their energy in the calorimeter and therefore reduce the
measured jet energy from its true value. Therefore, if no other jet
is above the trigger threshold, we remove the trigger jet from the
sample used to test the mistag prediction.  If there is an
additional jet above the trigger threshold, then all jets above
threshold are used. The opposite effect occurs in jets that are
measured well below trigger threshold.  In a di-jet event triggered,
for instance, with a 100 GeV threshold, a single recoil jet with
energy well below 100 GeV is likely to be significantly
mis-measured.  Such jets have an enhanced rate of $\sltmu$ tags
relative to jets in $W+$jets events because jet mis-measurement is
correlated with the population of $\sltmu$ tags through detector
cracks, hadronic punch-through of the calorimeter and real muon
content.  In addition to rejecting the trigger jet, we reject jets
in di-jet events if the recoil jet falls below the trigger
threshold.  For events with higher jet multiplicities we use only
tracks in jets that are separated from the trigger-jet axis by
$\Delta R$ between 0.7 and 2.6. These various criteria have been
chosen in order to provide, in the jet samples, a set of jets that
are similar in terms of $\sltmu$ tags to those found in $W+$jets
events~\cite{UlyThesis}.

To increase the number of jets available for the study we use, in
addition to the jet triggered data, events triggered on a single
photon candidate ($\gamma$+jets) with a threshold of 25, 50 or 70
GeV, and a $\Sigma\Et$-triggered sample, triggered on a four-jet
total energy of at least 100 GeV.  Jets in the $\gamma$+jets events
are selected in the same way as in the single jet triggered events.
All jets above 20 GeV are used in the $\Sigma\Et$ sample.

Since the mistag matrix is designed to predict the number of tags
from light-flavor jets, we must also suppress heavy-flavor jets in
our sample.  This is achieved by removing events in which any jet
has an identified secondary vertex~\cite{SecVtxPRD}, or in which the
mass of the tracks contained in a potential secondary vertex is
greater than 0.3 $\GeVcc$, or in which any jet contains a track with
an impact parameter significance~($d_0/\sigma_{d_0})\geq 2$.  This
is found~\cite{UlyThesis} to provide sufficient suppression of
heavy-flavor jets while leaving the remaining sample unbiased
against decays-in-flight inside the jet.

With the above jet selection, the systematic uncertainty is
determined using the difference between the number of $\sltmu$ tags
predicted by the mistag matrix and those observed in the data.  The
results are shown, as a function of the $\Et$ of the jet in
Tab.~\ref{tab:deltaetc}.
\begin{table}%[p]
\begin{center}
\begin{tabular}{l c c c}
\hline \hline Jet $E_T^{corr.}$ [$\GeV$] & Observed & Predicted  &
$\Delta$ [\%]  \\ \hline
20-30       & 1892  & 1641 $\pm$ 29 & -15.3 $\pm$ 3.3       \\
30-45       & 1561  & 1693 $\pm$ 45 & 7.8   $\pm$ 3.4   \\
45-65       & 701   & 768  $\pm$ 46 & 8.7   $\pm$ 6.4   \\
65-90       & 464   & 462  $\pm$ 46 & -0.5  $\pm$ 11.0  \\
$\geq$90    & 466   & 466  $\pm$ 76 & -0.1  $\pm$ 16.9  \\ \hline
$\geq$20    & 5084  & 5029 $\pm$ 219& -1.1  $\pm$ 4.6   \\
%\hline
%Weighted Sum    & 754 $\pm$ 13 & 758 $\pm$ 14  & 0.5 $\pm$ 2.6 \\ \hline
\hline \hline
\end{tabular}
\end{center}
\caption{Checks of the mistag matrix in different jet $\Et$ bins.
$\Delta=$(Pred.-Obs.)/Pred.  These values of $\Delta$ are weighted
using the $W$+3-or-more~jets distribution to determine a systematic
uncertainty on the mistag prediction.} \label{tab:deltaetc}
\end{table}

Finally, we use the $\Et$ spectrum of $W+$jets events from the Monte
Carlo to perform a weighted average over the deviations between
predicted and observed tags given in Tab.~\ref{tab:deltaetc}.  The
result is (Predicted $\sltmu$~-~Observed
$\sltmu$)/Pred.=($0.1\pm4.4$)\%.  We assign a systematic uncertainty
of 5\% on the prediction of the mistag matrix.

\subsection{Other Background Uncertainties}\label{sec:bkgsys}

\subsubsection{$W+$ Heavy Flavor Uncertainties}\label{sec:HFsys}
Three sources contribute to the uncertainty on the
$W\bbbar+W\ccbar$+$Wc$ background prediction: the choice of
$\Alpgen$ settings, the uncertainty associated with the scaling
factor that takes the heavy-flavor fraction in $\Alpgen$ to the
data, and the uncertainty on the tagging efficiency.  The
determination of the uncertainties on the $\Alpgen$ settings and the
scale factor are described in Section~\ref{sec:hfBkg}.  The
$\Alpgen$ settings contribute 23\% to the $W+$heavy-flavor
background uncertainty and the scale factor another 13\%. The
uncertainty on the heavy-flavor tagging efficiency is the same as
that for the $\ttbar$ tagging efficiency described in
Section~\ref{sec:AccEffSys}. The correlation between the efficiency
for the background determination and for the $\ttbar$ acceptance is
taken into account.

\subsubsection{QCD Background Uncertainties}\label{sec:QCDsys}
Uncertainties on the QCD background prediction are determined using
the level of agreement between predicted and measured events in
`Region~F', as described in Section~\ref{sec:Fqcd}.  We assign a
systematic uncertainty on the $F_{QCD}$ measurement of 11\% for
electrons and 120\% for muons, given conservatively by the worst
agreement of the Region~F prediction in each case.  We fold this in
with the statistical uncertainty on the $F_{QCD}$ determination, the
uncertainty on the correction factor $k$, both given in
Tab.~\ref{tab:QCDbkg}, and the 5\% systematic uncertainty due to the
application of the mistag matrix. The total QCD background
uncertainty is 19\% and 124\% for electrons and muons, respectively.
In the final determination of the QCD systematic, we add in
quadrature the separate effects on the cross section of the QCD
uncertainties for electrons and muons. The estimate of the QCD
background is correlated with the estimates of the mistags and
$W+$heavy-flavor
backgrounds~(equations~\ref{eq:fakeqcd},~\ref{eq:NHF},
and~\ref{eq:NQCD}). This is taken into account when determining the
effect on the $\ttbar$~cross section.  Together with a relatively
small QCD fraction of the events, the result is a rather small
effect on the cross section determination, despite the large
uncertainty on the QCD fraction itself.

\subsubsection{Other Background Uncertainties}\label{sec:othersys}
The systematic uncertainty on the small Drell-Yan background is
 determined by the statistical uncertainty of the estimate.
 Uncertainties on the Monte Carlo background predictions come from
 uncertainties in the cross sections for the various processes and
 from the event sizes of the Monte Carlo samples.  This uncertainty
 is reflected in the uncertainties quoted in
 Section~\ref{sec:mcBkg}.  The combined uncertainty on the Drell-Yan
 and Monte-Carlo-derived backgrounds is 11\%.

 \subsection{Summary of Systematic Uncertainties}\label{sec:sumsys}
The systematic uncertainties are summarized in Tab.~\ref{tab:sys}.
An additional systematic due to the uncertainty on the luminosity
determination (5.9\%~\cite{klimenko}), is treated separately.
%%%%%%%%%%%%%%%%%%
\begin{table*}[h]
\begin{center}
\begin{tabular}{l c c}
\hline \hline Source          & Fractional Sys. Uncert. (\%)  &
$\Delta\sigma_{\ttbar}$ (\%)  \\ \hline
Acceptance Modeling and & \multirow{2}{*}{7.7}      & $+8.3$    \\
SLT Tagging Efficiency  &               & $-7.5$    \\ \hline
Mistag Matrix Prediction& 5             & 3.6   \\
$W\bbbar+W\ccbar$+$Wc$
    Prediction  & 38                & 5.3   \\
QCD Prediction      & 19~(e) 124~($\mu$)        & 1.1   \\
Drell-Yan and other
    MC backgrounds  & 11             & 0.4   \\ \hline \hline
\multirow{2}{*}{Total Systematic Uncertainty}   &   & $+10.5$ \\
                        &   & $-10.0$ \\
 \hline \hline
\end{tabular}
\end{center}
\caption{Summary of systematic uncertainties.} \label{tab:sys}
\end{table*}
%%%%%%%%%%%%%%%%%%
\section{$\ttbar$ Production Cross Section}
\label{sec:xsCalc} Before calculating the cross section, the
estimated number of background events is corrected for $\ttbar$
events in the pretag sample using a simple iterative procedure. This
is required because we apply the mistag matrix to the events before
tagging to estimate the mistag and QCD backgrounds and also use the
pretag sample in the $W+$heavy-flavor background determination
assuming no $\ttbar$ content.  A summary of the number of observed
events and the background predictions, both before and after the
correction, as a function of the number of jets is given in
Tab.~\ref{tab:BkgSum}.  It is worth noting the excellent agreement
between the expected and observed tagged events in the $W+$1 and 2
jet samples, where the expectation is dominated by the mistag
contribution, and the $\ttbar$ contribution is negligible.  This is
a further validation of the mistag matrix.
\begin{table*}[htbp]
%\begin{adjustwidth}{-6em}{-6em}
\sans
\begin{center}
\renewcommand{\arraystretch}{1.25}
\begin{footnotesize}
\begin{tabular}{l c c c c c}
\hline \hline
& \multicolumn{2}{c}{$H_T\geq0~\GeV$} & \multicolumn{3}{c}{$H_T\geq200~\GeV$} \\ \hline
Background          & 1 jet & 2 jet & 3 jets & $\ge$ 4 jets &  $\ge$
3 jets \\ \hline Taggable events         & 75595        & 18264
& 2587           & 1120       & 3707       \\ \hline
Mistags             & 622$\pm$31   & 226$\pm$12   & 53.0$\pm$2.7 & 31.4$\pm$1.6 & 84.5$\pm$4.3 \\
$W\bbbar$+$W\ccbar$+$Wc$        & 145$\pm$55   & 66.6$\pm$25.2 &
15.3$\pm$5.8 & 8.5$\pm$3.2  & 23.0$\pm$8.7 \\ \hline
QCD multijet            & 91.9$\pm$16.5  & 44.9$\pm$10.4  & 7.0$\pm$1.5  & 4.1$\pm$0.9  & 11.1$\pm$2.4 \\
$WW$+$WZ$+$ZZ$          & 3.80$\pm$0.44    & 6.98$\pm$0.66    & 1.21$\pm$0.23  & 0.64$\pm$0.14  & 1.88$\pm$0.30  \\
Drell-Yan$\rightarrow\tau^+ \tau^-$ & 2.65$\pm$0.57    & 1.54$\pm$0.43    & 0.65$\pm$0.28  & 0.13$\pm$0.05  & 0.65$\pm$0.27  \\
Drell-Yan$\rightarrow\mu^+ \mu^-$   & 6.02$\pm$1.25    & 4.12$\pm$0.88    & 0.82$\pm$0.44  & 0.00$\pm$0.19  & 0.82$\pm$0.48  \\
Single top          & 4.36$\pm$0.39   & 9.00$\pm$0.66   & 2.14$\pm$0.18  & 0.57$\pm$0.06  & 2.71$\pm$0.23  \\
\hline
Total Background        & 876.5$\pm$53.6   & 359.0$\pm$24.0   & 80.2$\pm$5.4   & 45.3$\pm$3.0   & 124.6$\pm$8.2   \\
Corrected Background        & --           & --           & \multicolumn{2}{c}{79.5$\pm$5.3} & 79.5$\pm$5.3 \\
$\ttbar$ Expectation ($\sigma=$6.70) & 2.60$\pm$0.33 & 23.5$\pm$1.8   & 50.1$\pm$3.6   & 74.2$\pm$6.5   & 124.3$\pm$9.1   \\
\hline
Total Background + $\ttbar$     & 879.1$\pm$53.6 & 382.5$\pm$24.1 & \multicolumn{2}{c}{203.9$\pm$10.6} & 203.9$\pm$10.6 \\
\hline
  Tagged events         & 892          & 384          & 142            & 106        & 248         \\ \hline
\hline
\end{tabular}
\renewcommand{\arraystretch}{1.00}
\end{footnotesize}
\caption[Number of tagged events and the background summary.]{Number
of tagged events and the background summary.  The uncertainty on the
total background is not a simple sum in quadrature of the individual
backgrounds because of the correlation between the mistag,
$W+$heavy-flavor, and QCD background predictions.}
\label{tab:BkgSum}
\end{center}
%\end{adjustwidth}
\end{table*}

The cross section is calculated as
\begin{equation}
\sigma_{\ttbar}=\frac{N_{obs}-N_{bkg}}{A_{\ttbar}\cdot
\epsilon_{\ttbar}\cdot\int L dt}, \label{eq:xsec}
\end{equation}
where $N_{obs}$ is the number of events with $\geq3$~jets in which
at least one jet has an $\sltmu$ tag, $N_{bkg}$ is the corrected
background, $A_{\ttbar}$ and $\epsilon_{\ttbar}$ are the
$\ttbar$~event acceptance and tagging efficiency, and $\int L dt$ is
the integrated luminosity.  The acceptance and efficiency are
discussed in Section~\ref{sec:AccEff}, and summarized for the signal
region in Tab.~\ref{tab:Acc_wHt}. We measure a total $\ttbar$ cross
section of
\begin{equation}
\sigma(\ppbar\rightarrow\ttbar
X)=9.1\pm1.1^{+1.0}_{-0.9}\pm0.6~\mathrm{pb}, \label{eq:xsres}
\end{equation}
where the first uncertainty is statistical, the second is
systematic, and the third is from the luminosity.  This cross
section value uses acceptances and tagging efficiencies appropriate
for a top mass of $175~\GeVcc$.  The acceptances and efficiencies,
and therefore the calculated cross section, change slightly for
other assumed top masses.  The calculated cross section is 3\%
higher assuming a top mass of $170~\GeVcc$, and 4\% lower assuming a
top mass of $180~\GeVcc$.  As a check we also measure the cross
section separately for events in which the primary lepton is an
electron and in which it is a muon.  We measure $9.5\pm1.2$~pb when
the primary lepton is an electron and $8.5\pm1.2$~pb when it is a
muon. The uncertainties in both cases are statistical only.
\begin{table*}[h]
\begin{center}
\begin{tabular}{l c c c}
\hline \hline
            & Electrons     & CMUP Muons & CMX Muons \\ \hline
Acc. no Tag (\%)  & $3.71\pm0.01\pm 0.21$
                    & $2.05\pm0.01\pm 0.14$
                            & $0.946\pm0.004\pm 0.050$ \\ \hline
Event Tagging Eff. (\%) & $14.02\pm0.08\pm0.72$
                    & $13.07\pm0.10\pm0.67$
                            & $13.38\pm0.16\pm0.68$ \\ \hline
Acc. with Tag (\%)  & $0.520\pm0.003\pm0.039$
                    & $0.268\pm0.002\pm0.022$
                            & $0.127\pm0.002\pm0.009$ \\ \hline
Luminosity~(pb$^{-1}$)  & $2033.6\pm119.6$ & $2033.6\pm119.6$ &
$1992.5\pm117.2$ \\ \hline Denominator~(pb$^{-1}$) &
$10.58\pm0.07\pm0.80\pm0.62$  &
\multicolumn{2}{c}{$7.97\pm0.06\pm0.49\pm0.47$} \\ \hline Total
denominator~(pb$^{-1}$) & \multicolumn{3}{c}{
$18.56\pm0.09$(stat.)$\pm0.94$(sys.)$\pm1.09$(lum.)} \\ \hline
\hline
\end{tabular}
\caption{Summary of components of the denominator for the cross
section calculation.  The $\ttbar$ acceptance and tagging efficiency
for 3-or-more-jets events is determined using $\Pythia$ Monte
Carlo.} \label{tab:Acc_wHt}
\end{center}
\end{table*}

\begin{figure}[htbp]
\begin{center}
\includegraphics[width=3.375in]{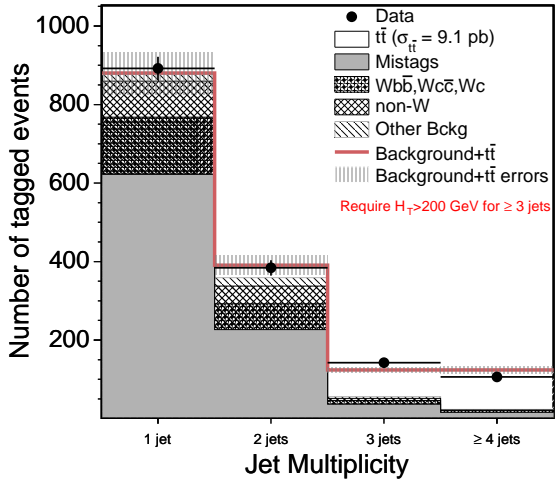}
\caption{The expected background and observed tags in $W+$1,~2,~3,
and 4-or-more jets events.  The expected $\ttbar$ contribution is
normalized to the measured cross section.} \label{fig:njets}
\end{center}
\end{figure}

Figure~\ref{fig:njets} shows, in bins of the number of jets in
$W+$jets candidates, the expected number of tagged background and
$\ttbar$ (normalized to the measured cross section) events together
with the number of observed $\sltmu$ tags.

\begin{figure}[htbp]
\begin{center}
\includegraphics[width=3.375in]{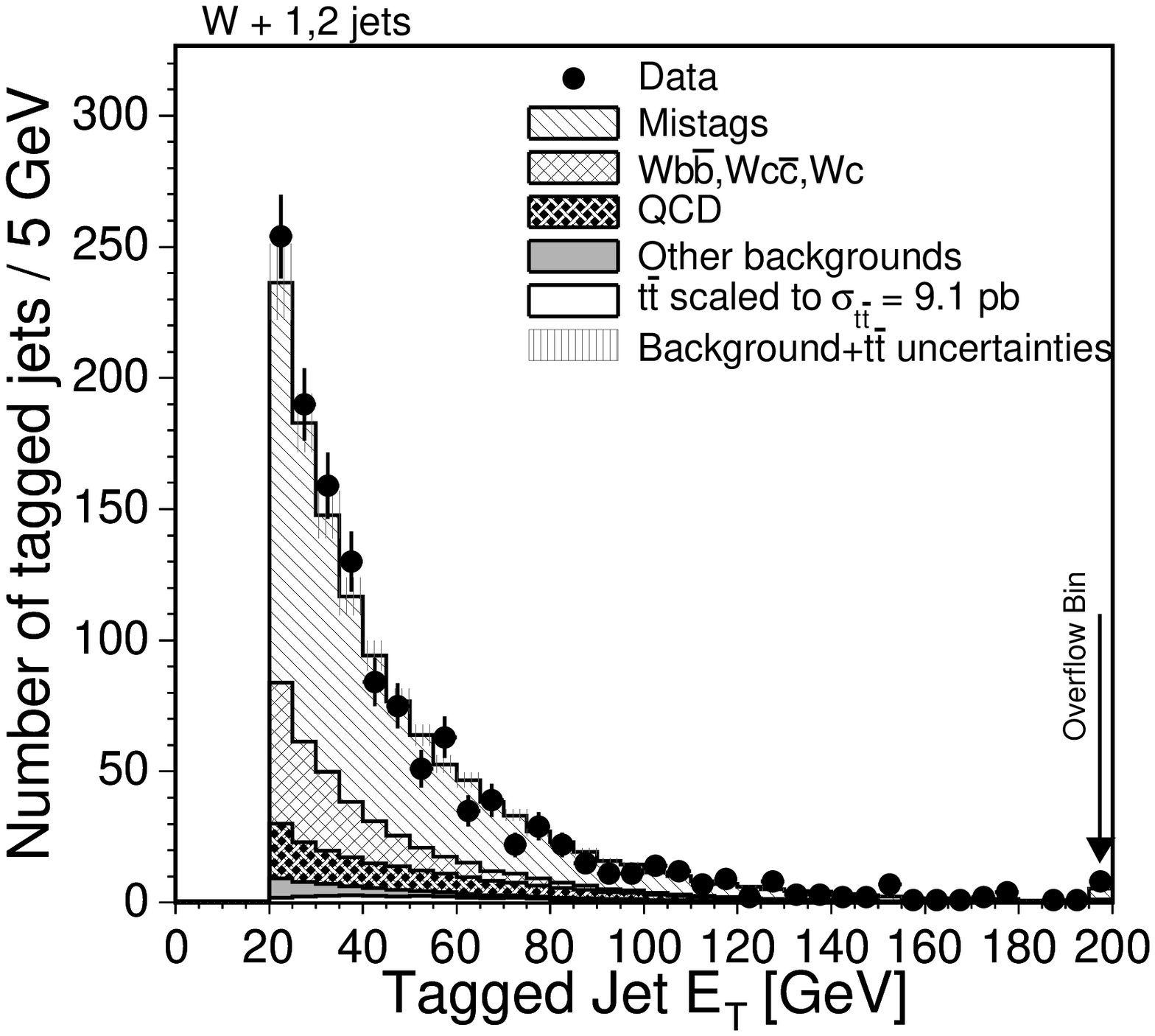}
\includegraphics[width=3.375in]{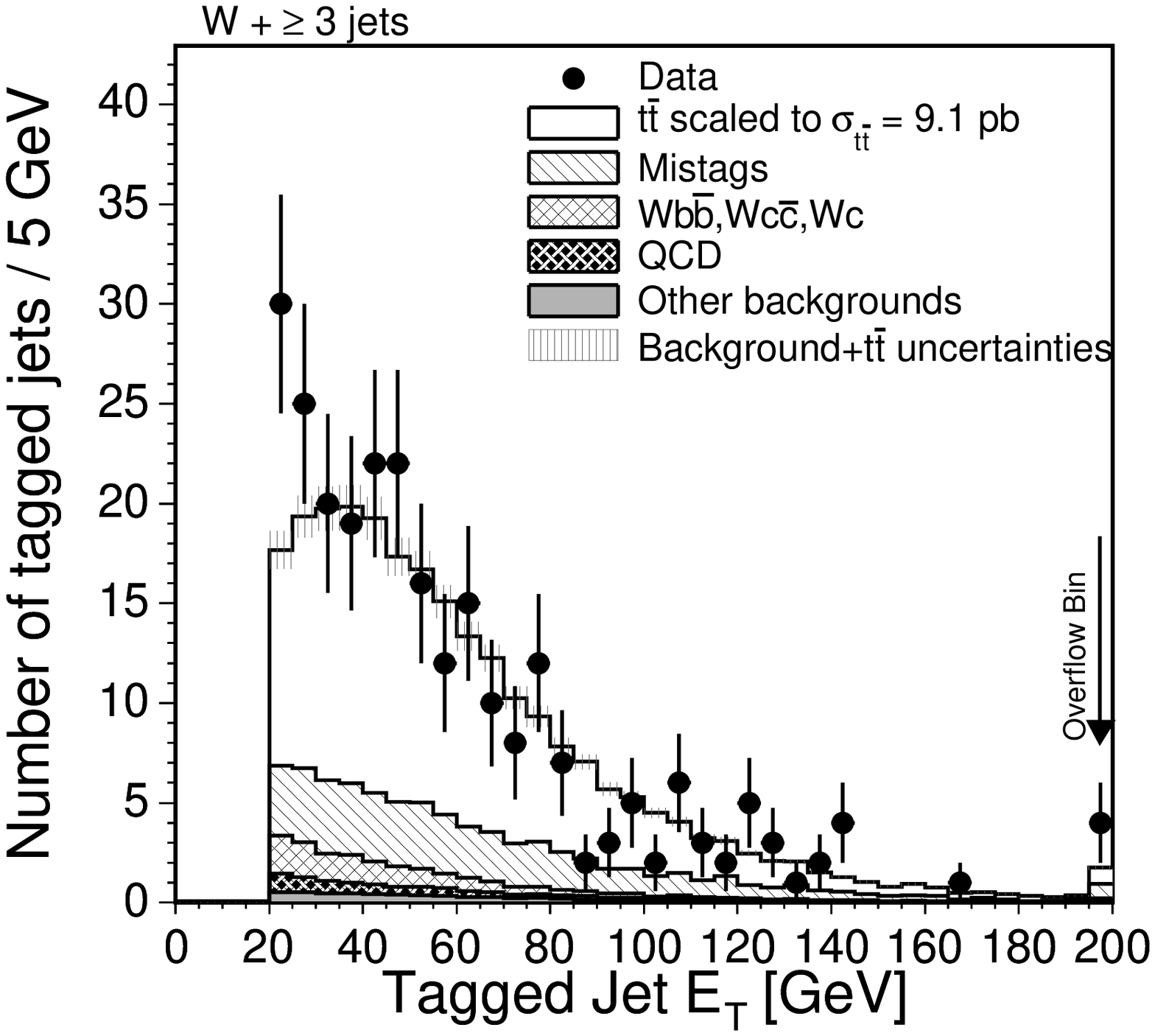}
\caption{Comparison of the jet $E_T$ distributions for tagged jets
and for expectations from mistags, $W+$heavy-flavor, QCD and
$\ttbar$ events. The upper plot is for $W+$1-~and 2-jet events and
the lower plot for $W+$3-or-more-jets events.} \label{fig:jetet}
\end{center}
\end{figure}

\begin{figure}[htbp]
\begin{center}
\includegraphics[width=3.375in]{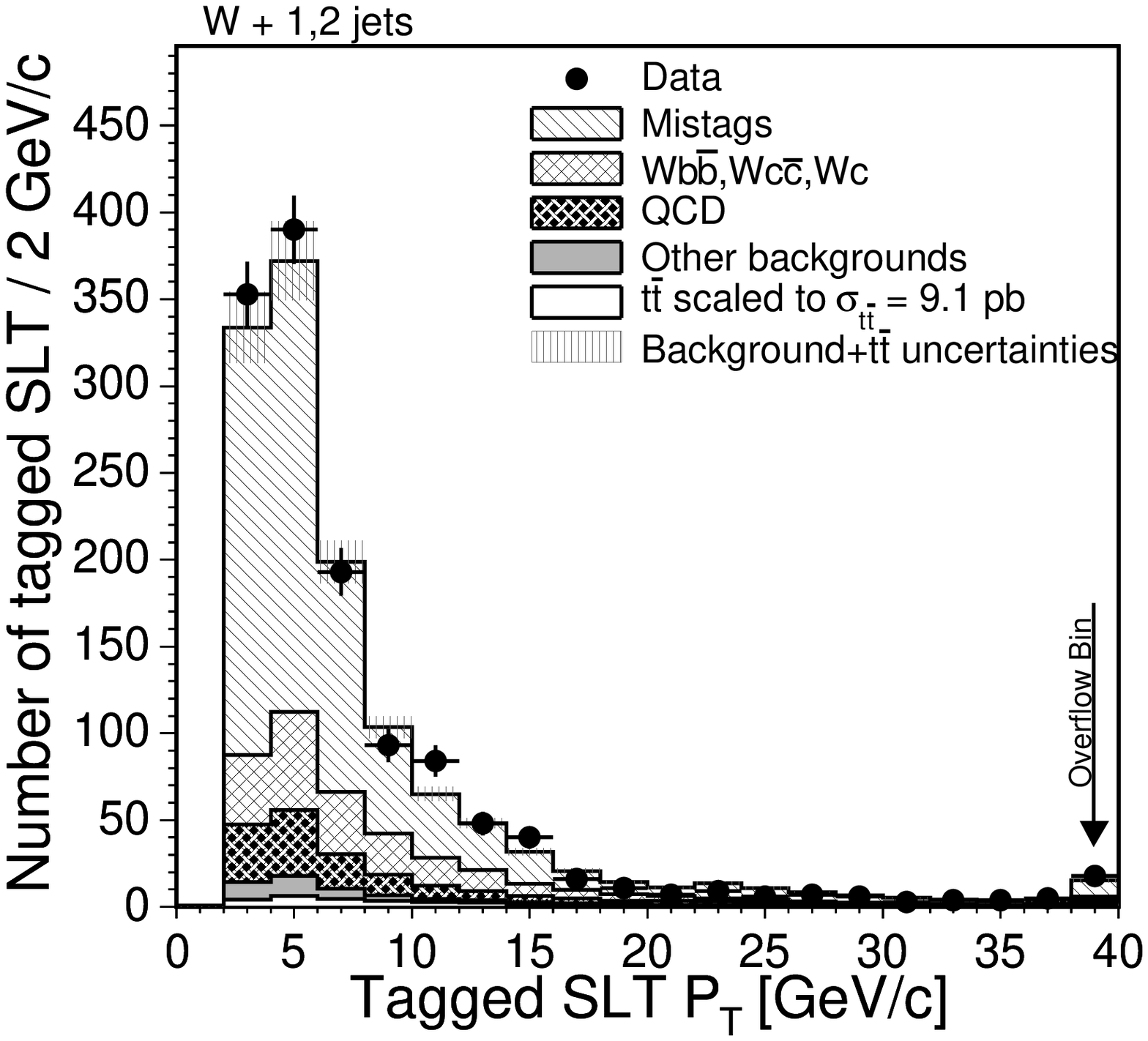}
\includegraphics[width=3.375in]{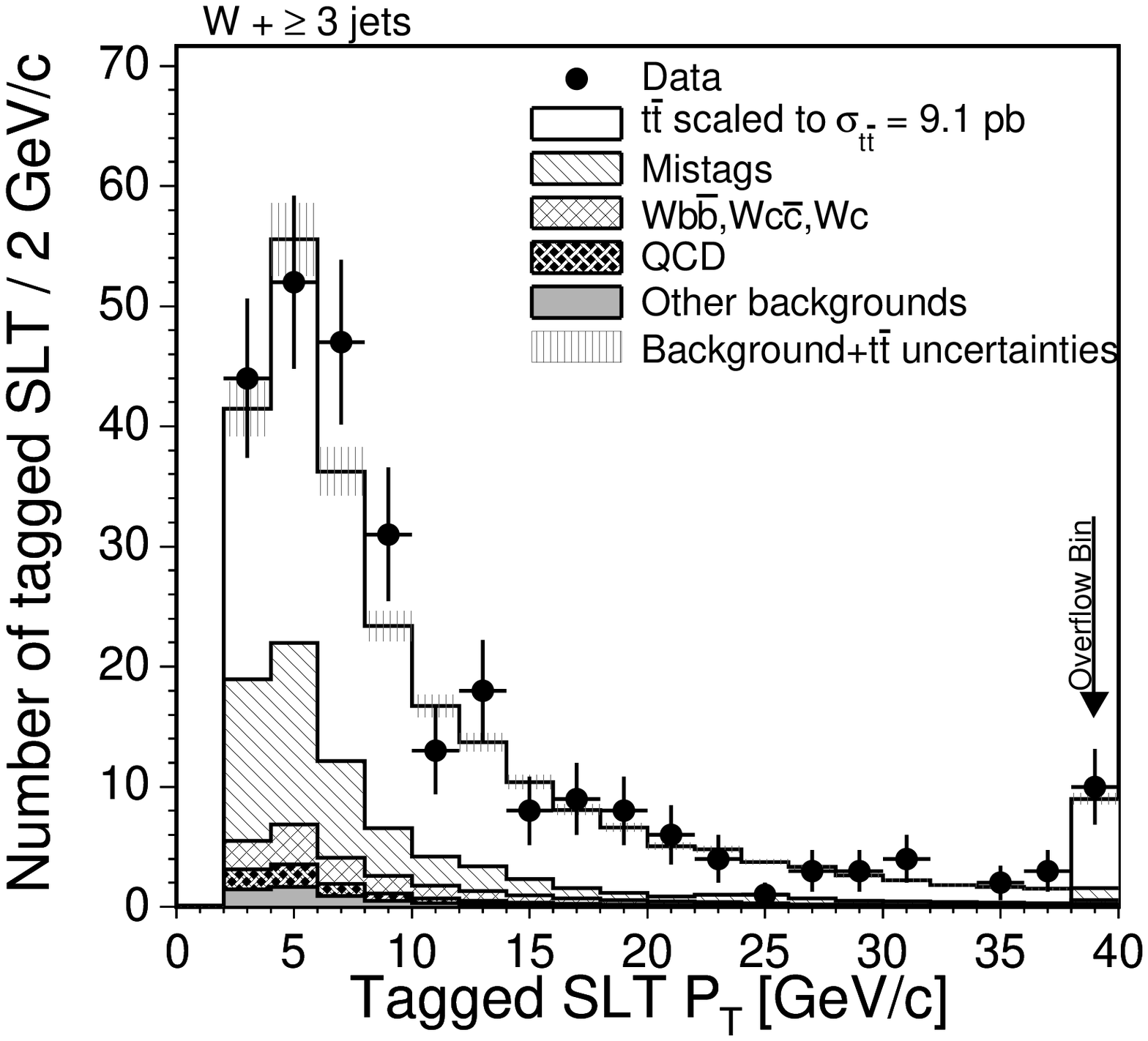}
\caption{$\Pt$ of the $\sltmu$ tags compared with expectations from
backgrounds and $\ttbar$. The upper plot is for $W+$1-~and 2-jet
events and the lower plot for $W+\geq3$-jet events.}
\label{fig:tagPt}
\end{center}
\end{figure}

\begin{figure}[htbp]
\begin{center}
\includegraphics[width=3.375in]{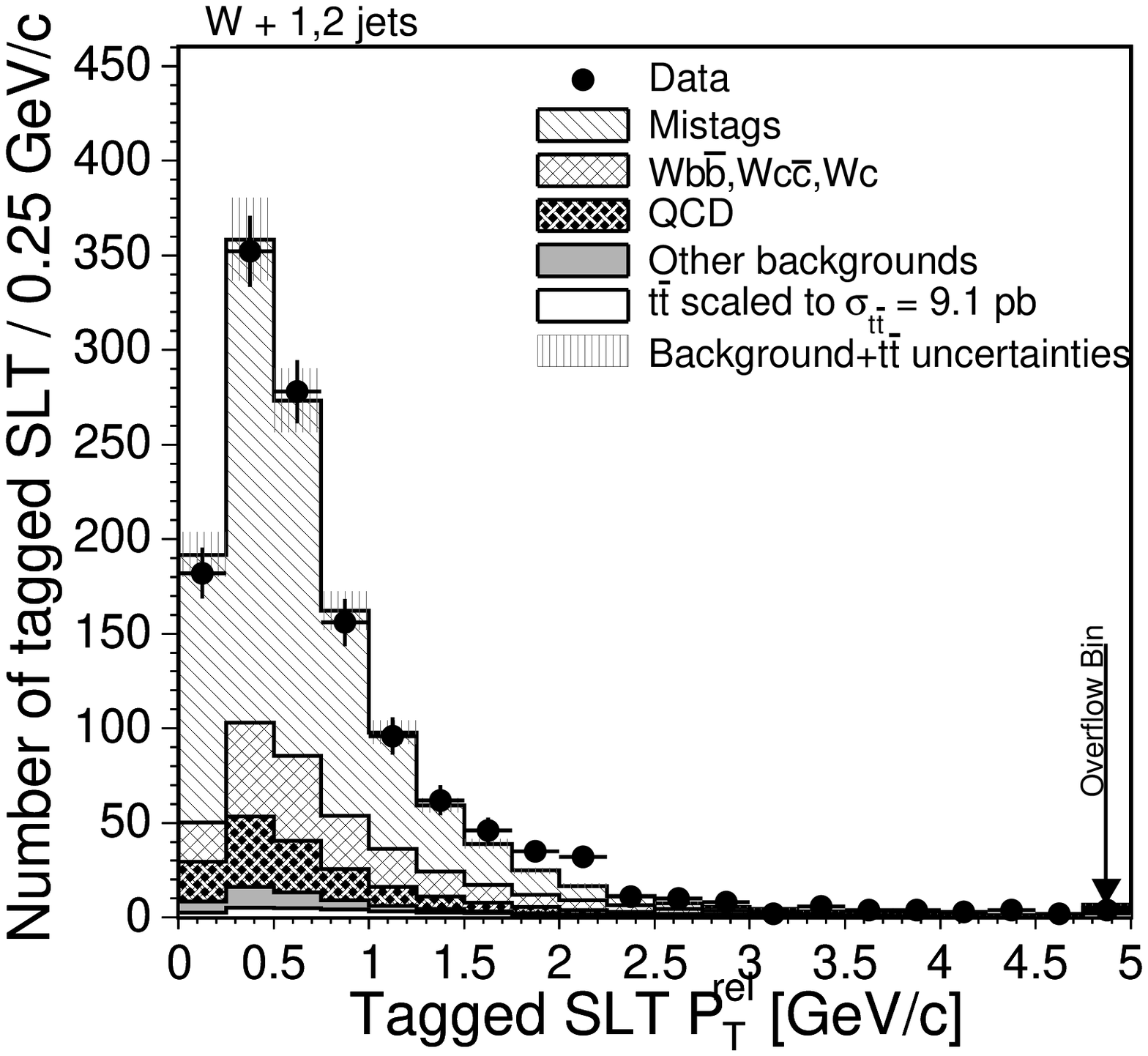}
\includegraphics[width=3.375in]{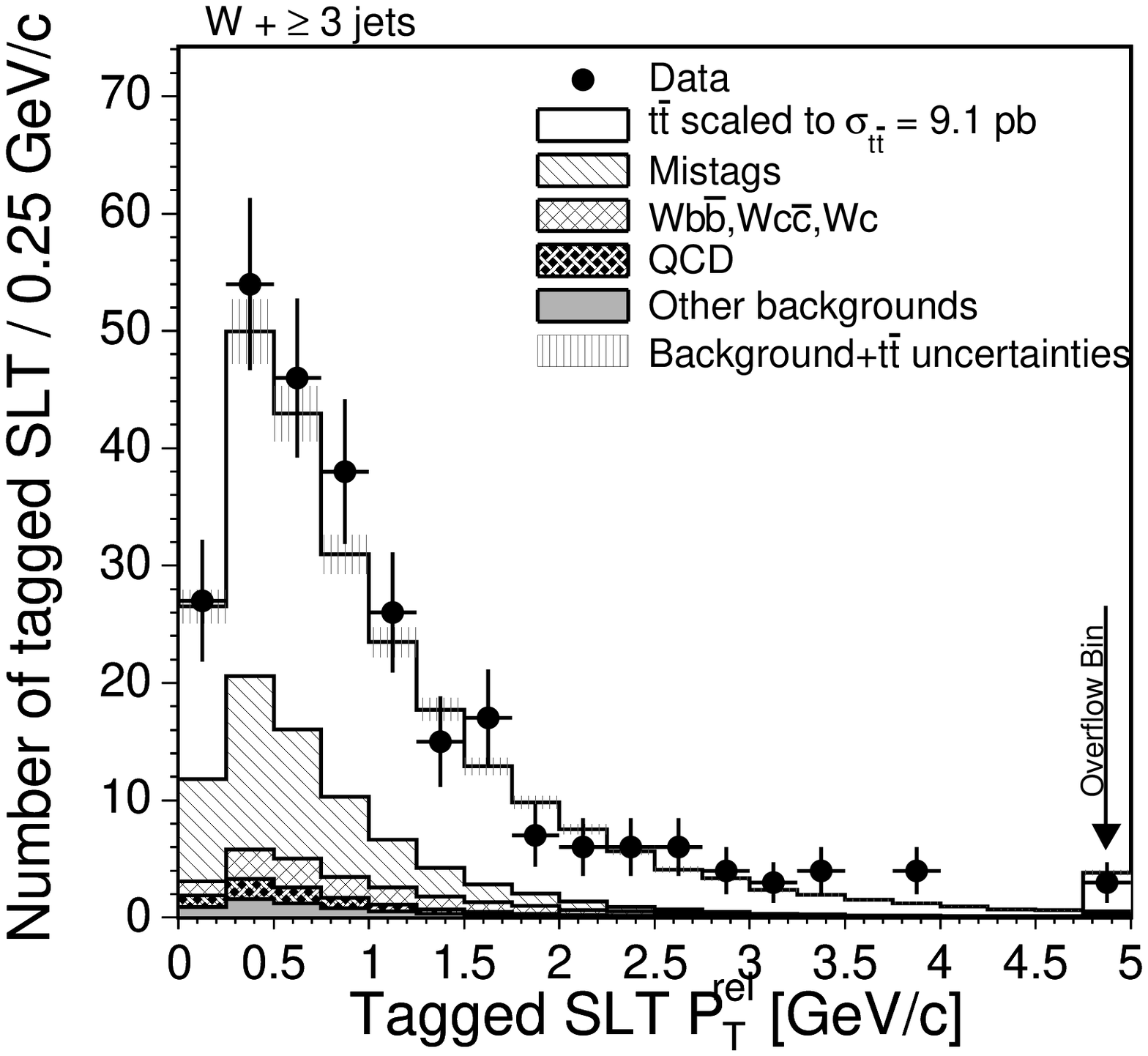}
\caption{The distribution of $\Pt$ relative to the jet axis
(${\Pt}^{rel}$) for tags in data, compared with expectations from
backgrounds plus $\ttbar$. The upper plot is for 1-~and 2-jet events
and the lower plot for $\geq3$-jet events.} \label{fig:sltPtRel}
\end{center}
\end{figure}

In Fig.~\ref{fig:jetet} through~\ref{fig:sltPtRel} we examine a few
kinematic features of the tagged events.  In each case the data are
compared to the expected backgrounds plus $\ttbar$, normalized to
the measured cross section.  The agreement between data and
expectation is good.  The only slight exceptions are a few bins at
low $\Et$ in the $W+\geq3$ jet events in Fig.~\ref{fig:jetet}, where
the number of observed tags exceeds somewhat the expectation. This
is consistent with the excess seen in the low $\Et$ jet data in
Tab.~\ref{tab:deltaetc}, which is folded into the systematic
uncertainty on the measurement.

\section{Conclusions}\label{sec:conc}

 Using 2~fb$^{-1}$ of integrated luminosity collected by the CDF II
detector, we have measured the total cross section for $\ttbar$
production in $\ppbar$ collisions with a center-of-mass energy,
$\sqrt{s}=1.96~\TeV$.  The measurement begins by selecting a dataset
of $W+$jets candidates.  We separate signal from background by
identifying candidate semileptonic decays of $b$-hadrons into muons.
This technique was first published in Reference~\cite{SLTPRD}.  This
measurement is an update that uses ten times the amount of data of
the previous measurement and a new technique for evaluating the
dominant background (see Section~\ref{sec:FakeMatrix}) of
misidentifying a jet from a light-flavor quark as one containing a
$b$-hadron. The measured $\ttbar$ cross section is
\begin{equation}
\sigma(\ppbar\rightarrow\ttbar
X)=9.1\pm1.1^{+1.0}_{-0.9}\pm0.6~\mathrm{pb}, \label{eq:xsconc}
\end{equation}
consistent with the expectation of $6.7^{+0.7}_{-0.9}$~pb for
standard model production and decay of top quark pairs with a mass
of $175~\GeVcc$.  The measurement agrees well with other CDF
measurements of the $\ttbar$ production cross
section~\cite{PubPage}. Assuming the cross section increases 0.2~pb
for every 1~$\GeVcc$ decrease in the top mass, then at the world
average top~mass of $172.4~\GeVcc$ the theoretical cross section is
approximately 7.2~pb.  Using a linear fit to the mass dependence,
the measured cross section was estimated at the world average
top~mass and is found to be $8.9\pm1.6$~pb. The kinematic
distributions of the tagged sample are also consistent with standard
model expectations. The observed number of tags in $W+1$-~and 2-jet
events is in excellent agreement with expectations from background,
indicating that the backgrounds are well understood.

\section{Acknowledgements}
We thank the Fermilab staff and the technical staffs of the
participating institutions for their vital contributions. This work
was supported by the U.S. Department of Energy and National Science
Foundation; the Italian Istituto Nazionale di Fisica Nucleare; the
Ministry of Education, Culture, Sports, Science and Technology of
Japan; the Natural Sciences and Engineering Research Council of
Canada; the National Science Council of the Republic of China; the
Swiss National Science Foundation; the A.P. Sloan Foundation; the
Bundesministerium f\"ur Bildung und Forschung, Germany; the Korean
Science and Engineering Foundation and the Korean Research
Foundation; the Science and Technology Facilities Council and the
Royal Society, UK; the Institut National de Physique Nucleaire et
Physique des Particules/CNRS; the Russian Foundation for Basic
Research; the Ministerio de Ciencia e Innovaci\'{o}n, and Programa
Consolider-Ingenio 2010, Spain; the Slovak R\&D Agency; and the
Academy of Finland.


\begin{thebibliography}{99}
    \bibitem{theory}
    M. Cacciari {\it et al.}, J. High Energy Phys. 0809, 127 (2008); N. Kidonakis and R. Vogt,
    Phys. Rev. D {\bf78} 074005 (2008); S. Moch and P. Uwer, Nucl. Phys. Proc. Suppl. {\bf 183} 75,
    (2008).\\

    \bibitem{MET} We use a ($z,\phi, \theta$) coordinate system where the $z$-axis
    is in the direction of the proton beam, and $\phi$ and $\theta$ are the azimuthal
    and polar angles respectively.  The pseudo-rapidity, $\eta$, is defined as
    $-\ln(\tan\frac{\theta}{2})$.  The transverse momentum of a charged particle is
    $\Pt = P\sin\theta$, where $P$ represents the measured momentum of the charged-particle
    track.  The analogous quantity using calorimeter energies, defined as $E_T = E\sin\theta$
    is called transverse energy. The missing transverse energy is defined as
    $\met = -\mid\sum_i E_T^i\hat{n}_i\mid$ where $E_T^i$ is the
    magnitude of the transverse energy contained in each calorimeter
    tower $i$ in the pseudo-rapidity region $\mid\eta\mid <3.6$ and
    $\hat{n}_i$ is the direction unit vector of the tower in the
    plane transverse to the beam direction.

      \bibitem{SecVtxPRD} A. Abulencia {\it et al.} (CDF Collaboration), Phys. Rev. Lett. {\bf
  97}, 082004 (2006); D. Acosta {\it et al.} (CDF Collaboration), Phys. Rev. D {\bf 71}, 052003
  (2005).

    \bibitem{SLTPRD} D. Acosta {\it et al.} (CDF Collaboration), Phys. Rev. D {\bf 72}, 032002
    (2005)

    \bibitem{UlyThesis} Ulysses Allen Grundler, ``A Measurement of the
    $\ttbar$ Production Cross Section in $\ppbar$ Collisions at
    $\sqrt{s}=1.96$ TeV Using Soft Muon Tagging", Ph.D Thesis,
    University of Illinois at Urbana-Champaign, 2008.
    FERMILAB-THESIS-2008-27.

    \bibitem{CDF}
    The CDF II Detector Technical Design Report,
    Fermilab-Pub-96/390-E;
    D. Acosta {\it et al.}, Phys. Rev. D {\bf 71}, 052003 (2005).

    \bibitem{klimenko} S. Klimenko, J. Konigsberg and T.M. Liss,
    Fermilab-FN-0741.

    \bibitem{Pythia}
    T. Sjostrand {\it et al.}, Comput. Phys. Commun. {\bf 135}, 238 (2001).

      \bibitem{Herwig}
   G. Corcella {\it et al.}, J. High Energy Phys. 01, 10 (2001).

  \bibitem{cteq5l}
  H. L. Lai {\it et al.}, Eur. Phys. J. C {\bf
                  12}, 375 (2000).

  \bibitem{evtgen}
  D. J. Lange {\it et al.},
  Nucl. Instrum. and Methods Phys. Res., Sect. A {\bf 462}, 152, (2001).

  \bibitem{Sherman} Daniel Sherman, ``Measurement of the Top Quark Pair Production Cross Section with 1.12
fb-1 of pp Collisions at $\sqrt{s}$ = 1.96 TeV", Ph.D. Thesis,
Harvard University.

  \bibitem{ALPGEN}
  M. Mangano {\it et al.}, J. High Energy Phys. 7, 1, (2003).

  \bibitem{MadEvent}
  F. Maltoni and T. Stelzer,  J. High Energy Phys. 02, 27 (2003).

  \bibitem{sim} E. Gerchtein and M. Paulini, ECONF {\bf C0303241},
    TUMT005 (2003), arXiv:physics/0306031.

    \bibitem{Chen:2003qe} D. Acosta {\it et al.} (CDF Collaboration), Phys. Rev. Lett. {\bf
    91}, 241804 (2003); A. Abulencia {\it et al.} (CDF Collaboration), Phys. Rev. D {\bf 74},
    031109 (2006).

    \bibitem{lambda} A. Abulencia {\it et al.} (CDF Collaboration), arXiv:hep-ex/0609021v1
    submitted to Phys. Rev. Lett.

  \bibitem{JetCorr} A. Bhatti {\it et al.}, Nucl. Instrum.
  and Methods Phys. Res., Sect. A {\bf 566}, 375 (2006).

  \bibitem{Campbell:1999ah} J. M.Campbell and R. K. Ellis, Phys.
  Rev. D {\bf 60}, 113006 (1999).\\
  The ZZ cross section of 3.4 pb is appropriate for the $\Pythia$ Monte Carlo sample we use that
  includes events with one off-shell Z.  The value
is arrived at by normalizing the $\Pythia$ sample to the Campbell
and Ellis cross section of 1.58 pb for two on-shell Z bosons.

  \bibitem{Acosta:2004uq} D. Acosta {\it et al.} (CDF Collaboration), Phys. Rev. Lett. {\bf
  94}, 091803 (2005).

  \bibitem{Sullivan:2004ie} Z. Sullivan, Phys. Rev. D {\bf 70},
  114012 (2004).

  \bibitem{mrst98} A.D. Martin, R.G. Roberts, W.J. Stirling and
R.S. Thorne, Eur. Phys. J. C {\bf 4} 463 (1998).

  \bibitem{PubPage} The CDF Collaboration, CDF Conference note 9448
  (2008);\\ \verb"http://www-cdf.fnal.gov/physics/new/top/confNotes/"
\end{thebibliography}
\end{document}